\documentclass[11pt]{article}
\usepackage{amsfonts,amssymb,amsmath,numbysec}
\usepackage{pdfsync,mathrsfs,fge}

\usepackage{graphics,graphicx,fancyhdr,color}
\usepackage{mathtools}

\hsize \textwidth
\advance \hsize by -\marginparwidth
\evensidemargin \oddsidemargin
\usepackage{amssymb}
\advance\hoffset by 5mm
\newcommand{\proof}{{\em Proof. }}
\newcommand{\qed}{$\Box$ }
\newcommand{\bbZ}{{\mathbb Z}}
\newcommand{\bbP}{{\mathbb P}}
\newcommand{\bbR}{{\mathbb R}}

\newcommand{\bbT}{{\mathbb T}}
\newcommand{\eps}{\varepsilon}

\newcommand{\si}{\sigma}
\newcommand{\Om}{\Omega}
\newcommand{\om}{\omega}
\newcommand{\omm}{\omega_{\rm min}}
\newcommand{\omM}{\omega_{\rm max}}

\newcommand{\ga}{\gamma}
\newcommand{\la}{\lambda}
\newcommand{\al}{\alpha}
\newcommand{\bbE}{\mathbb E}
\newcommand{\farc}{\frac}

\newtheorem{df}{Definition}[section]
\newtheorem{lm}[df]{Lemma}%[section]
\newtheorem{lemma}[df]{Lemma}
\newtheorem{thm}[df]{Theorem}%[section]
\newcommand{\fg}{\fgeeszett}

\newcommand{\red}{\textcolor{red}}
\newcommand{\blue}{\textcolor{blue}}
\newcommand{\ma}{\textcolor{magenta}}

\newcommand{\commentout}[1]{{}}

\makeatother
\begin{document}

\numberbysection

\title{High frequency limit  for a chain of harmonic oscillators with a point Langevin thermostat}

\author{Tomasz Komorowski\thanks{Institute of Mathematics, Polish Academy Of Sciences,
ul. \'{S}niadeckich 8,   00-956 Warsaw, Poland, 
e-mail: {\tt komorow@hektor.umcs.lublin.pl}}
\and
Stefano Olla\thanks{CEREMADE, UMR-CNRS, Universit\'e de Paris Dauphine, PSL Research University
{ Place du Mar\'echal De Lattre De Tassigny, 75016 Paris, France},
e-mail:{ \tt olla@ceremade.dauphine.fr}}
\and
 Lenya Ryzhik\thanks{{ Mathematics Department, Stanford University, Stanford, CA 94305, USA} ,
{email: {\tt ryzhik@stanford.edu}}}
\and
Herbert Spohn\thanks{{ Zentrum Mathematik and Physik Department, Technische Universit\"at Munchen,
{\footnotesize Boltzmannstrasse 3, 85747 Munich, Germany}} ,
{email:{ \tt spohn@ma.tum.de}}}
}

\maketitle

\begin{abstract}
We consider an infinite  chain of coupled harmonic oscillators with a Langevin thermostat at the origin.
In the high frequency limit, we establish the reflection-transmission coefficients for the wave energy for the
scattering of the thermostat. 
To our surprise, even though the thermostat fluctuations are time-dependent,
the scattering does not couple wave energy at various frequencies.
 \end{abstract}

\section{Introduction}
\label{intro}

Heat reservoirs with some given temperature $T$ are usually modelled at the microscopic level by 
the Langevin stochastic dynamics, or by other random mechanisms such as the renewal of velocities  at random times 
with Gaussian distributed velocities of variance $T$. This latter mechanism represents the interaction with an infinitely extended reservoir 
of independent particles in equilibrium at temperature $T$ and uniform density. 

{When such reservoirs are in contact with the system boundary and 
if energy diffuses on the macroscopic space-time scale, 
then it is expected that a thermostat enforces a local equilibrium at the boundary
at the temperature $T$.
The situation is much less clear for kinetic (hyperbolic) space-time scales. 
For instance, if the bulk evolution is governed by a discrete 
nonlinear wave equation, then in the kinetic (high frequency) limit the wave number density is governed by a phonon Boltzmann equation
 \cite{spohn-ph,BOS}.
If this system is coupled to a thermostat at the boundary, 
what are the appropriate macroscopic boundary 
conditions which have to be added to the kinetic equation?}

To make a study feasible, we very much simplify the set-up. We consider an infinite one-dimensional chain of harmonic oscillators, 
characterized by its dispersion relation $\omega(k)$, and couple it with a single Langevin thermostat at the origin. 
An efficient way to localize the distribution of the energy at 
wave number $k$ is to use the Wigner distribution. 
In
a space-time hyperbolic rescaling,  first ignoring the thermostat, the Wigner distribution converges 
to the solution $W(t,x,k)$ of a simple transport equation, namely
phonons of wavenumber $k$ have energy $\omega(k)$ and travel independently with group velocity $\omega'(k)/2\pi$. 
It will be proved that {when the dispersion relation is unimodal,
see Section \ref{sec:dynamics-the-main} for a precise definition}, 
in the scaling limit, the thermostat enforces the following reflection-transmission (and production) conditions at $x=0$:
phonons of wave number $k$  are generated with rate $ \fgeeszett(k)T$ 
and an incoming $k$-phonon is transmitted with probability $p_+(k)$,
reflected with probability $p_-(k)$, and absorbed with probability $\fgeeszett(k)$, see formulas  \eqref{033110} below. 
These coefficients are positive, depend on $\omega(\cdot)$, and satisfy 
$$
 p_+(k) + p_-(k) +\fgeeszett(k) = 1.
$$ 
With such boundary conditions the {stationary} solution of the transport equation is
the thermal equilibrium Wigner function $W(t,x,k) = T$.

The thermostat can be viewed  as a  ``scatterer" of a
time-varying strength: at the microscopic scale  a 
wave incident on the thermostat would produce
reflected and transmitted waves at all frequencies. 
It is remarkable that, after the scaling limit, the reflected and transmitted waves 
are of the same frequency as the incident wave, all other waves produced 
by the microscopic scattering are damped by oscillations in the macroscopic limit. The presence of oscillatory integrals, responsible for the damping mechanism, presents the main mathematical difficulty
of the model.
 To deal with the issue we consider the high frequency limit of the Laplace transform of the Wigner distribution.
The limit, see \eqref{010304} below, can be decomposed into the parts that correspond to the production, transmission and reflection of a phonon. The calculation of the production term is relatively straightforward, see Section \ref{sec:phonon-creation-term}. In contrast, the computations related to the scattering terms are remarkably difficult, see Sections \ref{sec:scatter}--\ref{w-s} for the proof. Moreover, the description of the limit is not intuitive and it is not clear to us how to obtain it by a simple heuristic argument.

{The multimodal case, that we shall not consider here,
can be also dealt with using the technique of  the present paper. In this situation
 the level set of   $\om(k)$ has generically $2N$
points (we assume that~$\om$ is even) for some positive integer $N$. 
The macroscopic description of the system is as follows: 
a $k$-phonon arriving at interface with group velocity $\om'(k)>0$
is transmitted  as a $k'$-phonon corresponding to the
solutions of $\om(k')=\om(k)$, with a positive group velocity.
 The  probabilities of transmission at a given $k'$ can be computed explicitly in
terms of the dispertion relation.
 On the other hand, it reflects as a $k'$-phonon corresponding to a  solution of $\om(k')=\om(k)$, with a
 group velocity $\omega'(k')< 0$. 
The probability of  absorption is the same as in the unimodal case.}

{We expect that introducing a rarefied random scattering in the bulk, in the same fashion as in \cite{BOS}, 
leads to a similar transport equation with a linear scattering term, 
without {modifying} the transmission properties at the interface with the thermostat \cite{KORS2}.
}

% The organization of the paper is as follows: the formulation of the model and our main result, see Theorem \ref{main:thm} below, are contained in Section \ref{sec:dynamics-the-main}. In Section \ref{sec:wave-fun} we compute the wave function that corresponds to the chain of oscillators. The proof of  the main result is contained in Section \ref{w-s}, preceded by the proofs of auxiliary results contained in Sections \ref{sec:phonon-creation-term}--\ref{sec:lem4.2}. Finally in  Section  \ref{sec8.3} we present the proof of identity \eqref{feb1402} satisfied by the coefficient describing the production rate of phonons.

There are rather few results on the high frequency limits of the Wigner transform in the presence of boundaries, interfaces
or sources. We mention~\cite{BCKP,CPR,Ger-Le,KPR,Miller} which, while highly non-trivial,
are all ultimately based on essentially explicit computations of the
Wigner transform near the interface. Our analysis also starts with computing the Wigner transform, but then passes to the limit
in the resulting expression. % An unusual aspect is that the passage to the limit is done in the Laplace transform 
% of the Wigner distribution, which explains the surprising appearance of the function $J(t)$ 
% and its Laplace transform in the expressions for the reflection and transmission coefficients. 
The thermal production of phonons can be seen quite straightforwardly in this limit. 
However the scattering terms are much more difficult to 
handle and they constitute the major part of our work. 
\noindent\medskip\\
{
{\bf Acknowledgement}. TK was partially supported by the NCN grant 2016/23/B/ST1/00492, 
SO by the French Agence Nationale Recherche grant LSD ANR-15-CE40-0020-01, and
LR  by an NSF grant DMS-1613603 and by ONR. 
This work was partially supported by the grant 346300 for IMPAN from the Simons Foundation 
and the matching 2015-2019 Polish MNiSW fund. 
The authors thank Stanford University and Universit\'e Paris Dauphine for hospitality during the preparation of this article. 
}

\section{The dynamics and the main result}
\label{sec:dynamics-the-main}

\subsection*{The infinite chain of harmonic oscillators}
%Let $\ell^2$ be the complex Hilbert space consisting of all square integrable sequences $\{\gamma(y),\,y\in\bbZ\}$  with the norm given by $\|\gamma\|_{\ell^2}^2=\sum_{y\in\bbZ}|\gamma(y)|^2$.  

%\subsubsection*{The background Hamiltonian system}

% The dynamics of the system of oscillators  can be written formally  as  a Hamiltonian system of differential equations 
% \begin{eqnarray}
% &&\dot {\frak q}_{y}(t)=\partial_{\frak p_y}{\cal H}({\frak p}(t),{\frak q}(t))
% \label{eq:bas}\\
% &&\nonumber\\
% && \dot {\frak p}_y(t)=-\partial_{\frak q_y}{\cal H}({\frak p}(t),{\frak q}(t)),\quad y\in\bbZ.\nonumber
% \end{eqnarray}
We consider the evolution of an infinite particle system governed by the Hamiltonian 
\begin{equation}
\label{011210}
{\cal H}({\frak p},{\frak q}):=\frac12\sum_{y\in\bbZ}{\frak p}_y^2+\frac{1}{2}\sum_{y,y'\in\bbZ}\alpha_{y-y'}{\frak q}_y{\frak q}_{y'}.
\end{equation}
Here, the  particle label is $y\in\bbZ$, $({\frak p}_y,{\frak q}_y)$ is the position and momentum of the $y$'s particle, respectively, and $({\frak q},{\frak p})=\{({\frak p}_y,{\frak q}_y),\,y\in\bbZ\}$ denotes the entire configuration. The coupling coefficients $\alpha_{y}$ are assumed to have exponential decay and chosen such that the energy is bounded from below.

A stochastically perturbed version of this system was considered first in~\cite{BOS},  where the long
time behavior of the wave energy was analyzed, and then in~\cite{KOR}, where the wave field itself was studied. 
 %The one dimensional torus $\bbT$  considered here is the interval $[-1/2,1/2]$ with identified endpoints.
%\subsubsection*{Stochastically perturbed dynamics with a Langevin thermostat} 
The stochasticity in~\cite{BOS,KOR} was introduced as a random exchange of momenta between particles at adjacent sites. 
Here, instead of random fluctuations ``in the bulk", we  couple the particle with label 0 to a Langevin thermostat at temperature~$T$ and with friction~$\gamma>0$. 
Then the   
Hamiltonian dynamics with stochastic source  \ma{is} governed by
\begin{eqnarray}
&&\dot{\frak q}_y(t)={\frak p}_y(t),
\label{eq:bas2}\\
%&&\nonumber\\
&& d{\frak p}_y(t) = -(\alpha\star {\frak q}(t))_ydt
+\big(-\ga {\frak p}_0(t)dt+\sqrt{2\ga T}dw(t)\big) \delta_{0,y},\quad y\in\bbZ.\nonumber
\end{eqnarray}
Here, $\{w(t),\,t\ge0\}$ is a standard Wiener process over a probability space $(\Om,{\cal F},\bbP)$. 
We use the notation 
\[
(f\star g)_y=\sum_{y'\in\bbZ}f_{y-y'}g_{y'}
\]
for the convolution of two functions on $\bbZ$.
%Our main interest here is to understand
%the wave scattering off the thermostat, thus, for simplicity we do not consider any randomness in the bulk.
%The initial condition $({\frak q}(0),{\frak p}(0))$ is assumed to belong to the Hilbert space $\ell^2(\bbZ)$. 

%\subsubsection*{The wave-function}

It is convenient to introduce the complex wave function 
\begin{equation}
\label{011307}
\psi_y(t) := (\tilde{\om} \star{\frak q}(t))_ y + i{\frak p_y}(t)
\end{equation}
%where  $({\frak q}(t),{\frak p}(t))$ satisfies \eqref{eq:bas2}. 
where $\{\tilde \om_y,\,y\in\bbZ\}$ is the inverse
 Fourier transform of the  dispersion relation  
\begin{equation}\label{mar2602}
 \om(k):=\sqrt{\hat \alpha (k)}.
\end{equation}
Hence $|\psi_y(t)|^2$ is the local energy of the chain at time $t$. The Fourier transform of the wave function is given by 
\begin{equation}
\label{011307a}
\hat\psi(t,k) := \om(k) \hat {\frak q}(t,k) + i\hat{\frak p}(t,k),
\end{equation}
so that
$$
\hat{\frak p}\left(t,k\right)=\frac{1}{2i}[\hat\psi(t,k)-\hat\psi^*(t,-k)],
~~{\frak p}_0(t)=\int_{\bbT} {\rm Im}\,\hat\psi(t,k) dk.
$$
Using \eqref{eq:bas2}, it is easy to check that the wave function evolves according to  
\begin{equation}
\begin{split}
 \label{basic:sde:2aa}
 d\hat\psi(t,k) &= \big(-i\om(k)\hat\psi(t,k)
 - i{\ga} {\frak p}_0(t) \big) dt
 +i\sqrt{2\ga T}dw(t).
 %\\
%\hat\psi(0,k) & = \hat\psi(k),
\end{split}
 \end{equation}
Above, the Fourier transform of  $f_x\in l^2(\bbZ)$ and the inverse Fourier transform of $\hat f\in L^2(\bbT)$ are
  \begin{equation}
  \label{fourier}
  \hat f(k)=\sum_{x\in\bbZ} f_x \exp\{-2\pi ixk\}, ~~
%\end{equation}
%The  inverse transform is given by
%\begin{equation}
%\label{inv-fourier}
f_x=\int_{\bbT} \hat f(k) \exp\{2\pi ixk\} dk, \quad x\in \bbZ,~~k\in\bbT.
\end{equation} 
For a function $G(x,k)$, we denote by $\tilde G:\bbR\times\bbZ\to\mathbb C$,  $\hat
G:\bbR\times\bbT\to\mathbb C$ the Fourier transforms of $G$
in the $k$ and $x$ variables, respectively,
$$
\tilde G(x,y):=\int_{\bbT}e^{-2\pi i ky}G(x,k)dk,\quad (x,y)\in\bbR\times\bbZ,\quad \hat G(\eta,k):=\int_{\bbR}e^{-2\pi i \eta x}G(x,k)dx,\quad (\eta,k)\in\bbR\times \bbT.
$$

%for any $\hat f$ belonging to $L^2(\bbT)$ - 
%the space of complex valued, square integrable functions.
 %with $\hat \psi\in L^2(\bbT)$.
%The adjoint equation is
%\begin{equation}
%  \label{sde:2aa*}
%  d\hat\psi^*(t,-k) =\left\{ i\om(k)\hat\psi^*(t,-k) + i{\ga} {\frak p}_0(t) \right\} dt
%  - i\sqrt{2\ga T}dw(t).
%\end{equation}
% We conclude therefore that
% \begin{equation}
%   \label{sde:2p}
%   d\hat\psi^*(t,-k) =\left\{ i\om(k)\hat\psi^*(t,-k) + i{\ga} {\frak p}_0(t) \right\} dt
%   - i\sqrt{2\ga T}dw(t).
% \end{equation}

\subsubsection*{The initial conditions}

For simplicity sake we restrict ourselves to initial configurations of finite energy. 
In addition, we assume that the initial energy density $|\psi_y|^2$ is finite  per unit length on the macroscopic scale $x\sim\eps y$,
where $\eps>0$ is the scaling parameter.
% The chain of harmonic oscillators  corresponds to the microscopic description of the system. 
% We will soon pass to the  spatial coordinates $x\sim\eps y$, on the macroscopic scale of the order $\eps^{-1}$. 
% We will assume that macroscopically the initial energy density $|\psi_y|^2$ is finite  per unit length. 
More precisely, given $\eps>0$, the initial wave function is distributed randomly, independent of the Langevin
noise $w(\cdot)$, according to a% Borel
probability measure $\mu_\eps$ on $\ell^2(\bbZ)$, and
\begin{equation}
\label{0001}
\sup_{\eps\in(0,1)}\sum_{y\in\bbZ}\eps\langle|\psi_y|^2\rangle_{\mu_\eps}=\sup_{\eps\in(0,1)}\eps\langle \|\hat \psi\|^2_{L^2(\bbT)}\rangle_{\mu_\eps}
<\infty,
\end{equation}
where $\langle\cdot\rangle_{\mu_\eps}$ denotes the expectation with respect to $\mu_\eps$. We will also assume that 
\begin{equation}
\label{null}
\langle\hat\psi(k)\hat\psi(\ell) \rangle_{\mu_\eps}=0,\quad k,\ell\in\bbT,
\end{equation}
Condition \eqref{null} can be replaced by $ \langle\hat\psi(k)\hat\psi(\ell) \rangle_{\mu_\eps} \sim 0$, as $\eps \to 0$ at the expense of
some additional calculations that we prefer not to perform in this article.

An additional hypothesis concerning the initial configuration will be stated later on, see formula \eqref{011812aa}.

%Equivalently, we have
% \begin{equation}
%\label{0001a}
%\sup_{\eps\in(0,1)}\eps\langle \|\hat \psi\|^2_{L^2(\bbT)}\rangle_{\eps}<+\infty.
%\end{equation}

\subsubsection*{The Wigner distribution}
%\label{sec:wigner-distribution}

To study the effect of the thermostat, we follow the
evolution of the chain on the macroscopic time scale $t'\sim\eps t$, and consider the
rescaled wave function $\psi^{(\eps)}_y(t)=\psi_y(t/\eps)$. 
A convenient tool to analyse the energy density is the Wigner distribution (or Wigner transform)
defined by its action on a test function 
$G \in {\cal S}(\bbR\times\bbT)$~as
\begin{equation}
\label{wigner}
\langle G,W^{(\eps)}(t)\rangle:=\frac{\eps}{2}\sum_{y,y'\in\bbZ}\bbE_\eps\left[\psi_y^{(\eps)}(t)
\left(\psi_{y'}^{(\eps)}\right)^*(t)\right]\tilde
G^*\left(\eps\frac{y+y'}{2},y-y'\right). 
\end{equation}
Here, $\bbE_\eps$ is the expectation with respect to the product measure $\mu_\eps\otimes\bbP$.

The Fourier transform of the Wigner distribution is
\begin{equation}
  \label{eq:20}
  \widehat{ W}_\varepsilon(t,\eta,k) \ :=\ 
 \frac{ \varepsilon}{2} \bbE_\eps \left[(\hat\psi^{(\eps)})^*( t , k- \frac{\varepsilon\eta}2) 
  \hat\psi^{(\eps)}( t , k +  \frac{\varepsilon\eta}2)\right],\quad (t,\eta,k)\in [0,\infty)\times\bbT_{2/\eps}\times\bbT,
\end{equation}
so that
\begin{equation}
\label{wigner1}
\langle G,W^{(\eps)}(t)\rangle=\int_{\bbT\times\bbR} \widehat{ W}_\varepsilon(t,\eta,k)\hat G^*(\eta,k)d\eta dk,\quad G \in {\cal S}(\bbR\times\bbT).
\end{equation}
We use the notation $\bbT_a=[-a/2,a/2]$ for the torus of size $a>0$, with identified endpoints. 

A straightforward calculation  
shows 
that the macroscopic energy grows at most
linearly in time,
  \begin{equation}
 \label{031709}
 d\|\hat\psi^{(\eps)}(t)\|^2_{L^2(\bbT)}
 = \Big[
 -\frac{\ga}{\eps}[{\frak p}^{(\eps)}_0(t)]^2
 +\frac{2\ga T }{\eps}\Big]dt
  +\sqrt{\frac{2\ga T}{\eps}}{\frak p}^{(\eps)}_0(t)dw(t),
 \end{equation}
with  ${\frak p}^{(\eps)}_0(t):={\frak p}_0(t/\eps)$. Thus, we have a uniform bound
%In conclusion, we obtain.
%\begin{prop}
%\label{prop012409}
%The following estimate holds
\begin{equation}
\label{011703-18}
\sup_{\eps\in(0,1]}\eps\bbE_\eps \|\hat\psi^{(\eps)}(t)\|^2_{L^2(\bbT)}\le \sup_{\eps\in(0,1]}\eps\bbE_\eps  \|\hat\psi^{(\eps)}(0)\|^2_{L^2(\bbT)}+2\ga T t,\quad t\ge0.
\end{equation}
%\end{prop}
Let us denote by
${\cal A}$ the completion of ${\cal S}(\bbR\times\bbT)$ in the norm
\begin{equation}
\label{norm-ta01}
\| G\|_{{\cal A}}:=\int_{\bbR}\sup_{k\in\bbT} |\hat G(\eta,k)|d\eta
\end{equation}
and by ${\cal A}'$ its dual.
%\bigskip
We conclude from (\ref{011703-18}) that (see~\cite{GMMP}) 
\begin{equation}\label{mar2302}
\sup_{t\in[0,T]}\|W^{(\eps)}(t)\|_{{\cal A}'}<\infty, \hbox{ for each $\tau>0$,}
\end{equation}
hence $W^{(\eps)}(\cdot)$ is sequentially weak-$\star$ compact over   
$(L^1([0,\tau];{\cal A}))^\star$ for any $\tau>0$. 
We will assume that the initial Wigner distribution 
\begin{equation}\label{eq:20x}
 \widehat{ W}_\varepsilon(\eta,k) :=\  \widehat{ W}_\varepsilon(0,\eta,k),\quad (\eta,k)\in \bbT_{2/\eps}\times\bbT
\end{equation} 
is a % pure \magenta{which means what?}
family that converges weakly in ${\cal A}'$ to a non-negative function  $W_0\in L^1(\bbR\times \bbT)\cap C(\bbR\times \bbT)$.  
%From \eqref{eq:10} we should be able to compute explicitly $W_\varepsilon(t,\eta,k)$ and its limit
%as $\varepsilon \to 0$:
%\begin{equation}
%  \label{eq:47}
%  \lim_{\veps\to 0} \hat W_\varepsilon(t,\eta,k) = \hat W(t,\eta,k)
%\end{equation}
%and the inverse Fourier transform in $\eta$
%\begin{equation}
%  \label{eq:48}
%  W(t,x,k) = \int  \hat W(t,\eta,k) e^{i2\pi \eta x} \; d\eta
%\end{equation}
%We will also compute directly the differential equation that the limit will satisfy.
We will also assume that
%\begin{itemize}
%\item[I1)] 
%\begin{equation}
%\label{null}
%\langle\hat\psi(k)\hat\psi(\ell) \rangle_{\mu_\eps}=0,\quad k,\ell\in\bbT,
%\end{equation}
%\item[(I1)] for any $G\in {\cal A}$ we have
%\begin{equation}
%\label{wigner0}
%\lim_{\eps\to0}\langle G,W^{(\eps)}(0)\rangle=\int_{\bbR\times \bbT}W_0(x,k)G^*(x,k)dxdk,
%\end{equation}
%where $W_0\in L^1(\bbR\times \bbT)\cap C(\bbR\times \bbT)$ is non-negative, 
% \item[I3)]
% for any $\rho>0$ there exists $K>0$ such that
% \begin{align}
% \label{010706}
% \limsup_{\eps\to0+}\int_{\bbT^K_{2/\eps}\times \bbT}|\widehat W_\eps(\eta,k)|d\eta dk\le \rho,
% \end{align}
% where $\bbT^K_{2/\eps}:=[\eta\in \bbT_{2/\eps}:|\eta|\ge K]$.
% In addition, for any $\rho,K>0$ there exists $\delta>0$ such that
% \begin{align}
% \label{010706a}
% \limsup_{\eps\to0+}\int_{A}|\widehat W_\eps(\eta,k)|d\eta dk\le \rho
% \end{align}
% for any measurable subset $A\subset \bbT^K_{2/\eps}\times \bbT$ such
% that $m_2(A)<\delta$. Here $m_2$ is the two dimensional Lebesgue
% measure.
% \item[I3)]
% there exist $C,\kappa>0$ such that
% \begin{align}
% \label{010706-1}
% |\widehat W_\eps(\eta,k)|\le \frac{C}{1+|\eta|^{1+\kappa}},\quad
% (\eta,k)\in\bbT_{2/\eps}\times \bbT,\quad \eps\in(0,1],
% \end{align}
%\item[I3)] 
there exist $C,\kappa>0$ such that 
\begin{equation}
\label{011812aa}
|\widehat W_\eps(\eta,k)|\le 
C\varphi(\eta),\quad (\eta,k)\in\bbT_{2/\eps}\times \bbT, \,\eps\in(0,1],
\end{equation} 
where
\begin{equation}
\label{011812c}
\varphi(\eta):=\frac{1}{(1+\eta^2)^{3/2+\kappa}}.
\end{equation} 
%\textcolor{red}{\em note that this condition implies I3)}
%\end{itemize}

\subsubsection* {Assumptions on the dispersion relation and its basic properties} 

%\subsection{Assumptions on $\alpha_y$}

We assume, as in \cite{BOS},  that
 $\alpha_y$ is a real-valued even function of~$y\in\bbZ$,
and there exists $C>0$ so that 
\[
|\alpha_y|\le Ce^{-|y|/C}, \hbox{ for all $y\in \bbZ$,}
\]
thus $\hat\alpha\in C^{\infty}(\bbT)$. 
We also assume that  $\hat\alpha(k)>0$ for $k\not=0$, and if $\hat \alpha(0)=0$ then  $\hat\alpha''(0)>0$, so that
% The above conditions imply that $\hat\alpha\in C^{\infty}(\bbT)$,
%and if $\hat\alpha(0)=0$, then 
$\hat\alpha(k)=\sin^2(\pi k)\hat\al_0(k)$ for some strictly positive even function  $\hat\al_0\in C^{\infty}(\bbT)$.
It follows that the dispersion relation $\omega(k)=\sqrt{\hat{\alpha}(k)}$ is also an
even and continuous function in $C^\infty(\bbT\setminus\{0\})$.   
We assume that $\om$ is increasing on $[0,1/2]$, and denote its 
unique minimum attained at $k=0$ by~$\omm\ge 0$, its unique maximum, 
attained at~$k=1/2$, by $\omM$, and the two branches of the inverse of
$\om(\cdot)$ as~$\om_-:[\omm,\omM]\to[-1/2,0]$ and
$\om_+:[\omm,\omM]\to[0,1/2]$.  
%
%\red{This was moved from later on, should be fixed. Since the function $\om(k)$ is even and unimodal, we know that it attains its minimum at $k_{\rm min}=0$ and its maximum at 
%$k_{\rm max}=1/2$. Assume that the dispersion relation is smooth on $\bbT$ (\red{what about the acoustic case? Shouldn't this be in the
%introduction anyway?}) 
%and the  branches of its inverse
% $\om_+:[\om_{\rm min},\om_{\rm max}]\to[0,1/2]$ and $\om_-:=-\om_+$  
They satisfy $\om_-=-\om_+$, $\om_+(\om_{\rm min})=0$, $\om_+(\om_{\rm max})=1/2$ and in the case $\om\in C^{\infty}(\bbT)$:
\begin{equation}\label{opt}
\om_\pm'(w)=\pm (w-\om_{\rm min})^{-1/2}\chi_*(w),~~ w-\om_{\rm min}\ll1,
\end{equation}
and 
\begin{equation}\label{opt-b}
\om_\pm'(w) = \pm (\om_{\rm max}-w)^{-1/2}\chi^*(w),~~\om_{\rm max}-w\ll1,
\end{equation}
with $\chi_*,\chi^*\in C^\infty(\bbT)$ that are strictly positive. 
When $\om$ is not differentiable at $0$ (the acoustic case) instead of \eqref{opt} we assume
\begin{equation}\label{opt-a}
\om_\pm'(w)=\pm \chi_*(w),~~ w-\om_{\rm min}\ll1,
\end{equation}
leaving condition \eqref{opt-b} unchanged.

% Here unimodality of $\omega$ is only assumed to simplify the proof. For a general dispersion 
% relation, at incident wave number $k$ the scattered wave numbers would be determined by  $\{k'|\omega(k') = \omega(k)\}$.
%{\color{red} Question: If one drops the monotonicity of $\omega$, 
%then the energy delta function will have more solutions than $\pm k$, hence more scattering channels. 
%If you agree one should mention. No proof I would think.}

%{\color{green} Answer: yes, for non unimodal $\omega$, for an incident wave $k$ there will be scattered  
%waves of mode $k'$ according to the solutions of $\omega(k') = \omega(k)$. Maybe it is enough to mention this without 
%computing here the rates.}

An important role in the analysis will be played by the function
\begin{equation}
  \label{eq:bessel0}
  J(t) = \int_{\bbT}\cos\left(\omega(k) t\right) dk,
\end{equation}
its Laplace transform
\begin{equation}
  \label{eq:2}
\tilde J(\la):=\int_0^{\infty}e^{-\la t}J(t)dt= \int_{\bbT}  \frac{\lambda}{\lambda^2 + \omega^2(k)} dk,\quad {\rm Re}\,\la>0,
\end{equation} 
and the function
\begin{equation}
\label{tg}
\tilde g(\lambda) := ( 1 + \gamma \tilde J(\lambda))^{-1}.
\end{equation} 
%Let $\mathbb C_+:=[\la\in \mathbb C:\, {\rm Re}\,\la>0]$.
Note that ${\rm Re}\,\tilde J(\la)>0$ for $\la\in \mathbb C_+:=[\la\in \mathbb C:\, {\rm Re}\,\la>0]$, therefore 
\begin{equation}
\label{012410}
|\tilde g(\lambda)|\le 1,\quad \la\in \mathbb C_+.
\end{equation} 
The function $\tilde g(\cdot)$ is analytic on $ \mathbb C_+$ so, 
by the Fatou theorem, see e.g. p. 107 of \cite{koosis},
we know that
\begin{equation}
\label{nu}
\nu(k) :=\lim_{\eps\to0}\tilde g(\eps-i\om(k))
\end{equation}
exists a.e. in $\bbT$ and in any $L^p(\bbT)$ for
$p\in[1,\infty)$.

%Properties of $\nu(\cdot)$ are discussed in detail in Section
%\ref{sec8.3} below.  For a moment we only observe that
%$\nu\in L^\infty(\bbT)$ thanks to \eqref{012410}.
% \begin{itemize}
% \item[$\om$2)]
% we have
% \begin{align}
% \label{010706aaa}
% \lim_{\eps\to0+}\eps^{1/2}\log\left(\frac{1}{\eps}\right)\int_{\bbT_{2/\eps}\times
%   \bbT}\frac{|\widehat W_\eps(\eta,k)|d\eta dk}{\om(k-\eps \eta/2)+\om(k+\eps \eta/2) }=0.
% \end{align}
% %\textcolor{red}{\em satisfied under assumption I3')}
%\end{itemize}

To state our main result we need some additional notation.
Let us introduce the group velocity
\[
\bar\om'(k):=\om'(k)/(2\pi)
\]
and
\begin{equation}
\label{033110}
\wp(k):=\frac{\ga \nu(k)}{2|\bar\om'(k)|},\quad \fgeeszett(k):=\frac{\ga|\nu(k)|^2}{|\bar\om'(k)|},\quad p_+(k):=\left|1-\wp(k)\right|^2 ,\quad 
p_-(k):=|\wp(k)|^2  .
\end{equation}
%and 
%$$
%p_+(k):=\left|1-\wp(k)\right|^2 ,\quad 
%p_-(k):=|\wp(k)|^2  .
%$$
%Note that
%$$
%p_+(k)=\left|1-\wp(k)\right|^2\ge 0.
%$$
We will show  in Section~\ref{sec8.3} that
\begin{equation}
\label{feb1402}
{\rm Re}\,\nu(k)=\left(1+\frac{\ga}{2|\bar\om'(k)|}\right)|\nu(k)|^2.
\end{equation}
It follows that 
\begin{equation}\label{mar1528}
p_+(k)+p_-(k)=1- \fgeeszett(k)
\le 1 ,
\end{equation}
so that, in particular, we have
\begin{equation}\label{feb1420}
0\le \fgeeszett(k)\le 1.
\end{equation}

\subsection*{The main result} 
Our main result reads as follows. 
%We use the notation $1_{[0,a]}$ to denote the characteristic function of the
%interval $[0,a]$ if $a>0$ and $[a,0]$ if $a<0$.
\begin{thm}
\label{main:thm}
Suppose that the initial conditions and the dispersion relation
satisfy the above assumptions. 
Then, for any $\tau>0$ and $G\in L^1\left([0,\tau];{\cal A}\right)$ we have 
\begin{equation}
\label{022410}
\lim_{\eps\to0}\int_0^\tau\langle G(t),W_\eps(t)\rangle dt=\int_0^\tau
dt\int_{\bbR\times\bbT}G^*(t,x,k)W(t,x,k)dxdk,
\end{equation}
where
\begin{align}
\label{010304}
&W\left(t,x,k\right)
=  
W_0\left(x-\bar{\om}'(k)t,k\right) 1_{[0,\bar{\om}'(k)t]^c}(x) + \fgeeszett(k) T1_{[0,\bar{\om}'(k)t]}(x)\vphantom{\int_0^1}
\\
&
+p_+(k)W_0\left(x-\bar{\om}'(k)t,k\right)1_{[0,\bar{\om}'(k)t]}(x)
+p_-(k) W_0\left(-x+\bar\om'(k) t,-k\right)1_{[0,\bar \om'(k)t]}(x).\nonumber
\end{align}
%The symbol $1_{[0,\bar\om'(k)t]}(x)$ should be understood as the
%indicator function of the interval $[0,\bar\om'(k)t]$, when
%$\bar\om'(k)>0$ and the indicator function of $[\bar \om'(k)t,0]$, in
%case $\bar \om'(k)<0$, then $1_{[0,\bar\om'(k)t]^c}(x)$ is the
%indicator function of the respective completion of $[0,\bar\om'(k)t]$.
\end{thm}
%The proof of this result is given in Section \ref{sec:nul-init-cond}.

The limit dynamics has an obvious interpretation. 
The first term is the ballistic transport of those phonons which did not cross $\{x=0\}$ up to time $t$. 
The second term on the RHS of \eqref{010304} describes the phonon production of the thermostat.
The third and the fourth term correspond respectively to the transmission and reflection of the phonons at the boundary point $\{x=0\}$.
More precisely,  $\fgeeszett(k)T$ is the phonon production rate,   $p_-(k)$ is the probability of reflection, and $p_+(k)$ is the probability of 
transmission at $\{x=0\}$. Notice that the phonons are absorbed by the thermostat with probability
$1 - p_+(k) - p_-(k) = \fgeeszett(k)$. The scattering at the origin depends only on the friction coefficient $\gamma$. At zero temperature the production of 
phonons is turned off, while the scattering remains unmodified.
%Notice also that the contribution due to the thermal creation 
%and the one from the initial distribution are completely separate. That means that the scattering depends only 
%on the damping and not on the temperature $T$. }

From (\ref{feb1402}) it follows that
\[
\fgeeszett(k)=\hbox{Re}\; \nu(k)-|\nu(k)|^2,
\]
and we also know that $\nu(k_0)= 0$ at the points where $\omega'(k_0)=0$. This means that the thermostat
does not generate phonons with zero velocity, which otherwise would have led to an accumulation
of energy at the boundary. 

 Our main theorem is for the averaged Wigner distribution. In general, one expects a suitable law of large numbers for the quantity on the left 
 of \eqref{022410} with respect to $\mu_\eps\otimes\bbP$.
 
Our result can be written as a boundary value problem, which is a simple but useful exercise. In case the bulk is governed by a wave equation with a small nonlinearity, one would expect 
a nonlinear transport equation for the bulk, but the boundary terms would be dominated by the linear equation, hence of the form 
as written below. Firstly
$W(t,x,k)$ solves the homogeneous transport equation
\begin{equation}\label{feb1406}
\partial_tW(t,x,k)+ \bar\om'(k) \partial_x W(t,x,k)=0,
\end{equation}
away from the boundary point $\{x=0\}$. Secondly we denote the right and left limits of $W$ by
\[
W_-(t,k):=W(t,0^-,k),~~W_+(t,k):=W(t,0^+,k).
\]
%respectively:
%\begin{equation}\label{feb1412}
%W_-(t,k)=W_0(-\bar\om'(k)t,k),~~\hbox{ for $0\le k\le 1/2$},
%\end{equation}
%and 
%\begin{equation}\label{feb1414bis}
%W_+(t,k)=W_0(\bar\om'(k)t,-k),~~\hbox{ for $-1/2\le k\le 0$},
%\end{equation}
Then at $\{x=0\}$ the outgoing phonons are related to the incoming phonons as
\begin{equation}\label{feb1408}
W_+(t,k)=p_-(k)W_+(t,-k)+p_+(k)W_-(t,k)+\fgeeszett(k)T,\hbox{ for $0\le k\le 1/2$},
\end{equation}
and
\begin{equation}\label{feb1410}
W_-(t,k)=p_-(k)W_-(t,-k)+p_+(k)W_+(t,k)+\fgeeszett(k)T,
\hbox{ for $-1/2\le k\le 0$},
\end{equation}
where in both equations the term for the production of phonons has been added.

%In other words, $p_-(k)$ is the reflection coefficient, $p_+(k)$ is the
%transmission coefficient and $T\fgeeszett(k)$ is the phonon production rate  
%by the thermostat. Note that there is no "macroscopic"
%production at the thermostat at zero temperature. We also remark that
%(\ref{feb1420}) implies that the macroscopic phonon production rate is smaller 
%than a naive guess that would omit the factor $\fgeeszett(k)$.

%\blue{The thermostat is a time-dependent "scatterer" -- hence, one might expect that a 
%wave incident on the thermostat would produce
%reflected and transmitted waves at all frequencies. 
%It is rather remarkable that this is not the case -- the reflected and transmitted waves 
%are of the same frequency as the incident wave, as seen from (\ref{feb1408}) and (\ref{feb1410}).
%}

By equipartition, the equilibrium Wigner distribution is given by
%Note that \eqref{mar1528} implies in particular that the equilibrium solution
$$
W(t,x,k)\equiv T,
$$
which indeed satisfies (\ref{feb1408})-(\ref{feb1410}), as one should expect. % A localized initial Wigner function is dispersed 

%equilibratesthe -- if the initial phonon energy matches the
%thermostat temperature, the system is in an equilibrium. 

{
It is interesting to consider what happens when the strength of the thermostat $\gamma\to+\infty$, so that the
oscillations of the particle in contact with the thermostat are sped up
 by a factor of $\gamma$. Then $\tilde g(\la)\sim \gamma^{-1}$, and~$\nu(k)\sim \gamma^{-1}$, 
hence~$\fg(k)\to 0$ (see \eqref{033110}), 
and there is no phonon production or absortion by the thermostat as the particle at the thermostat moves ``too incoherently". 
However, there is still non-trivial reflection and transmission
at the interface. }

%\blue{The result does not depend on the particular model of the thermostat: consider  a Gaussian random renewal 
%of the velocity at $0$, i.e. a Poisson process of intensity $\gamma$ and a sequence of i.i.d. random variables $\{\tilde p_n\}$
%normally distributed with variance $T$, and the dynamics of the chain given by
%\begin{eqnarray}
%&&\dot{\frak q}_y(t)={\frak p}_y(t),
%\label{eq:bas2N}\\
%&&\nonumber\\
%&& d{\frak p}_y(t) = -(\alpha\star {\frak q}(t))_y
%+ \delta_{0,y}\left(\tilde p_{N(t)} - p_0(t^-)\right) dN(t),\quad y\in\bbZ.\nonumber
%\end{eqnarray}
%}
%{\red{\emph{Actually I am not sure this is true anymore...}}

% The Schwartz class ${\cal S}(\bbR\times\bbT)$ consists of all
% $C^\infty$ smooth functions $G:\bbR\times\bbT\to\mathbb C$ such that 
%$$
%\sup_{(x,k)\in\bbR\times
%  \bbT}|G^{(m)}_x(x,k)|(1+|x|)^n<+\infty,\quad\mbox{ for any natural
%  $m,n$}.
%$$

% A word on notation: for two functions $f,g:D\to \bbR$ we say that $f\preceq g$ if there
% exists $C>0$ such that $f(x)\le C g(x)$, $x\in D$. We shall use the
% notation $f\approx g$ if $f\preceq g$ and $g\preceq f$.

The paper is organized as follows. In Section~\ref{sec:wave-fun}, 
we define the Fourier-Laplace transform of the wave function and explain
how the functions $J(t)$ and $\tilde g(\lambda)$ appear in this context. 
% An explicit expression for the Wigner transform 
% (\ref{032112d})-(\ref{063110}) is obtained in
% Section~\ref{sec:nul-init-cond}.
The Wigner transform 
can be decomposed into the ballistic part coming from the initial condition with no scattering, 
the thermostat production part (which is independent from the initial condition) and the scattering part.
 It is quite straightforward to analyze the former two
terms and pass to the limit $\eps\to 0$ in the corresponding expressions. 
Passage to the limit in the scattering term is much more difficult. 
It is outlined in Section~\ref{sec:scatter}, where one of the
scattering terms is analyzed in Lemma~\ref{lem-feb1504}, 
and the asymptotics for the other one is stated in Lemma~\ref{lem-feb1502}.
The scattering terms are put together in Section~\ref{sec:4.3}. 
The bulk of the remainder of the paper, Sections~\ref{sec:6},~\ref{sec:lem5.1} and~\ref{sec:lem4.2},
is essentially devoted to the proof of that Lemma. 
The critical steps are outlined in Lemmas~\ref{lem-mar202}-\ref{lem-mar1202}.
Each of these statements is quite intuitive on the formal level but a rigorous justification is, unfortunately,  
rather lengthy and with little room to spare in the estimates. 
In Section~\ref{sec:end-proof}, we remove an extra assumption that the initial
condition is supported away from the non-propagating modes,
made to simplify the proof. Finally, in Section~\ref{sec8.3} we prove relation (\ref{feb1402}).

%\bigskip
%\noindent
%{\bf Remark.}
%Directly from \eqref{010304} we conclude that
%\begin{align}
%\label{032112e1}
%&\partial_tW(t,x,k)+ \bar\om'(k) \partial_x W(t,x,k)
%= T|\bar\om'(k)|\fgeeszett(k)\delta_0(x)\nonumber\\
%&
%+ |\bar\om'(k)|\delta_0(x)\left\{[p_+(k)-1]W_0\left(-\bar\om'(k)t,k\right)+\vphantom{\int_0^1}p_-(k) W_0\left(\bar\om'(k)t,-k\right)\right\},\\
%&
%W(0,x,k)=W_0(x,k).\nonumber
%\end{align}
%Expressing  $W_0\left(x\pm\bar{\om}'(k)t,\mp k\right)$ by
%$W\left(t,0^\pm,k\right)$, see \eqref{010304x} and \eqref{010304xy} below,  we obtain the
%following autonomous in time equation governing the evolution of $W(t,x,k)$:
%\begin{align}
%\label{010512}
%&\partial_tW(t,x,k)+ \bar\om'(k) \partial_x W(t,x,k)
%=T|\bar\om'(k)|\fgeeszett(k)\delta_0(x)
%\nonumber\\
%&
%+|\bar\om'(k)|\delta_0(x) \sum_{\iota=\pm}1_{(0,+\infty)}(\iota\bar\om'(k))\left\{\vphantom{\int_0^1}[p_+(k)-1]W(t,0^{-\iota},k) \vphantom{\int_0^1} +p_-(k)W(t,0^{\iota},-k)\right\} ,\\
%&
%W(0,x,k)=W_0(x,k).\nonumber
%\end{align}
%Obviously, due to the fact that  $W(t,x,k)$ has discontinuity at $0$ (cf \eqref{010304}),
% this equation has to be properly interpreted via the usual testing against smooth function. 

\section{The Laplace-Fourier transform of the wave function and of the Wigner distribution}
\label{sec:wave-fun}

In this section, we obtain an explicit expression for  the Laplace-Fourier transform of the wave function. 
We use the mild formulation of   \eqref{basic:sde:2aa}: 
\begin{equation}
  \label{eq:sol1}
  \begin{split}
    \hat\psi(t,k) = e^{-i\omega(k) t} \hat\psi(0,k) - i\gamma \int_0^t e^{-i\omega(k) (t-s)} {\frak p}_0(s) ds
    + i \sqrt{2\ga T} \int_0^t e^{-i\omega(k) (t-s)} dw(t).
  \end{split}
\end{equation}
Integrating both sides in the $k$-variable and taking the imaginary part in both sides, we obtain a closed equation 
for ${\frak p}_0(t)$:
\begin{equation}
  \label{eq:p0}
  \begin{split}
    {\frak p}_0(t) &= {\frak p}^0_0(t) - \gamma  \int_0^t J(t-s) {\frak p}_0(s) ds +  
    \sqrt{2\ga T} \int_0^t J(t-s) dw(s),
  \end{split}
\end{equation}
where $J(t)$ is given by \eqref{eq:bessel0}
and 
\begin{equation}
  \label{eq:1}
   {\frak p}^0_0(t) = \int_{\bbT} {\rm Im}\left(\hat\psi(0,k) e^{-i\omega(k) t}\right) dk,
\end{equation}
is the  momentum at   $y=0$   for the free evolution 
with $\gamma = 0$ (without the thermostat). 
Taking the Laplace transform
$$
\tilde {\frak p}_0 (\lambda) =\int_0^{+\infty}e^{-\la t} {\frak p}_0(t) dt,\quad {\rm Re}\,\la>0,
$$
 in \eqref{eq:p0} we obtain
 \begin{equation}
  \label{eq:3}
  \tilde {\frak p}_0 (\lambda) =  \tilde g(\lambda) \tilde   {\frak p}^0_0(\lambda) + 
    \sqrt{2\ga T} \tilde g(\lambda) \tilde J(\lambda) \tilde w(\lambda).
\end{equation}
Here, $\tilde g(\lambda)$ is given by \eqref{tg},
and $\tilde{\frak p}_0^0(\lambda)$ and $ \tilde J(\lambda)$ are  the Laplace transforms of ${\frak p}_0^0(t)$ and 
$J(t)$, respectively, and  
$\tilde w(\lambda)$  is the Laplace transform of the Wiener process.
It is a zero mean Gaussian
process with the covariance
\begin{equation}
  \label{eq:4}
  \bbE [ \tilde w(\lambda_1) \tilde w(\lambda_2) ]= \frac {1}{\lambda_1 + \lambda_2},\quad {\rm Re}\,\la_1,\,{\rm Re}\,\la_2>0.
\end{equation}
%Note that
%\begin{equation}
%  \label{eq:2}
%  \tilde J(\lambda) 
%\end{equation}

%If we start with Gaussian initial distribution, independent of $w(\cdot)$, then $ \tilde {\frak p}_0 (\lambda) $ is 
%a Gaussian process with covariance 
%\begin{equation}
%  \label{eq:5}
%  \bbE \left(\tilde {\frak p}_0 (\lambda_1) \tilde {\frak p}_0 (\lambda_2) \right) =
%  \tilde g(\lambda_1) \tilde g(\lambda_2) 
%\left(\bbE \left(\tilde {\frak p}^0_0 (\lambda_1) \tilde {\frak p}^0_0 (\lambda_2) \right) 
%  + 2\gamma T \frac{\tilde J(\lambda_1) \tilde J(\lambda_2) }{\lambda_1 +\lambda_2}\right),
%\end{equation}
%where 

%\emph{Notice that this is the same function (2) that appears in Spohn's note (2006).}

Next, taking the Laplace transform of both sides of  \eqref{eq:sol1} and using \eqref{eq:3},  we arrive at an explicit formula for the Fourier-Laplace 
transform of $\psi_y(t)$:
\begin{equation}
  \label{eq:6}
  \begin{split}
    \tilde \psi(\lambda,k) &= 
\frac{\hat\psi(0,k) - i\gamma \tilde{\frak p}_0(\lambda) + i\sqrt{2\gamma T} \tilde w(\lambda)}
{\lambda + i\omega(k)} \\
& 
= \frac{\hat\psi(0,k) - i\gamma \tilde g(\lambda) (\tilde{\frak p}_0^0(\lambda)  
+ \sqrt{2\gamma T}  \tilde J(\lambda)\tilde w(\lambda)) + i  \sqrt{2\ga T} \tilde w(\lambda)}
{\lambda + i\omega(k)}  \\
& = \frac{\hat\psi(0,k) - i\gamma \tilde g(\lambda) \tilde{\frak p}_0^0(\lambda)  
+  i \tilde g(\lambda) \sqrt{2\gamma T} \tilde w(\lambda)}
{\lambda + i\omega(k)} .
  \end{split}
\end{equation}
%This is quite an explicit formula, since $\tilde{\frak p}_0^0(\lambda)$ concern the evolution for 
%$\gamma = 0$, so it is independent from $\tilde w$.
Note  that \eqref{eq:6} implies, in
particular, that, even at the zero temperature, and if the
initial wave function is monochromatic, that is,
$\hat\psi(0,k)=\delta_0(k-k_0) $ for some $k_0$,
scattering at the thermostat generates various modes $k\neq k_0$, 
due to the damping at $y=0$. This is a microscopic phenomenon not observed
on the macroscopic level, as seen from the discussion following 
Theorem~\ref{main:thm}.

Let us momentarily assume that $\tilde g(\lambda)$ is the Laplace transform
of a signed locally finite measure~$g(d\tau)$. Then, the term
$(\lambda + i \omega(k))^{-1} \tilde g(\lambda) \tilde{\frak p}_0^0(\lambda)$,
that appears in (\ref{eq:6}),
is the Laplace transform~of
\begin{equation}
  \label{eq:11}
  \int_0^t ds \int_0^{t-s}  e^{-i\omega(k)(t-s-\tau)} g(d\tau) {\frak p}_0^0(s).
\end{equation}
% \begin{equation}
%   \label{eq:7}
%   (\lambda + i \omega(k))^{-1} \tilde g(\lambda) =  
%   (\lambda + i \omega(k))^{-1}\left[\tilde g(i\omega(k)) - \tilde g(\lambda + 2i\omega(k))\right]
% \end{equation}
% Notice that Herbert's function $\nu(k)$ is 
% \begin{equation}
%   \label{eq:8}
%   \nu(k) = \int_0^\infty e^{-i\omega(k) t} g(t) dt = \tilde g(i\omega(k)) .
% \end{equation}
% Consequently we can rewrite the \eqref{eq:6} as
% \begin{equation}
%   \label{eq:9}
%   \tilde \psi(\lambda,k) = 
% \frac{\hat\psi(0,k) - i \left[\tilde g(i\omega(k)) - \tilde g(\lambda + 2i\omega(k))\right]
%  \tilde{\frak p}_0^0(\lambda)  
% +  i \nu(k) \sqrt{2\gamma T} \tilde w(\lambda)}
% {\lambda + i\omega(k)} 
% \end{equation}
Now, the Laplace inversion of  \eqref{eq:6} gives an explicit expression for
$\hat\psi(t,k)$: 
\begin{equation}
  \label{eq:10}
  \begin{split}
    \hat\psi(t,k) =& e^{-i\omega(k) t} \hat\psi(0,k) - 
     i \gamma  \int_0^t ds \int_0^{t-s}   e^{-i\omega(k)(t-s-\tau)} g(d\tau) {\frak p}_0^0(s)\\
   &+ i \sqrt{2\gamma T} \int_0^t ds \int_0^{t-s} e^{-i\omega(k)(t-s-\tau)} g(d\tau) dw(s)\\
   =& e^{-i\omega(k) t} \hat\psi(0,k) - 
     i \gamma  \int_0^t \phi(t-s,k) {\frak p}_0^0(s)\; ds + i \sqrt{2\gamma T} \int_0^t \phi(t-s,k) \; dw(s),
  \end{split}
\end{equation}
where 
\begin{equation}
  \label{eq:12}
   \phi(t,k) = \int_0^{t}e^{-i\omega(k)(t-\tau)} g(d\tau) .
\end{equation}
Likewise, we conclude from \eqref{eq:3} that
\begin{equation}
  \label{eq:100}
    {\frak p}_0(t) = \int_0^t   {\frak p}_0^0(t-s)g(ds)
     +  \sqrt{2\gamma T} \int_0^t dw(s) \int_0^{t-s} \;  J(t-s-\tau) g(d\tau) .
\end{equation}

%\bigskip
%
%\noindent {\bf  Remark 2.}
%Notice that for the dispersion relation $\omega(k) = |\sin(\pi k)|$, $J(t)$ is the first Bessel function 
%$$
%J(t)=\sum_{m=0}^{+\infty}\frac{(it)^{2m}}{(2^mm!)^2}= \sqrt{\frac{2}{\pi t}}\left[\cos\left(t-\frac{\pi}{4}\right)+O\left(\frac{1}{t}\right)\right],\quad t\gg1 .
%$$
%
%
%\bigskip
%
%\noindent {\bf  Remark 3.}

In order to understand how $g(d\tau)$ looks like, note that
a function $g_*(t)$ that has the Laplace transform
\[
\tilde g_*(\lambda):=\tilde g(\lambda)-1=
-\frac{\ga\tilde J(\la)}{1+\ga\tilde J(\la)},
\]
%let $g_*\in L^1_{loc}[0,+\infty)$ be 
is the solution of  the Volterra equation
\begin{equation}
\label{g*}
g_*(t)+\ga J\star g_*(t)=-\ga J(t),\quad t\ge0.
\end{equation}
Here, we denote by
\[
f_1\star f_2(t)=\int_0^t f_1(t-s)f_2(s) ds
\]
the convolution of  $f_1,f_2\in L^1_{loc}[0,+\infty)$.
The solution $g_*$ of (\ref{g*}) is  
 given by the convolution series
\begin{equation}
\label{050611}
g_*(t)=\sum_{n=1}^{+\infty}(-\ga)^nJ^{\star,n}(t).
\end{equation}
Here, $J^{\star,n}(t)$ is the $n$-time convolution of $J$ with itself. 
As $|J(t)|\le 1$, we see that~$g_*\in C^\infty[0,+\infty)$ and
$|g_*(t)|\le e^{\ga t}$, $t\ge0$. 
%Thus, one can define the Laplace transform
%$\tilde g_*(\la)$, at least when ${\rm Re}\,\la>\ga$, and
%\begin{equation}
%\label{050611a}
%\tilde  g_*(\la)=-\frac{\ga\tilde J(\la)}{1+\ga\tilde J(\la)},
%\end{equation}
%which, can be analytically continued
%to ${\rm Re} \,\la>0$, as this set is contained in 
%the domain of analyticity of $\tilde J(\la)$. 
%%The latter in fact implies that for any $\rho>0$ there exists $C_\rho>0$, for which
%%\begin{equation}
%%\label{rho}
%%|g_*(t)|\le C_{\rho}e^{\rho t}, \quad t\ge0.
%%\end{equation}
%Note that, the Laplace transform 
%\begin{equation}
%  \label{eq:49}
%  \tilde g(\lambda) = \int_0^\infty e^{-\lambda t} g(dt).
%\end{equation}
%of  the distribution
Then, we can represent $g(d\tau)$ as
\begin{equation}
\label{g}
g(dt)=\delta_0(dt)+g_*(t)dt,\quad t\ge0.
\end{equation}
Here, $\delta_0$ is the Dirac distribution.

{Observe that the existence of $g(dt)$ with the above properties implies that
  \begin{equation}
    \label{eq:19}
    \int_0^t e^{i\omega(k) \tau} g(d\tau) = e^{i\omega(k)t} \phi(t,k) \mathop{\longrightarrow}_{t\to\infty} \nu(k)
  \end{equation}
in the sense that for $\text{Re} \lambda>0$ the limit defined by \eqref{nu} implies
\begin{equation}
  \label{eq:23}
  \lim_{\eps\to 0} \int_0^{+\infty} e^{-\lambda t} \int_0^{\eps^{-1}t} e^{i\omega(k) \tau} g(d\tau) = \frac{\nu(k)}{\lambda}
\end{equation}
}

{
We let
\begin{equation}
\label{LF}
\widehat{
  w}_\varepsilon(\la,\eta,k):=\int_0^{+\infty}e^{-\la t}\widehat{
  W}_\varepsilon(t,\eta,k)dt,\quad {\rm Re}\,\la>0
\end{equation}
be the Laplace-Fourier transform of the Wigner distribution defined in \eqref{wigner}.
The claim of Theorem~\ref{main:thm} is equivalent to the following: 
%We shall prove that  
for any test function $G\in {\cal S}(\bbR\times\bbT)$ we have
\begin{equation}
\label{023110}
\lim_{\eps\to0+}\int_{\bbR\times\bbT}\hat G^*(\eta,k)\widehat{
  w}_\varepsilon(\la,\eta,k)d\eta dk=\int_{\bbR\times\bbT}\hat G^*(\eta,k)\widehat{
  w}(\la,\eta,k)d\eta dk,
\end{equation}
where
\begin{align}
\label{013110}
\widehat{w}(\la,\eta,k):=&\frac{T|\bar\om'(k)|\fgeeszett(k)}{\la(\la+i\om'(k)\eta)}  + \frac{\widehat{W}_0(\eta,k)}{\la+i\om'(k)\eta} \\
 &+\frac{|\bar\om'(k)|(p_+(k)-1)}{\la+i\om'(k)\eta}\int_{\bbR}\frac{\widehat W_0(\eta',k)d\eta'}{\la+i\om'(k)\eta'} 
+\frac{|\bar\om'(k)|p_-(k)}{\la+i\om'(k)\eta}\int_{\bbR}\frac{\widehat W_0(\eta',-k)d\eta'}{\la-i\om'(k)\eta'}.\nonumber
\end{align}
and
$p_{\pm}(k)$ and $ \fgeeszett(k)$
are given by \eqref{033110}. 
%That is, the weak-$\star$ limit of $W_\eps(\cdot)$, as $\eps\to0$, 
%with the function $W:[0,+\infty)\times\bbR\times\bbT\to[0,+\infty)$
%is given by
The Fourier-Laplace transform of (\ref{013110}) leads to  \eqref{010304}. The rest of the paper is devoted to the derivation
of (\ref{013110}). 
}

\section{The phonon creation term}
\label{sec:phonon-creation-term}

{
Since the contribution to the energy given by the thermal term and the initial energy are completely separate,
we can derive the first term of \eqref{013110} assuming $\widehat W_0 = 0$. In this case  $\hat\psi(0,k) = 0$
and \eqref{eq:10} reduce to a stochastic integral:
\begin{equation}
  \label{eq:10T}
  \begin{split}
    \hat\psi(t,k) = i \sqrt{2\gamma T} \int_0^t \phi(t-s,k) \; dw(s),
  \end{split}
\end{equation}
To shorten notations denote $\tilde\phi(t, k) = \int_0^{t} e^{i\omega(k)\tau} g(d\tau) = e^{i\omega(k)t}\phi(t,k)$,
and
\begin{equation}
\label{053110}
\delta_{\eps}\om(k,\eta):=\frac{1}{\eps}\left[\om\left(k+\frac{\eps
      \eta}{2}\right)-\om\left(k-\frac{\eps \eta}{2}\right)\right].
\end{equation}
We can compute directly
\begin{equation*}
  \widehat W_\eps(t,\eta,k) = \gamma T \int_0^t 
  e^{-i \delta_{\eps}\om(k,\eta) s} \tilde\phi(s/\eps, k+\eps\eta/2) \tilde\phi^*(s/\eps, k+\eps\eta/2) ds
\end{equation*}
The Laplace transform of $\tilde\phi(\eps^{-1} t, k)$ is given by $\tilde g(\eps\lambda - i\omega(k))$.
Then we can compute directly the Laplace-Fourier tranform of the Wigner distribution and obtain
\begin{equation}
  \label{eq:24}
  \begin{split}
    \widehat{w}_\eps(\la,\eta,k) = \gamma T \int_0^\infty dt e^{-\lambda t} \int_0^t ds e^{-i \delta_\eps\omega(k,\eta) s}
    \tilde\phi\left(\eps^{-1} s, k+\frac{\eps\eta}2\right) \tilde\phi^*\left(\eps^{-1} s, k-\frac{\eps\eta}2\right) \\
    = \frac{\gamma T}{\lambda}  \int_0^\infty ds e^{-(\lambda +i \delta_\eps\omega(k,\eta)) s} 
    \tilde\phi\left(\eps^{-1} s, k+\frac{\eps\eta}2\right) \tilde\phi^*\left(\eps^{-1} s, k-\frac{\eps\eta}2\right) \\
  \end{split}
\end{equation}
and by using the inverse Laplace formula for the product of functions we have, for $c>0$, we obtain
\begin{equation}
  \label{eq:16}
  \begin{split}
    \widehat{w}_\eps(\la,\eta,k)  = \frac{\gamma T}{\lambda} \frac 1{2\pi i} \lim_{\ell\to\infty} \int_{c-i\ell}^{c+i\ell} 
    \frac{\tilde g(\eps \sigma - i\omega(k+\frac{\eps\eta}{2}) ) 
      \tilde g^*(\eps(\lambda + i \delta_\eps\omega(k,\eta) - \sigma) - i\omega(k-\frac{\eps\eta}{2}) )}
    {\sigma\left(\lambda + i \delta_\eps\omega(k,\eta) - \sigma\right)} \ d\sigma
  \end{split}
\end{equation}
Since $\tilde g$ is bounded and $\text{Re} \lambda >0$, there is no problem in taking the limit as $\eps \to 0$ obtaining
\begin{equation}
  \label{eq:17}
  \frac{\gamma T |\nu(k)|^2}{\lambda\left(\lambda + i \omega'(k) \eta\right)}.
\end{equation}
}

\section{The scattering terms}
\label{sec:scatter}

If the thermal production at $0$ was easy to prove, the scattering terms are much more challenging.
Since the terhmal part will not affect the scattering, we can set $T=0$ and consider  
a non-zero initial energy. 

% In this section, we obtain an explicit expression for the Wigner transform for $\eps>0$ that can
% be interpreted as a decomposition into the ballistic term that experiences neither scattering at the thermostat
% nor production at the thermostat, the thermostat term that is produces purely by the the thermostat, and, finally, the
% scattering term that is produced by an interaction between the incident waves and the thermostat. We also
% analyze the ballistic and the thermostat terms. 

% \subsection*{The decomposition into ballistic, thermostat and scattering terms}

We will first prove (\ref{013110}) 
under a stronger assumption than
\eqref{011812aa}: { we will assume no energy is concentrated around modes that have null velocity, more precisely}
 that there exist $C,\delta>0$ and $\kappa>0$ such that 
\begin{equation}
\label{011812}
|\widehat W_\eps(\eta,k)|\le 
C\varphi(\eta)\chi\left(k-\frac{\eps \eta}{2}\right) \chi\left(k+\frac{\eps \eta}{2}\right),\quad (\eta,k)\in\bbT_{2/\eps}\times \bbT, \,\eps\in(0,1],
\end{equation} 
here  $\varphi(\cdot)$ is given by \eqref{011812c}
and
$\chi\in C(\bbT)$ is non-negative and satisfies 
\begin{equation}
\label{011812b}
\chi(k)\equiv 0\mbox{ for } k\in L(\delta),
\end{equation}
with
\begin{equation}
\label{L}
L(\delta):=[k:{\rm dist}(k,\Om_*)<\delta] 
\end{equation}
and 
$\Om_*:=[k\in\bbT:\,\om'(k)=0]\subset \{0,1/2\}$. 
The proof of Theorem \ref{main:thm} under the weaker assumption~\eqref{011812aa} 
is presented in Section \ref{w-s} below.

We could have continued to compute   $\widehat{w}_\eps$ directly from the expression of the wave function,
as we did for the termal part. We find more practical to use the time evolution of $\widehat W_\eps(t,\eta,k)$

A straightforward computation starting from \eqref{basic:sde:2aa} and  
\eqref{eq:20} shows
that the Wigner transform obeys, {for $T=0$}, an evolution equation
 \begin{eqnarray}
\label{exp-wigner-eqt}
&&\partial_t\widehat W_\eps(t,\eta,k)=-i\delta_{\eps}\om(k, \eta)\widehat W_\eps(t,\eta,k)
% +\ga T
   \nonumber\\
&&
\\
&&+i\ga\left\{\int_{\bbT}\bbE_\eps\left[\hat\psi^{(\eps)}\left(t,k+\frac{\eps \eta}{2}\right)(\hat{\frak p}^{(\eps)})^*(t,k')\right]dk'
-\int_{\bbT}\bbE_\eps\left[(\hat\psi^{(\eps)})^*\left(t,k-\frac{\eps \eta}{2}\right)\hat{\frak p}^{(\eps)}(t,k')\right]dk'\right\},\nonumber
\end{eqnarray}

Performing the Laplace transform in both sides of
\eqref{exp-wigner-eqt}, we obtain  
\begin{align}
\label{032112d}
\left(\la+ i\delta_\eps\om(k,
\eta) \right)w_{\eps}(\la,\eta,k)
=\widehat W_{\eps}(\eta,k) % +\frac{\ga T}{\la}
  -\frac{\ga}{2}\left[{\frak d}_\eps\left(\la,k-\frac{\eps \eta}{2}\right)+{\frak d}_\eps^\star\left(\la,k+\frac{\eps \eta}{2}\right)\right],
\end{align}
where
\begin{equation}
\label{063110}
{\frak d}_\eps(\la,k):=i\int_0^{+\infty}e^{-\la t}\bbE_\eps\left[
 (\hat\psi^{(\eps)})^*\left(t,k\right){\frak
  p}_0^{(\eps)}\left(t\right)\right] dt.
\end{equation}
As $\delta_\eps\om(k,\eta)\to \omega'(k)\eta$ as $\eps\to 0$, to get (\ref{013110}), we need to understand the limit of ${\frak d}_\eps(\la,k)$. 
{
Using~\eqref{eq:10} for $T=0$, % and \eqref{eq:100},
we may write
\[
{\frak d}_\eps\left(\la,k\right)=  {\frak  d}_\eps^1\left(\la,k\right)+{\frak  d}_\eps^2\left(\la,k\right)
% +{\frak d}_\eps^3\left(\la,k\right).
\]
Here, ${\frak d}_\eps^j\left(\la,k\right)$, $j=1,2$ are the  
Laplace transforms of  $I_\eps(t/\eps)$, $I\!I_\eps(t/\eps)$%and $I\!I\!I(t/\eps)$
, where
\begin{equation}
\label{012703}
I_\eps(t,k):= i e^{i \om(k)t} \int_0^t\left\langle {\frak p}^0_0(t-s)\hat\psi^*(k)\right\rangle_{\mu_\eps}g(ds),
\end{equation}
\begin{eqnarray*}
&&
I\!I_\eps(t,k):= 
    - \gamma  \int_0^{t}  g\left(ds'\right) \int_0^{t} \phi^*\left(t-s,k\right)\langle{\frak p}_0^0(s){\frak p}_0^0(t-s')\rangle_{\mu_\eps}ds,
\end{eqnarray*}
% and
% \begin{eqnarray*}
% &&
% I\!I\!I(t,k):=2\gamma T\int_0^t\phi^*(s,k) (J\star g)(s)ds.
% \end{eqnarray*}
}

Now, with (\ref{023110}) in mind, we can introduce %\red{This should go to the beginning of this sectio%n.}
\begin{equation}\label{feb1416}
{\frak L}_\eps(\la):=\int_{\bbR\times\bbT}\hat G^*(\eta,k)\widehat{w}_\varepsilon(\la,\eta,k)
d\eta dk ={\frak L}_{init}^\eps(\lambda)
% +{\frak L}_{th}^\eps(\lambda)
+{\frak L}_{scat}^\eps(\la).
\end{equation}  
The first term in the right side is
\begin{equation}
\label{073110bis}
{\frak L}_{init}^\eps(\lambda):=
\int_{\bbR\times\bbT}\hat G^*(\eta,k)\frac{\widehat W_{\eps}(\eta,k) }{\la+ i\delta_\eps\om(k,
\eta)}d\eta dk. 
\end{equation}
% The thermostat production term in the right side of (\ref{feb1416}) is
% \begin{equation}\label{feb1418}
% {\frak L}_{th}^\eps(\la)={\frak L}_{th,1}^\eps(\la)
% +{\frak L}_{th,2}^\eps(\la),
% \end{equation}
% with
% \begin{equation}
% {\frak L}_{th,1}^\eps(\la):=\frac{\ga T}{\la} \int_{\bbR\times\bbT}\frac{\hat G^*(\eta,k) }{\la+ i\delta_\eps\om(k,
% \eta)}d\eta dk
% \end{equation}
% and
% \begin{equation}
% {\frak L}_{th,2}^\eps:=-\frac{\ga}{2}
% \int_{\bbR\times\bbT}\frac{\hat G^*(\eta,k) }{\la+ i\delta_\eps\om(k,
% \eta)}\left[{\frak d}_\eps^3\left(\la,k-\frac{\eps \eta}{2}\right)+
% ({\frak d}_\eps^3)^\star\left(\la,k+\frac{\eps \eta}{2}\right)\right]d\eta dk.
% \end{equation}
The scattering term in the right side of (\ref{feb1416}) is 
\begin{equation}\label{feb1414}
{\frak L}_{scat}^\eps(\la)={\frak L}_{scat,1}^{\eps}+{\frak L}_{scat,2}^{\eps},
\end{equation}
with
\begin{equation}
{\frak L}_{scat,j}^{\eps}:=-\frac{\ga}{2}
\int_{\bbR\times\bbT}\frac{\hat G^*(\eta,k) }{\la+ i\delta_\eps\om(k,
\eta)}\left[{\frak d}_\eps^j\left(\la,k-\frac{\eps \eta}{2}\right)
+({\frak d}_\eps^j)^\star\left(\la,k+\frac{\eps \eta}{2}\right)\right]d\eta dk,
~~j=1,2.
\end{equation}
%It is helpful to decompose ${\frak L}_\eps(\la)$ as follows.
  %The last term  in the calculation of the limit in \eqref{023110} is
%\begin{equation}
%\label{L-feb14}
%{\frak L}(\la)=\lim_{\eps\to 0^+}{\frak L}_\eps(\la),~~
%{\frak L}_\eps(\la):=-\frac{\ga}{2}
%\int_{\bbR\times\bbT}\frac{\hat G^*(\eta,k) }{\la+ i\delta_\eps\om(k,
%\eta)}\left[{\frak d}_\eps\left(\la,k-\frac{\eps \eta}{2}\right)+{\frak d}_\eps^\star\left(\la,k+\frac{\eps \eta}{2}\right)\right]d\eta dk,
%\end{equation}
%and is the only one to cause any difficulty. 
% This also gives the decomposition
% expression under the limit in the right hand side of \eqref{L-feb14} can
%be then written accordingly as 
%\[
%{\frak L}_\eps(\lambda)= {\frak L}_\eps^1(\la)+{\frak L}_\eps^2(\la)
%+{\frak L}_\eps^3(\la),
%\]
%with each ${\frak L}_\eps^j(\la)$ corresponding to the 
%respective ${\frak  d}_\eps^j$, $j=1,2,3$. 
%Using \eqref{032112d}, we can calculate the limit in the left side
%of \eqref{023110}.
%The terms ${\frak L}_\eps^1(\la)$ and ${\frak L}_\eps^3(\la)$ are relatively easy
%to understand, but ${\frak L}_\eps^2(\la)$ is more involved. 

\subsubsection*{The ballistic term}

We note that thanks to assumption I2), we can easily show that
\begin{equation}
\label{073110}
{\frak L}_{init}^\eps(\lambda):=
\int_{\bbR\times\bbT}\hat G^*(\eta,k)\frac{\widehat W_{\eps}(\eta,k) }{\la+ i\delta_\eps\om(k,
\eta)}d\eta dk\to
\int_{\bbR\times\bbT}\hat G^*(\eta,k)\frac{\widehat W_0(\eta,k) }{\la+ i\om'(k)
\eta}d\eta dk, \hbox{ as $\eps\to 0$},
\end{equation}
which is the {second} term in the right side of (\ref{013110}).

\subsection{The limit of the first scattering term}
%Calculation of  $\lim_{\eps\to0+}{\frak L}_\eps^1(\la)$}

Here, we use the notation
\begin{equation}
\label{ompm}
\begin{aligned}\delta_\eps^+\om(k,
\eta):=\delta_\eps\om\Big(k+ \frac{\eps\eta}{2}\Big)=\frac{1}{\eps}\Big[\omega(k+{\eps\eta}  )-
\omega(k)\Big],\\
%\end{equation}
%\begin{equation}
%\label{ompm}
\delta_\eps^-\om(k,
\eta):=\delta_\eps\om\Big(k-\frac{\eps\eta}{2}\Big)=\frac{1}{\eps}\Big[\omega(k)-
\omega(k -{\eps\eta} )\Big].
\end{aligned}
\end{equation}

We now compute the limit of ${\frak L}_{scat,1}^\eps(\la)$ in (\ref{feb1414})
that we can re-write, after a simple change of variables as 
%Recall that
\begin{eqnarray}
\label{010811}
&&
{\frak L}_{scat,1}^\eps(\la)
%=-\frac{\ga}{2}\int_{\bbR\times\bbT}\frac{\hat G^*(\eta,k) }{\la+ i\delta_\eps\om(k,
%\eta)}\left[{\frak d}_\eps^1\left(\la,k-\frac{\eps
%   \eta}{2}\right)+\left({\frak
%   d}_\eps^1\right)^\star\left(\la,k+\frac{\eps
%   \eta}{2}\right)\right]d\eta dk\nonumber\\
%&&
%\\
%&&
=-\frac{\ga}{2}\int_{\bbR\times\bbT}\Big[\frac{\hat G^*(\eta,k+\eps\eta/2) }{\la+ i\delta_\eps^+\om(k,
\eta)}{\frak d}_\eps^1\left(\la,k\right)+\frac{\hat G^*(\eta,k-\eps\eta/2) }{\la+ i\delta_\eps^-\om(k,
\eta)}\left({\frak
   d}_\eps^1\right)^*\left(\la,k\right)\Big]d\eta dk.\nonumber
\end{eqnarray}
%The main purpose of the present section is to 
We will show the following.
\begin{lm} \label{lem-feb1504}
For any test function $G\in {\cal S}(\bbR\times\bbT)$ and
  $\la>0$ we have
\begin{equation}
\label{020811}
\lim_{\eps\to0+}{\frak L}_{scat,1}^\eps(\la)=
 -\ga \int_{\bbR\times\bbT}{\rm Re} [\nu(k)]
 \frac{\widehat W_\eps(\eta',k)}{\la+i\om'(k)\eta'} 
 \left\{\int_{\bbR}\frac{G^*(\eta,k)}{\la+i\om'(k)\eta}d\eta\right\}dk d\eta'.
\end{equation}
\end{lm}
\proof
From \eqref{012703} and \eqref{eq:1} we get
\begin{equation}
\label{Ibis}
I_\eps(t,k)= \frac{1}{2} \int_0^{t} g\left(ds\right) \int_{\bbT} \left\{\langle\hat\psi^\star(k)\hat\psi(\ell) \rangle_{\mu_\eps}e^{i(\omega(k)-\om(\ell))t+i\om(\ell) s}-\langle\hat\psi^\star(k)\hat\psi^\star(\ell) \rangle_{\mu_\eps}e^{i(\omega(k)+\om(\ell))t-i\om(\ell) s}\right\} d\ell.
\end{equation}
Using assumption I1) (see \eqref{null}) and \eqref{Ibis} we conclude that 
\begin{eqnarray*}
&&
{\frak d}_\eps^1(\la,k)
=\frac12\int_{\bbT}\eps \langle\hat \psi^\star (k)\hat \psi(\ell) \rangle_{\mu_\eps}d\ell\int_0^{+\infty}\exp\left\{i\om(\ell)s\right\}g(ds)
\int_s^{+\infty}e^{-\la\eps t}
\exp\left\{i(\om(k)-\om(\ell))t\right\}dt\\
&&
=\frac12\int_{\bbT}\frac{\eps \langle\hat \psi^\star (k)\hat \psi(\ell) \rangle_{\mu_\eps}d\ell}{\la\eps+i(\om(\ell)-\om(k))}\int_0^{+\infty}g(ds)
e^{-\la\eps s}
\exp\left\{i\om(k)s\right\}ds\\
&&
=\int_{\bbT}\frac{(\eps/2) \langle\hat \psi^\star (k)\hat \psi(\ell) \rangle_{\mu_\eps}d\ell}{\la\eps+i(\om(\ell)-\om(k))}\tilde g(\eps \la-i\om(k)).
\end{eqnarray*}
{For any test function $G\in {\cal S}(\bbT\times\bbR)$ we can write, therefore, (cf. \eqref{ompm})
\begin{eqnarray}
\label{020706aa}
&&
\int_{\bbR\times\bbT}\frac{\hat G^\star(\eta,k+\eps\eta/2){\frak d}_\eps^1(\la,k)}{\la+i\delta_\eps^+\om(k,\eta)}dkd\eta
\\
&&
=\int_{\bbR\times\bbT}\frac{\hat G^\star(\eta,k+\eps\eta/2)}{\la+i\delta_\eps^+\om(k,\eta)}dkd\eta\left\{\int_{\bbT}\frac{(\eps/2) \langle\hat \psi^\star (k)\hat \psi(\ell) \rangle_{\mu_\eps}\tilde g(\eps \la-i\om(k))}{\la\eps+i(\om(\ell)-\om(k))}d\ell\right\}.\nonumber
\end{eqnarray}
Changing variables
$
k:=k'-\eps \eta'/2,$ $\ell:=k'+\eps \eta'/2$
the right hand side of \eqref{020706aa} can be rewritten in the form
\begin{equation}
\label{020706a}
\int_{\bbR}d\eta\left\{\int_{T_\eps}\frac{\widehat
   W_\eps(\eta',k')\tilde g(\eps \la-i\om(k'-\eps \eta'/2))\hat
   G^*(\eta,k'+\eps\eta/2-\eps
   \eta'/2)}{[\la+i\delta_\eps^+\om(k'-\eps
   \eta'/2,\eta)][\la+i\eps^{-1}(\om(k'+\eps \eta'/2)-\om(k'-\eps
   \eta'/2))]}dk'd\eta'\right\}.
\end{equation}
Here, $T_\eps\subset
\bbT_{2/\eps}\times \bbT$ is the image of $\bbT^2$ under the inverse map
$
k':=(\ell+k)/2,$ $\eta':=(\ell-k)/\eps$. Note that
%~$T_\eps\subset \bbT_{2/\eps}\times \bbT$, and
$$
\lim_{\eps\to0}\frac{\tilde g(\eps \la-i\om(k'-\eps \eta'/2))\hat
   G^\star(\eta,k'+\eps\eta/2-\eps
   \eta'/2)}{[\la+i\delta_\eps^+\om(k'-\eps
   \eta'/2,\eta)][\la+i\eps^{-1}(\om(k'+\eps \eta'/2)-\om(k'-\eps
   \eta'/2))]}=\frac{\nu(k')\hat
   G^\star(\eta,k')}{[\la+i\om'(k')\eta][\la+i\om'(k')\eta']}
$$
a.e. in $(\eta,\eta',k')$. Using bounds \eqref{011812} and
\eqref{012410} we can argue, via the dominated convergence theorem
that the limit of \eqref{020706a}, as $\eps\to0$, is the same as that
of  
\begin{equation}
\label{020706b}
\int_{\bbR^2\times \bbT} \frac{\widehat
   W_\eps(\eta',k') \nu(k')
   \hat
   G^*(\eta,k') d\eta d\eta' dk'}{[\la+i\om'(k')\eta][\la+i\om'(k')\eta']}.
\end{equation}
Summarizing, the above argument proves that
\begin{equation}
\label{020706}
\lim_{\eps\to0+}\int_{\bbR\times\bbT}\frac{\hat G^*(\eta,k+\eps\eta/2){\frak d}_\eps^1(\la,k)}{\la+i\delta_\eps^+\om(k,\eta)}dkd\eta
= \int_{\bbR\times \bbT}\frac{\nu(k)\widehat
  W_0(\eta',k)}{\la+i\om'(k)\eta'} \left\{\int_{\bbR}\frac{\hat
    G^*(\eta,k)}{\la+i\om'(k)\eta}d\eta\right\}d\eta' dk,
\end{equation}
for any test function $G\in {\cal S}(\bbT\times\bbR)$.
Similarly, we have
\begin{equation}
\label{020706b-bis}
\lim_{\eps\to0+}\int_{\bbR\times\bbT}\frac{\hat G^*(\eta,k-\eps\eta/2)({\frak d}_\eps^1)^*(\la,k)}{\la+i\delta_\eps^-\om(k,\eta)}dkd\eta
= \int_{\bbR\times \bbT}\frac{\nu^*(k)\widehat
  W_0^*(\eta',k)}{\la-i\om'(k)\eta'} \left\{\int_{\bbR}\frac{\hat
    G^*(\eta,k)}{\la+i\om'(k)\eta}d\eta\right\}d\eta' dk.
\end{equation}
As
$
\widehat W_\eps^*(\eta,k)=\widehat W_\eps(-\eta,k),
$
we conclude that \eqref{020811}
holds.~\qed

%\textcolor{red}{\bf DOTAD}

\subsection{Asymptotics of the second scattering term}
%  ${\frak L}_{scat,2}^\eps(\la)$}

Let us split ${\frak L}_{scat,2}^\eps(\la)$ as
\begin{eqnarray}
\label{041511}
&&{\frak L}_{scat,2}^\eps(\la)=-\frac{\ga}{2}\int_{\bbR\times \bbT}\left[{\frak
   d}_\eps^2\left(\la,k-\frac{\eps \eta}{2}\right)+({\frak
   d}_\eps^2)^*\left(\la,k+\frac{\eps
   \eta}{2}\right)\right]\frac{\hat
   G^*(\eta,k)}{\la+i\delta_\eps\om(k,\eta)}d\eta dk\\
   &&=-\frac{\ga}{2}\int_{\bbR\times
  \bbT} \Big[{\frak
   d}_\eps^2\left(\la,k\right) \frac{\hat
   G^*(\eta,k+\eps\eta/2)}{\la+i\delta_\eps^+\om(k,\eta)}+
   ({\frak d}_\eps^2)^*\left(\la,k\right)\frac{\hat
   G^*(\eta,k-\eps\eta/2)}{\la+i\delta_\eps^-\om(k,\eta)}\Big]d\eta dk\nonumber\\
&&   ={\frak L}_{scat,21}^\eps(\la)+{\frak L}_{scat,22}^\eps(\la),\nonumber
\end{eqnarray}
with the two terms corresponding to writing 
\begin{equation}\label{feb1514}
{\frak d}_\eps^2={\rm Re}{\frak d}_\eps^2+i{\rm Im}{\frak d}_\eps^2.
\end{equation} 
%here
%\begin{equation}
%\label{041511a}
%{\frak L}_{scat,21}^\eps(\la)=-\frac{\ga}{2}\int_{\bbR\times
%  \bbT}{\rm Re}\,{\frak
%   d}_\eps^2\left(\la,k\right) \left[ \frac{\hat
%   G^\star(\eta,k+\eps\eta/2)}{\la+i\delta_\eps^+\om(k,\eta)}+\frac{\hat
%   G^\star(\eta,k-\eps\eta/2)}{\la+i\delta_\eps^-\om(k,\eta)}\right]d\eta dk
%\end{equation}
%and
%\begin{equation}
%\label{041511b}
%{\frak L}_{scat,22}^\eps(\la)=-\frac{i\ga}{2}\int_{\bbR\times
%  \bbT}{\rm Im}\,{\frak
%   d}_\eps^2\left(\la,k\right) \left[ \frac{\hat
%   G^\star(\eta,k+\eps\eta/2)}{\la+i\delta_\eps^+\om(k,\eta)}-\frac{\hat
%   G^\star(\eta,k-\eps\eta/2)}{\la+i\delta_\eps^-\om(k,\eta)}\right]d\eta dk.
%\end{equation}
We recall that
$$
{\frak d}_\eps^2(\la,k)=\eps\int_0^{+\infty}e^{-\la\eps t}I\!I_\eps\left(t,k\right)dt
= -\gamma \eps\int_0^{+\infty}e^{-\la\eps t}dt \left\{\int_0^{t}
  e^{i\om(k)(t-s)}\left\langle g\star {\frak p}_0^0(s)g\star {\frak p}_0^0(t)
\right\rangle_{\mu_\eps}\right\} ds .
$$
We will prove the following. %\red{The first term is transmission, the second is reflection.}
\begin{lemma}\label{lem-feb1502}
For any $\la>0$ and $G\in {\cal S}(\bbR\times \bbT)$ we have
\begin{eqnarray}
\label{021511}
&&\lim_{\eps\to 0}{\frak L}_{scat,2}^\eps(\la)=
\frac{\gamma}{4}\int_{\bbR\times\bbT}\frac{\fgeeszett(k)\widehat
  W(\eta',k)d\eta' dk}{\la+i\om'(k)\eta'} \int_{\bbR}
  \frac{\hat G^*(\eta,k)d\eta}{\la+i\om'(k)\eta}\\
&&
+\frac{\gamma}{4}\int_{\bbR\times \bbT}\frac{\fgeeszett(k)\widehat
  W(\eta',-k)d\eta' dk}{\la-i\om'(k)\eta'} 
  \int_{\bbR}\frac{\hat G^*(\eta,k)d\eta}{\la+i\om'(k)\eta}.\nonumber
\end{eqnarray}
\end{lemma}
%{\bf Remark.} According to \eqref{020104} we have $|\nu|^2/|\om'|\in
%L^\infty(\bbT)$ so the right hand side of \eqref{021511} is well
%defined for any $G\in  {\cal S}(\bbR\times \bbT)$ and $\la>0$.
%
%\bigskip
The conclusion of this lemma is the consequence of the following two limits
\begin{eqnarray}
\label{021511a}
&&\lim_{\eps\to0+}{\frak L}_{scat,21}^{\eps}(\la)=
\frac{\gamma}{4}\int_{\bbR\times\bbT}\frac{\fg(k)
\widehat W(\eta',k)d\eta' dk}{\la+i\om'(k)\eta'} 
\int_{\bbR}\frac{\hat G^*(\eta,k)d\eta}{\la+i\om'(k)\eta}\\
&&
+\frac{\gamma}{4}\int_{\bbR\times \bbT}\frac{\fg(k)
\widehat W(\eta',-k)d\eta' dk}{ \la-i\om'(k)\eta'} \int_{\bbR}
\frac{\hat G^*(\eta,k)d\eta}{\la+i\om'(k)\eta},\nonumber
\end{eqnarray}
and
\begin{equation}
\label{021511b}
\lim_{\eps\to0+}{\frak L}_{scat,22}^\eps(\la)=0.
\end{equation}

\subsection{The limit of the full scattering term}\label{sec:4.3}

Putting together the results of Lemmas~\ref{lem-feb1504} and \ref{lem-feb1502},
we see that
\begin{eqnarray} 
&&\lim_{\eps\to 0}{\frak L}_{scat}^{\eps}(\la)=
-\ga \int_{\bbR\times\bbT}{\rm Re} [\nu(k)]
\frac{\widehat W_\eps(\eta',k)}{\la+i\om'(k)\eta'} 
\left\{\int_{\bbR}\frac{G^*(\eta,k)}{\la+i\om'(k)\eta}d\eta\right\}dk d\eta'
\nonumber\\
&&
+\frac{\gamma}{4}\int_{\bbR\times\bbT}\frac{\fg(k)
\widehat W(\eta',k)d\eta' dk}{\la+i\om'(k)\eta'} 
\int_{\bbR}\frac{\hat G^*(\eta,k)d\eta}{\la+i\om'(k)\eta}
+\frac{\gamma}{4}\int_{\bbR\times \bbT}\frac{\fg(k)
\widehat W(\eta',-k)d\eta' dk}{ \la-i\om'(k)\eta'} \int_{\bbR}
\frac{\hat G^*(\eta,k)d\eta}{\la+i\om'(k)\eta}\nonumber\\
&&=\int [{\cal W}_{tr}(\eta,k)+{\cal W}_{ref}(\eta,k)]\hat G^*(\eta,k)
\frac{|\bar\omega'(k)|d\eta dk}{\lambda+i\omega'(k)\eta}d\eta dk,
\label{feb1506}
\end{eqnarray}
with the transmission term
\begin{eqnarray}\label{feb1508}
{\cal W}_{tr}(\eta,k)=\farc{\gamma}{|\bar\omega'(k)|}
\Big[- {\rm Re}[\nu(k)]+\frac{\fg(k) }{4}\Big]
\int\frac{ \widehat W(\eta',k)d\eta' dk}{\la+i\om'(k)\eta'}=
(p_+(k)-1)\int\frac{ \widehat W(\eta',k)d\eta' dk}{\la+i\om'(k)\eta'}.
\end{eqnarray}
We used (\ref{feb1402}) in the last step. The other term in (\ref{feb1506}),
corresponding to reflection, is
\begin{eqnarray}\label{feb1510}
{\cal W}_{ref}(\eta,k)=\frac{\gamma\fg(k)}{4|\bar\omega'(k)| }
\int_{\bbR\times \bbT}\frac{ 
\widehat W(\eta',-k)d\eta' dk}{ \la-i\om'(k)\eta'} =
p_-(k) \int_{\bbR\times \bbT}\frac{ 
\widehat W(\eta',-k)d\eta' dk}{ \la-i\om'(k)\eta'}.
\end{eqnarray}
Combining the scattering terms in
(\ref{feb1506})-(\ref{feb1510}), together with the ballistic term in
(\ref{073110})% and the thermostat term in (\ref{feb1430})
,
we obtain (\ref{013110}). Thus, the proof of Theorem~\ref{main:thm}
is reduced to the computation in Lemma~\ref{lem-feb1502}.

%We split \eqref{041511} as ${\frak L}_\eps^2(\la)=
%{\frak L}_{\eps,1}^2(\la)+ {\frak L}_{\eps,2}^2(\la)$, 
 %X\subsection{Proof of \eqref{021511a}}

\section{The proof of Lemma~\ref{lem-feb1502}: 
the limit of ${\frak L}_{scat,21}^{\eps}(\la)$}\label{sec:6}

We now turn to the proof of Lemma~\ref{lem-feb1502}. In this section,
we begin the rather long and technical computation leading to
(\ref{021511a}). 

\subsubsection*{A calculation of ${\rm Re}\,{\frak d}_\eps^2$}

Recall that ${\frak L}_{scat,21}^{\eps}(\la)$ comes from the contribution
to ${\frak L}_{scat,2}^{\eps}(\la)$ that appears from ${\rm Re}\,{\frak d}_\eps^2$.
Our first task, therefore, is to compute ${\rm Re}\,{\frak d}_\eps^2$.
We have
\begin{eqnarray*}
&&
2{\rm Re}\,{\frak d}_\eps^2(\la,k)
=-2\gamma \eps\left\langle\int_0^{+\infty}e^{-\la\eps t}dt \left\{\int_0^{t}  
\cos(\om(k)s)(g\star{\frak p}_0^0)(s)ds\right\} 
\cos(\om(k)t) (g\star{\frak p}_0^0)(t)
\right\rangle_{\mu_\eps}\\
&&
-2\gamma \eps\left\langle\int_0^{+\infty}e^{-\la\eps t}dt \left\{\int_0^{t}  \sin(\om(k)s)(g\star{\frak p}_0^0)(s)ds\right\} \sin(\om(k)t) 
(g\star {\frak p}_0^0)(t)\right\rangle_{\mu_\eps}\\
&&
=-\gamma \eps\left\langle\int_0^{+\infty}e^{-\la\eps t}\frac{d}{dt}\left\{\left[\int_0^{t}  \cos(\om(k)s)g*{\frak p}_0^0(s)ds\right]^2+\left[\int_0^{t}  \sin(\om(k)s)g*{\frak p}_0^0(s)ds\right]^2
\right\}dt \right\rangle_{\mu_\eps}.
\end{eqnarray*}
Integrating by parts, we obtain
\begin{eqnarray}
&&
2{\rm Re}\,{\frak d}_\eps^2(\la,k)=
-\gamma \eps^2\la\int_0^{+\infty}e^{-\la\eps t} dt\left\langle\left\{\int_0^{t}  \cos(\om(k)s)(g\star{\frak p}_0^0)(s)ds\right\}^2
\right\rangle_{\mu_\eps}\nonumber
\\
&&
-\gamma \eps^2\la\int_0^{+\infty}e^{-\la\eps t} dt\left\langle\left\{\int_0^{t}  \sin(\om(k)s)(g\star{\frak p}_0^0)(s)ds\right\}^2
\right\rangle_{\mu_\eps}:=C_\eps(\la,k)+S_\eps(\la,k).\label{feb1518}
\end{eqnarray}
%Denote the terms in the right hand side by $C_\eps(\la,k)$ and
%$S_\eps(\la,k)$ respectively.
The first term in the right side is
\begin{equation}
\label{012803}
C_\eps(\la,k)=
-\gamma \eps^2\la\int_0^{+\infty}e^{-\la\eps t} dt\int_0^{t} \int_0^t ds ds'  \cos(\om(k)s)\cos(\om(k)s')\langle g*{\frak p}_0^0(s)g*{\frak p}_0^0(s')\rangle_{\mu_\eps}
\end{equation}
Using \eqref{null} and \eqref{eq:1} gives
\begin{eqnarray}
&&
\!\!\!\eps\langle (g\star{\frak p}_0^0)(s)(g\star{\frak p}_0^0)(s')\rangle_{\mu_\eps}
=\frac14\int_0^s\int_0^{s'}g(d\tau) g(d\tau')\int_{\bbT^2}d\ell d\ell' 
e^{-i\om(\ell) (s-\tau)}e^{i\om(\ell') (s'-\tau')} 
\eps\langle \hat\psi(\ell)\hat\psi^*(\ell')\rangle_{\mu_\eps}
\nonumber\\
&&
\!\!+\frac14\int_0^s\int_0^{s'}g(d\tau) g(d\tau')\int_{\bbT^2}d\ell d\ell' e^{i\om(\ell) (s-\tau)}e^{-i\om(\ell') (s'-\tau')} 
\eps\langle \hat\psi^*(\ell)\hat\psi(\ell')\rangle_{\mu_\eps}.\label{feb1522}
\end{eqnarray} 
Now, symmetry implies that the two terms above make an identical
contribution to $C_\eps(\la,t)$, hence
%which leads to a decomposition
%\begin{equation}\label{feb1520}
%C_\eps(\la,t)=C_\eps^1(\la,t)+C_\eps^2(\la,t),
%\end{equation}
%according to the two terms in (\ref{feb1522}).
%%Let us denote by $C_\eps^j(\la,t)$, $j=1,2$ the respective terms that arise
%%after substituting from the above into \eqref{012803}.
%The first term can be re-written as  
\begin{eqnarray}
\label{012803a}
&&
C_\eps(\la,k)=
-\frac{\gamma}{2} \eps\la\int_0^{+\infty}e^{-\la\eps t} dt\int_0^{t} \int_0^t ds ds'  \cos(\om(k)s)\cos(\om(k)s')\int_0^s\int_0^{s'}g(d\tau) g(d\tau')\int_{\bbT^2}d\ell d\ell' \nonumber\\
&&
\times e^{-i\om(\ell) (s-\tau)}e^{i\om(\ell') (s'-\tau')} 
\eps\langle \hat\psi(\ell)\hat\psi^*(\ell')\rangle_{\mu_\eps}\\
%\end{eqnarray}
%Using the relation
%$
%(2\pi)^{-1}\int_{\bbR}e^{i\beta t}d\beta=\delta(t)
%$
%we can write
%\begin{eqnarray*}
&&
=-\frac{\gamma}{4\pi} \eps\la\int_{\bbR}d\beta \int_{\bbT^2}d\ell d\ell' \eps\langle \hat\psi(\ell)\hat\psi^*(\ell')\rangle_{\mu_\eps}\int_0^{+\infty} \int_0^{+\infty}e^{i\beta(t-t')}e^{-\la\eps (t+t')/2} dt dt' \nonumber\\
&&
\times \int_0^{t} \int_0^{t'} ds ds'  \cos(\om(k)s)\cos(\om(k)s') \int_0^s\int_0^{s'}g(d\tau) g(d\tau')e^{-i\om(\ell) (s-\tau)}e^{i\om(\ell') (s'-\tau')}
\nonumber\\
&&
=-\frac{\gamma}{4\pi} \eps\la\int_{\bbR}d\beta \int_{\bbT^2}\eps\langle \hat\psi(\ell)\hat\psi^*(\ell')\rangle_{\mu_\eps} \left|\Om(\ell,\ell',k,\la)\right|^2 d\ell d\ell' ,\nonumber
\end{eqnarray}
with
$$
\Om(\ell,\ell',k,\la):=\int_0^{+\infty}\cos(\om(k)s)ds\left\{\int_0^s e^{-i\om(\ell) (s-\tau)}g(d\tau)\int_{s}^{+\infty}e^{(-\la\eps/2+i\beta) t}dt\right\}.
$$
Integrating out first the $t$ variable, and then the $s$ varable,
we obtain 
\begin{eqnarray*}
&&
\Om(\ell,\ell',k,\la)
=\frac{1}{\la\eps/2-i\beta}\int_0^{+\infty}e^{i\om(\ell)
  \tau}g(d\tau)\int_{\tau}^{+\infty}
\cos(\om(k)s)e^{[-\la\eps/2+i(\beta-\om(\ell))] s}ds
\\
&&
=\frac{1}{2(\la\eps/2-i\beta)}\int_0^{+\infty}g(d\tau)e^{i\om(\ell) \tau} \left\{\frac{e^{[-\la\eps/2+i[\beta+\om(k)-\om(\ell)]] \tau}}{\la\eps/2-i(\beta+\om(k)-\om(\ell))}+\frac{e^{[-\la\eps/2+i(\beta-\om(k)-\om(\ell))] \tau}}{\la\eps/2-i[\beta+\om(k)+\om(\ell)]}\right\}\\
&&
=\frac{1}{2(\la\eps/2-i\beta)}\left\{\frac{\tilde g\left(\la\eps/2-i[\beta+\om(k)]\right) }{\la\eps/2-i(\beta+\om(k)-\om(\ell))}+\frac{\tilde g\left(\la\eps/2-i[\beta-\om(k))]\right)}{\la\eps/2-i[\beta+\om(k)+\om(\ell)]}\right\}.
\end{eqnarray*}
% Similarly we conclude that
% \begin{eqnarray*}
% &&
% \int_0^{+\infty}ds'\cos(\om(k)s')\int_0^{s'} g(d\tau')e^{i\om(\ell') (s'-\tau')}\int_{s'}^{+\infty}e^{(-\la\eps/2-i\beta) t'}dt'\\
% &&
% =\frac{1}{2(\la\eps/2+i\beta)}\left\{\frac{\tilde g\left(\la\eps/2+i[\beta+\om(k)]\right) }{\la\eps/2+i(\beta+\om(k)-\om(\ell'))}+\frac{\tilde g\left(\la\eps/2+i[\beta+\om(k)]\right)}{\la\eps/2+i[\beta+\om(k)+\om(\ell')]}\right\}.
% \end{eqnarray*}
Hence, after a change of variables $\beta:=\eps\beta'$ we get
\begin{eqnarray}\label{feb1524}
&&
C_\eps(\la,k)
=-\frac{\gamma\la}{16\cdot\pi\eps^2}\int_{\bbR}\frac{d\beta}{(\la/2)^2+\beta^2} \int_{\bbT^2}d\ell d\ell' \eps\langle \hat\psi(\ell)\hat\psi^*(\ell')\rangle_{\mu_\eps}\\
&&
\times \left\{\frac{\tilde g\left(\la\eps/2-i[\eps\beta+\om(k)]\right) }{\la/2-i\{\beta+\eps^{-1}[\om(k)-\om(\ell)]\}}+\frac{\tilde g\left(\la\eps/2-i[\eps\beta-\om(k)]\right)}{\la/2-i\{\beta+\eps^{-1}[\om(k)+\om(\ell)]\}}\right\}
\nonumber\\
&&
\times \left\{\frac{\tilde g\left(\la\eps/2+i[\eps\beta+\om(k)]\right) }{\la/2+i\{\beta+\eps^{-1}[\om(k)-\om(\ell')]\}}+\frac{\tilde g\left(\la\eps/2+i[\eps \beta+\om(k)])\right)}{\la/2+i\{\beta+\eps^{-1}[\om(k)+\om(\ell')]\}}\right\}.\nonumber
\end{eqnarray}
%By symmetry we obtain $C_\eps^1(\la,k)=C_\eps^2(\la,k)$ and 
%as a result we conclude
%\begin{eqnarray*}
%&&
%C_\eps(\la,k)
%=-\frac{\gamma\la}{2^4\pi\eps^2}\int_{\bbR}\frac{d\beta}{(\la/2)^2+\beta^2} \int_{\bbT^2}d\ell d\ell' \eps\langle \hat\psi(\ell)\hat\psi^*(\ell')\rangle_{\mu_\eps}\\
%&&
%\times \left\{\frac{\tilde g\left(\la\eps/2-i[\eps\beta+\om(k)]\right) }{\la/2-i\{\beta+\eps^{-1}[\om(k)-\om(\ell)]\}}+\frac{\tilde g\left(\la\eps/2-i[\eps\beta-\om(k)]\right)}{\la/2-i\{\beta+\eps^{-1}[\om(k)+\om(\ell)]\}}\right\}\\
%&&
%\times \left\{\frac{\tilde g\left(\la\eps/2+i[\eps\beta+\om(k)]\right) }{\la/2+i\{\beta+\eps^{-1}[\om(k)-\om(\ell')]\}}+\frac{\tilde g\left(\la\eps/2+i[\eps \beta+\om(k)]\right)}{\la/2+i\{\beta+\eps^{-1}[\om(k)+\om(\ell')]\}}\right\}.
%\end{eqnarray*}
%We have
%\begin{equation}
%\label{012803s}
%S_\eps(\la,k)=
%-\gamma \eps^2\la\int_0^{+\infty}e^{-\la\eps t} dt\int_0^{t} \int_0^t ds ds'  \sin(\om(k)s)\sin(\om(k)s')\langle g*{\frak p}_0^0(s)g*{\frak p}_0^0(s')\rangle_{\mu_\eps}
%\end{equation}
A similar calculation leads to  %to the ones done for $C_\eps(\la,k)$ 
%we conclude that
\begin{eqnarray}\label{feb1526}
&&
S_\eps(\la,k)
=\frac{\gamma\la}{2^4\pi\eps^2}\int_{\bbR}\frac{d\beta}{(\la/2)^2+\beta^2} \int_{\bbT^2}d\ell d\ell' \eps\langle \hat\psi(\ell)\hat\psi^*(\ell')\rangle_{\mu_\eps}\\
&&
\times \left\{\frac{\tilde g\left(\la\eps/2-i[\eps\beta+\om(k)]\right) }{\la/2-i\{\beta+\eps^{-1}[\om(k)-\om(\ell)]\}}-\frac{\tilde g\left(\la\eps/2-i[\eps\beta-\om(k)]\right)}{\la/2-i\{\beta+\eps^{-1}[\om(k)+\om(\ell)]\}}\right\}
\nonumber\\
&&
\times \left\{\frac{\tilde g\left(\la\eps/2+i(\eps\beta-\om(k)])\right) }{\la/2+i\{\beta-\eps^{-1}[\om(k)+\om(\ell')]\}}-\frac{\tilde g\left(\la\eps/2+i[\eps\beta+\om(k)]\right)}{\la/2+i\{\beta+\eps^{-1}[\om(k)-\om(\ell')]\}}\right\}.
\nonumber
\end{eqnarray}
Putting (\ref{feb1518}), (\ref{feb1524}) and (\ref{feb1526}) together, gives
\begin{equation}
\label{010104}
2{\rm Re}\,{\frak d}_\eps^2(\la,k)
=R_\eps(\la,k)+\rho_\eps(\la,k),
\end{equation}
with
\begin{eqnarray}
\label{R-eps}
&&
R_\eps(\la,k):=-\frac{\gamma\la}{8\pi\eps^2}\int_{\bbR}\frac{d\beta}{(\la/2)^2+\beta^2} \int_{\bbT^2}d\ell d\ell' \eps\langle \hat\psi(\ell)\hat\psi^*(\ell')\rangle_{\mu_\eps}\\
&&
\times \frac{|\tilde g\left(\la\eps/2-i[\eps\beta+\om(k)]\right)|^2 }{\la/2-i\{\beta+\eps^{-1}[\om(k)-\om(\ell)]\}}
\times \frac{1
   }{\la/2+i\{\beta+\eps^{-1}[\om(k)-\om(\ell')]\}}\nonumber
\end{eqnarray}
and
\begin{eqnarray}
\label{rho-e}
&&
\rho_\eps(\la,k):=-\frac{\gamma\la}{16\cdot \pi\eps^2}
\int_{\bbR}\frac{d\beta}{(\la/2)^2+\beta^2} 
\int_{\bbT^2}d\ell d\ell' \eps\langle \hat\psi(\ell)
\hat\psi^*(\ell')\rangle_{\mu_\eps}
%\\
%&&
%\times
\Big\{ \frac{\tilde g\left(\la\eps/2-i[\eps\beta+\om(k)]\right) }{\la/2-i\{\beta+\eps^{-1}[\om(k)-\om(\ell)]\}}\nonumber\\
&&\times
 \Big\{\frac{\tilde g\left(\la\eps/2+i[\eps\beta+\om(k)]\right)
   }{\la/2+i\{\beta+\eps^{-1}[\om(k)+\om(\ell')]\}}+\frac{\tilde
   g\left(\la\eps/2+i(\eps\beta-\om(k)])\right)
   }{\la/2+i\{\beta-\eps^{-1}[\om(k)+\om(\ell')]}\Big\} \nonumber\\
&&
+\frac{\tilde
   g\left(\la\eps/2-i[\eps\beta-\om(k)]\right)}{\la/2-i\{\beta+\eps^{-1}[\om(k)+\om(\ell)]\}}\\
   &&\times
 \Big\{\frac{\tilde g\left(\la\eps/2+i[\eps \beta+\om(k)]\right)}{\la/2+i\{\beta+\eps^{-1}[\om(k)+\om(\ell')]\}}+\frac{\tilde g\left(\la\eps/2+i(\eps\beta-\om(k)])\right) }{\la/2+i\{\beta-\eps^{-1}[\om(k)+\om(\ell')]}\Big\}\Big\}.\nonumber
\end{eqnarray}
The main contribution to ${\frak L}_{scat,21}^\eps(\la)$
will come from the term $R_\eps(\lambda,k)$ due to the
difference $\omega(k)-\omega(\ell)$ in the denominator that can become small.
As $\rho_\eps(\la,k)$ only has the sum $\omega(k)+\omega(\ell)$, we expect its
contribution to be small in the limit. 
More precisely, we will show the following result for the limit
of $R_\eps(\la,k)$. %\red{Say which one is transmission and which one is reflection}
\begin{lemma}\label{lem-feb1508}
Let
\begin{equation}\label{Hpm}
{\cal H}_\pm(\la,\eps):=\int_{\bbR\times
  \bbT}R_\eps\left(\la,k\right) \frac{\hat
   G^*(\eta,k\pm \eps\eta/2) }{\la+i\delta_\eps^\pm\om(k,\eta)}
d\eta dk
\end{equation}
and
\begin{equation}\label{mar1402}
{\cal I}_{tr}(\la):=-2\gamma \pi\int_{\bbR\times\bbT}\frac{|\nu(k)|^2\widehat
  W(\eta',\pm k)d\eta' dk}{|\om'(k)|[\la+ i\om'(k)\eta']} \int_{\bbR}\frac{\hat G^*(\eta,k)d\eta}{\la+i\om'(k)\eta},
\end{equation}
and
\begin{equation}\label{mar1404}
{\cal I}_{ref}(\la):=-2\gamma \pi\int_{\bbR\times\bbT}\frac{|\nu(k)|^2\widehat
  W(\eta',\pm k)d\eta' dk}{|\om'(k)|[\la- i\om'(k)\eta']} \int_{\bbR}\frac{\hat G^*(\eta,k)d\eta}{\la+i\om'(k)\eta},
\end{equation}
then
\begin{equation}
\label{021511c}
\lim_{\eps\to0+}{\cal H}_\pm(\la,\eps)=\frac12(\vphantom{\int_0^1}{\cal I}_{tr}(\la)+{\cal I}_{ref}(\la)).
\end{equation}
\end{lemma}
On the other hand, $\rho_\eps(\la,k)$ vanishes in the limit.
\begin{lemma}
\label{lm011311}
For each $\la>0$ we have
\begin{equation}
\label{011311}
\lim_{\eps\to0+}\int_{\bbT}|\rho_\eps(\la,k)|dk=0.
\end{equation}
\end{lemma}
These two lemmas, together with (\ref{041511})-(\ref{feb1514}) and (\ref{010104}) 
imply (\ref{021511a}). 

\subsection*{Proof of Lemma~\ref{lm011311}}

% We first prove Lemma~\ref{lm011311}.

A word on notation: for two functions $f,g:D\to \bbR$ we say that $f\preceq g$ if there
exists $C>0$ such that $f(x)\le C g(x)$, $x\in D$. We shall use the
notation $f\approx g$ if $f\preceq g$ and $g\preceq f$.

Opening the parentheses in \eqref{rho-e}, we can
write 
\[
\rho_\eps(\la,k)=\sum_{j=1}^4 \rho_\eps^j(\la,k).
\]
We will only show that 
\begin{equation}
\label{0113111}
\lim_{\eps\to0}\int_{\bbT}|\rho_\eps^1(\la,k)|dk=0,
\end{equation}
as the other terms are analyzed in a similar fashion.
To verify \eqref{0113111}, it suffices to show that
\begin{eqnarray}
\label{021311}
&&
\lim_{\eps\to0}\frac{1}{\eps}\int_{\bbR}\frac{d\beta}{(\la/2)^2+\beta^2}
   \int_{\bbT^3} \left|\langle
   \hat\psi(\ell)\hat\psi^*(\ell')\rangle_{\mu_\eps}\right|\\
&&
\times \left|\frac{\tilde g\left(\la\eps/2-i[\eps\beta+\om(k)]\right) }{\la/2-i\{\beta+\eps^{-1}[\om(k)-\om(\ell)]\}}\right|
\times
\left|\frac{\tilde g\left(\la\eps/2+i[\eps\beta+\om(k)]\right)
   }{\la/2+i\{\beta+\eps^{-1}[\om(k)+\om(\ell')]\}}\right|dkd\ell d\ell'=0.\nonumber
\end{eqnarray}
Change variables 
\begin{equation}
\label{ell-k}
\ell=:k'+\frac{\eps \eta'}{2},\quad \ell'=:k'-\frac{\eps \eta'}{2}
\end{equation}
and let
\begin{equation}
\label{T2e}
T^2_\eps:=\left[(\eta',k'):\,|\eta'|\le
\frac{1}{\eps},\,|k'|\le\frac{1-\eps |\eta'|}{2}\right]
\subset\bbT_{2/\eps}\times \bbT,
\end{equation}
be the image of $\bbT^2$ under the inverse map, as below (\ref{020706a}). 
%$$
%k':=\frac{\ell+\ell'}{2},\quad \eta':=\frac{\ell-\ell'}{\eps}.
%$$
%We have
%\begin{equation}
%\label{T2e}
%T^2_\eps:=\left[(\eta',k'):\,|\eta'|\le
%  \frac{1}{\eps},\,|k'|\le\frac{1-\eps |\eta'|}{2}\right],
%\end{equation}
%with the equivalence relation $(\eta'_1,k'_1)\equiv (\eta'_2,k'_2)$,
%given by
%$(\eta'_1-\eta'_2,k'_1-k'_2)=(\iota_1 \eps^{-1},\iota_2 1)$ with
%$\iota_1,\iota_2=\pm$. 
%Note that $T^2_\eps \subset\bbT_{2/\eps}\times \bbT$.
The expression under the limit in \eqref{021311} can then be estimated by
\begin{eqnarray}
\label{011411}
&&
\frac{\|\tilde g\|_\infty^2}{\eps}\int_{\bbR}\frac{d\beta}{(\la/2)^2+\beta^2}
   \int_{\bbT^2\times \bbT_{2/\eps}} \frac{|\hat W_\eps(\eta',k')| }
   {|\la/2-i\{\beta+\eps^{-1}[\om(k)-\om(k'+\eps \eta'/2)]\}|}
\nonumber\\
&&
\times
\frac{dkd k' d\eta'
   }{|\la/2+i\{\beta+\eps^{-1}[\om(k)+\om(k'-\eps \eta'/2)]\}|}\le I_{1,\eps}+I_{2,\eps},
\end{eqnarray}
where
\begin{eqnarray*}
&&
 I_{1,\eps}:=\frac{\|\tilde g\|_\infty^2}{\eps}\int_{\bbR}\frac{d\beta}{(\la/2)^2+\beta^2}
   \int_{\bbT^2\times \bbT_{2/\eps}} \frac{|\hat W_\eps(\eta',k')| }{|\la+i\eps^{-1}\left(\om(k'-\eps \eta'/2)+\om(k'+\eps
   \eta'/2)\right)|}
\\
&&
\times \frac{dkd k' d\eta'}{|\la/2-i\{\beta+\eps^{-1}[\om(k)-\om(k'+\eps
  \eta'/2)]\}|}
\end{eqnarray*}
and
\begin{eqnarray*}
&&
 I_{2,\eps}:=\frac{\|\tilde g\|_\infty^2}{\eps}\int_{\bbR}\frac{d\beta}{(\la/2)^2+\beta^2}
   \int_{\bbT^2\times \bbT_{2/\eps}} \frac{|\hat W_\eps(\eta',k')| }{|\la+i\eps^{-1}\left(\om(k'-\eps \eta'/2)+\om(k'+\eps
   \eta'/2)\right)|}
\\
&&
\times \frac{dkd k' d\eta'}{|\la/2+i\{\beta+\eps^{-1}[\om(k)+\om(k'-\eps
   \eta'/2)]\}|}.
\end{eqnarray*}
We used here the identity
\[
\farc{1}{(\lambda/2-ia)(\lambda/2+ib)} =
\Big(\frac{1}{\lambda/2-ia}+\frac{1}{\lambda/2+ib}\Big)\frac{1}{\lambda-i(a-b)}.
\]
%Consider first the case $\om_{\rm min}>0$.
%We can estimate then 
%\begin{align}
%\label{011511}
% &
%I_{1,\eps} \preceq
%\frac{\|\tilde g\|_\infty^2}{\omm}\int_{\bbR}\frac{d\beta}{(\la/2)^2+\beta^2}
%   \int_{\bbT^2\times \bbT_{2/\eps}} \frac{|\hat W_\eps(\eta',k')| 
%   dkd k' d\eta'}{|\la/2-i\{\beta+\eps^{-1}[\om(k)-\om(k'+\eps
%  \eta'/2)]\}|}\nonumber\\
%&
%\preceq \int_{\bbR}\frac{d\beta}{(\la/2)^2+\beta^2}
%   \int_{\bbT^2\times \bbT_{2/\eps}} \frac{\varphi(\eta')\psi(k') dkd k' d\eta'}{|1+|\beta+\eps^{-1}[\om(k)-\om(k'+\eps
%  \eta'/2)|]|}.
%\end{align}
%The right hand side of \eqref{011511} tends to $0$, with $\eps\to0$,
%by virtue of the dominated convergence theorem.
%
Now, we can   estimate $I_{1,\eps}$ as follows:
$$
 I_{1,\eps}\le \eps\Gamma_\eps\|\tilde g\|_\infty^2\int_{\bbR}\frac{d\beta}{(\la/2)^2+\beta^2}
   \int_{\bbT\times \bbT_{2/\eps}} \frac{|\hat W_\eps(\eta',k')|d k' d\eta' }{\om(k'-\eps \eta'/2)+\om(k'+\eps
   \eta'/2)},
$$
with 
$$
 \Gamma_\eps :=\sup_{A\in\bbR}\int_{\bbT}\frac{dk}{|\eps\la/2-i\left(\om(k)-A\right)|}\le\Gamma_{\eps}^++\Gamma_{\eps}^-,
$$
%To estimate  $\Gamma_\eps$  we change variables $u:=\om(k)$.
%Then
%$\Gamma_\eps=\Gamma_{\eps}^++\Gamma_{\eps}^-$ with
with
$$
  \Gamma_{\eps}^\pm:=\sup_{A\in\bbR}\int_{0}^{\om_{\rm max}}\frac{du}{|\eps\la/2-i(u-A)||\om'(\om_\pm(u))|}.
 $$
Recall that $\om_-$, $\om_+$ are the decreasing and increasing
branches of the inverse function of the dispersion relation
$\om(\cdot)$. 
Our assumptions on the dispersion relation imply that  
\[
\om'(\om_\pm(u))\approx (\om_{\rm max}-u)^{1/2},
\hbox{ for $\om_{\rm max}-u\ll1$.}
\]
The consideration near the minimum of $\omega$ is 
identical unless $\omega_{\rm min}=0$,
in which case $|\omega'(k)|$ stays uniformly positive near the minimum.
Therefore, we have
$$
\Gamma_{\eps}^\pm\preceq \sup_{A\in [0,1]}\int_0^1\frac{du}{[\eps+|u-A|]\sqrt{u}}
\preceq \eps^{-1/2}\log\eps^{-1}.
 $$
%The expression in the right hand side  is
%of the order $\eps^{-1/2}\log\eps^{-1}$. 
We obtain therefore 
$$
 I_{1,\eps}\preceq \eps^{1/2}\log\eps^{-1}
   \int_{\bbT_{\eps}\times \bbT} \frac{|\hat W_\eps(\eta',k')| d\eta' d k' }{\om(k'-\eps \eta'/2)+\om(k'+\eps
   \eta'/2)}\preceq  
   \int_{0}^{+\infty}\int_0^1 
   \frac{\eps^{1/2}\log\eps^{-1} dq d u }{(u+\eps (q\wedge 1))(1+q^{3+2\kappa})}\to 0,
$$
as $\eps\to 0$, due to \eqref{011812} and (\ref{011812c}), and since 
if $\omega_{\rm min}=0$, then $\om(k)$ behaves as $|k|$ near the minimum~$k=0$. 
One can easily verify that the right hand side
vanishes, with $\eps\to0$. 
Similarly we obtain also that
\[
\lim_{\eps\to0+}I_{2,\eps}=0,
\]
which finishes the proof of Lemma~\ref{lm011311}. \qed 
%implies that \eqref{021311} holds in this case. \qed

\section{The proof of Lemma \ref{lem-feb1508} }\label{sec:lem5.1}

\subsection{Outline of the proof}

We now turn to the proof of Lemma \ref{lem-feb1508}, the main ingredient in the computation of
the limit of~${\frak L}_{scat,21}^{\eps}(\la)$.
We will only consider the term ${\cal H}_+(\la,\eps)$, as the computation of the limit of
${\cal H}_-(\la,\eps)$ is essentially the same. We will focus on the harder case when the dispersion relation~$\omega(k)$ is smooth both at its
maximum~$k=1/2$ and its minimum $k=0$, so that the inverse function has a square root singularity at each of these points.
That is, the two  branches of its inverse
 $\om_+:[\om_{\rm min},\om_{\rm max}]\to[0,1/2]$ and $\om_-:=-\om_+$  
satisfy 
\[
\om_\pm'(w)=\pm (w-\om_{\rm min})^{-1/2}\chi_*(w),~~ w-\om_{\rm min}\ll1,
\]
and 
\[
\om_\pm'(w)=\pm (\om_{\rm max}-w)^{-1/2}\chi^*(w),~~\om_{\rm max}-w\ll1,
\]
with $\chi_*,\chi^*\in C^\infty(\bbT)$ that are strictly positive. 

%Let $k_{\rm min}$ and $k_{\rm max}$ be the minimizer and the maximizer 
%of the dispersion relation $\om(k)$, respectively. 
%Since the function $\om(k)$ is even and unimodal, we know that it attains its minimum at $k_{\rm min}=0$ and its maximum at 
%$k_{\rm max}=1/2$. Assume that the dispersion relation is smooth on $\bbT$ (\red{what about the acoustic case? Shouldn't this be in the
%introduction anyway?}) 
%and the  branches of its inverse
% $\om_+:[\om_{\rm min},\om_{\rm max}]\to[0,1/2]$ and $\om_-:=-\om_+$  
%satisfy 
%\[
%\om_\pm(w)=\pm (w-\om_{\rm min})^{1/2}\chi_*(w),~~ w-\om_{\rm min}\ll1,
%\]
%and 
%\[
%\om_\pm(w)\approx\pm (\om_{\rm max}-w)^{1/2}\chi_*(w),~~\om_{\rm max}-w\ll1,
%\]
%with $\chi_*,\chi^*\in C^\infty(\bbT)$ that are strictly positive.
%

Using \eqref{R-eps}  and the change of variables \eqref{ell-k} we can write  
\begin{eqnarray}
\label{011811}
&&
{\cal H}_+(\la,\eps)
=-\frac{\gamma\la}{4\pi\eps}\int_{\bbR\times \bbT_{2/\eps}}\frac{d\beta d\eta}{(\la/2)^2+\beta^2}
\int_{\bbT\times T_\eps^2}  \frac{\widehat W_\eps(\eta',k')  dk d\eta' dk'
}{\la/2-i\{\beta+\eps^{-1}[\om(k)-\om(k'+\eps \eta'/2)]\}}\\
&&
\times \frac{|\tilde g\left(\la\eps/2-i[\eps\beta+\om(k)]\right)|^2 }{\la/2+i\{\beta+\eps^{-1}[\om(k)-\om(k'-\eps \eta'/2)]\}}\times\frac{\hat
   G^*(\eta,k+\eps\eta/2) }{\la+i\delta_\eps^+\om(k,\eta)}. \nonumber
\end{eqnarray}
In fact, we may discard the contribution due to large $\eta'$, thanks
  to assumption \eqref{011812}. More precisely, let $\widetilde{\cal
    H}_+(\la,\eps)$ be the expression analogous to ${\cal
    H}_+(\la,\eps)$ corresponding to integration over $\eta'$ and $k'$ over
\begin{equation}
\label{T2el}
T^2_\eps:=\left[(\eta',k'):\,|\eta'|\le
\frac{\delta}{2^{100}\eps},\,|k'|\le\frac{1-\eps |\eta'|}{2}\right]
\subset\bbT_{2/\eps}\times \bbT,
\end{equation}
with $\delta$ as in \eqref{011812b}. Due to \eqref{011812} and \eqref{011812c} we have
\begin{eqnarray}
\label{011811-l}
&&
\left|{\cal H}_+(\la,\eps)-\widetilde {\cal H}_+(\la,\eps)\right|
\preceq \frac{1}{\eps^2}\int_{|\eta'|\ge
   \delta/(2^{100}\eps)}\frac{d\eta'}{(1+(\eta')^2)^{3/2+\kappa}}=\eps^{2\kappa}\to0,\quad
   \mbox{ as $\eps\to0$}.
\end{eqnarray}
In what follows we restreict ourselves therefore to studying the limit
of $\widetilde {\cal H}_+(\la,\eps)$.

The main contribution to the limit comes from the regions where $\omega(k)\approx\omega(k')$, that is, where 
either~$k\approx k'$ -- this generates the transmission term, or $k\approx -k'$ -- this is responsible for the reflection term in the limit.
The smallness of these regions will compensate for the factor $1/\eps$ in front of the integral in (\ref{011811}).
To distinguish   the contributions of these two regions, we decompose 
\begin{equation}\label{mar1406}
\widetilde{\cal H}_+(\la,\eps)=\sum_{\iota\in\{-,+\}}{\cal
  I}_{\iota}(\la,\eps).
\end{equation}
Here ${\cal I}_{\iota}(\la,\eps)$ correspond to the integration over 
the domains $\tilde T_{\eps,\iota}^3$, $\iota=\pm$:
\begin{eqnarray*}
&&
\tilde T^3_{\eps,\iota}:=\Big[(k,\eta',k')\in \bbT\times \tilde T_\eps^2:\,{\rm sign}\,
   k=\iota{\rm sign} (k'+\eps\eta'/2)\vphantom{\int_0^1}\Big],
\end{eqnarray*}
 so that the integration over $\tilde T_{\eps,+}^3$ will generate the transmission term, and over $\tilde T_{\eps,-}^3$ the reflection. 
Changing variables $k':=\iota k+\eps\eta''/2$,  and using the fact that $\omega(k)$ is even, gives
 \begin{eqnarray}
\label{011811aab}
&&
{\cal I}_{\iota}(\la,\eps)
=-\frac{\gamma\la}{8\pi}\int_{\bbR\times \bbT_{2/\eps}}\frac{d\beta d\eta}{(\la/2)^2+\beta^2}
\int_{ T_{\eps,\iota}^3}  \frac{\widehat W_\eps(\eta',\iota k+\eps\eta''/2)  dk d\eta' d\eta''
}{\la/2-i\{\beta+\eps^{-1}[\om(k)-\om(k+\iota (\eps \eta'/2+\eps \eta''/2))]\}} \nonumber\\
&&
\times \frac{|\tilde g\left(\la\eps/2-i[\eps\beta+\om(k)]\right)|^2
   }{\la/2+i\{\beta+\eps^{-1}[\om(k)-\om(k+\iota(-\eps \eta'/2+
\eps\eta''/2))]\}}\times \frac{\hat
   G^*(\eta,k+\eps\eta/2) }{\la+i\delta_\eps^+\om(k,\eta)}.
\end{eqnarray}
Here, $T_{\eps,\iota}^3\subset
\bbT\times \bbT_{2/\eps}\times \bbT_{6/\eps}$ is the pre-image of $\tilde T_{\eps,\iota}^3$ under the mapping $(k,\eta',\eta'')\mapsto
(k,\eta',\iota k+\eps\eta''/2)$:
\begin{equation}
\label{T2ek1}
T^3_{\eps,\iota}:=\Big[(k,\eta',\eta''):\,k\in\bbT,\,|\eta'|\le
  \frac{\delta}{2^{50}\eps},\,\Big|k+\iota\frac{\eps\eta''}{2}\Big|\le\frac{1-\eps |\eta'|}{2},\, \,{\rm sign}\,
   k={\rm sign} \left(k+\iota(\eps\eta'/2+\eps\eta''/2)\right)\Big].
   %\subset\bbT\times \bbT_{2/\eps}\times \bbT_{6/\eps}.
\end{equation}

We will pass to the limit $\eps\to 0$ in expression (\ref{011811aab}) in several steps. The first step will be to replace
the quotient $\eps^{-1}[\om(k)-\om(k+\iota (\eps \eta'/2+\eps \eta''/2))]$ in the first denominator by $-\iota\omega'(k)(\eta'+\eta'')/2$.
That is, we will show the following.
\begin{lemma}\label{lem-mar202}
We have
\begin{equation}
\label{012811}
\lim_{\eps\to0} \{{\cal I}_\iota(\la,\eps)- {\cal
    I}_\iota ^{(1)}(\la,\eps) \}=0, \quad \iota\in\{-,+\},
\end{equation}
where
\begin{align}
\label{022112}
&  {\cal I}_{ \iota}^{(1)}(\la,\eps):=- \frac{\gamma\la }{8\pi }\int_{\bbR\times \bbT_{2/\eps}}\frac{d\beta
d\eta}{(\la/2)^2+\beta^2}\int_{
  T_{\eps,\iota}^3} \frac{\widehat W_\eps\left(\eta',\iota k+\eps \eta''/2\right) }{\la/2-i\{\beta-\iota \om'(k)( \eta'+ \eta'')/2\}}\\
&
\times 
\frac{|\tilde g\left(\la\eps/2-i[\eps\beta+\om(k)]\right)|^2
  }{\la/2+i\{\beta+\eps^{-1}[\om(k)-\om(k+\iota(- \eps \eta'/2+\eps
  \eta''/2))]\}}\times \frac{\hat
   G^\star(\eta,k+\eps\eta/2) }{\la+i\delta_\eps^+\om(k,\eta)}dkd\eta'
  d\eta''.\nonumber
\end{align}
\end{lemma}
Next, we will replace a similar term in the second denominator by $\iota \omega'(k)(\eta'-\eta'')/2$. 
\begin{lemma}\label{lem-mar204}
We have
\begin{equation}
\label{012811x}
\lim_{\eps\to0}\{{\cal I}_{\iota}^{(1)}(\la,\eps)- {\cal
    I}_{\iota}^{(2)}(\la,\eps)\vphantom{\int_0^1}\}=0,\quad \iota\in\{-,+\},
\end{equation}
where
\begin{align}
\label{022112bis}
& {\cal I}_{\iota}^{(2)}(\la,\eps):=- \frac{\gamma\la }{8\pi }\int_{\bbR\times \bbT_{2/\eps}}\frac{d\beta
d\eta}{(\la/2)^2+\beta^2}\int_{
  T_{\eps,\iota}^3} \frac{\widehat W_\eps\left(\eta',\iota k+\eps \eta''/2\right) }{\la/2-i\{\beta-\iota \om'(k)( \eta'+ \eta'')/2\}}\\
&
\times 
\frac{|\tilde g\left(\la\eps/2-i[\eps\beta+\om(k)]\right)|^2
  }{\la/2+i\{\beta+\iota\om'(k)( \eta'-\eta'')/2\}}\times \frac{\hat
   G^\star(\eta,k+\eps\eta/2) }{\la+i\delta_\eps^+\om(k,\eta)}dkd\eta'
  d\eta''.\nonumber
\end{align}
\end{lemma}
The third step will be to replace the term
$|\tilde g\left(\la\eps/2-i[\eps\beta+\om(k)]\right)|^2$ in (\ref{022112bis})
by its limit $|\nu(k)|^2$.
\begin{lemma}\label{lem-mar602}
We have
\begin{equation}\label{mar620}
\lim_{\eps\to 0}|{\cal I}_\iota^{(2)}(\la,\eps)-{\cal I}_\iota^{(3)}(\la,\eps)|=0,
~~\iota\in\{-,+\},
\end{equation}
%\begin{equation}
%\label{I}
%{\cal I}_{+}(\la,\eps) =\tilde{\cal I}_{+,+}(\la,\eps)+\tilde {\cal I}_{+,-}(\la,\eps)+o(1),
%\end{equation}
with
\begin{eqnarray}
\label{011811xx}
&&
{\cal I}_{\iota}^{(3)}(\la,\eps)
=-\frac{\gamma\la}{8\pi}\int_{\bbR\times \bbT_{2/\eps}}\frac{d\beta d\eta}{(\la/2)^2+\beta^2}
\int_{ T_{\eps,\iota}^3}  \frac{\widehat W_\eps(\eta',\iota k+\eps\eta''/2)  dk d\eta' d\eta''
}{\la/2-i\{\beta-\iota\om'(k) (\eta'+\eta'')/2\}} \nonumber\\
&&
\times \frac{|\nu(k)|^2
   }{\la/2+i\{\beta+\iota\om'(k) (\eta'-\eta'')/2\}} \times \frac{\hat
   G^\star(\eta,k+\eps\eta/2) }{\la+i\delta_\eps^+\om(k,\eta)}.
\end{eqnarray}
\end{lemma}

Next, we will approximate
the Wigner transform $\widehat W_\eps(\eta',\iota k+\eps\eta''/2)$ by 
$\widehat W_\eps(\eta',\iota k)$, 
the test function~$ G^*(\eta,k+\eps\eta/2)$ by $ G^*(\eta,k)$, 
and $\delta_\eps^+\omega(k,\eta)$ by $\omega'(k)\eta$,
respectively.
\begin{lemma}\label{lem-mar1202}
We have
\begin{equation}\label{mar1202}
\lim_{\eps\to 0}|{\cal I}_\iota^{(3)}(\la,\eps)-{\cal I}_\iota^{(4)}(\la,\eps)|=0,
~~\iota\in\{-,+\},
\end{equation}
%\begin{equation}
%\label{I}
%{\cal I}_{+}(\la,\eps) =\tilde{\cal I}_{+,+}(\la,\eps)+\tilde {\cal I}_{+,-}(\la,\eps)+o(1),
%\end{equation}
with
\begin{eqnarray}
\label{mar1204}
&&
{\cal I}_{\iota}^{(4)}(\la,\eps)
=-\frac{\gamma\la}{8\pi}\int_{\bbR\times \bbR}\frac{d\beta d\eta}{(\la/2)^2+\beta^2}
\int_{ \bbT\times\bbR^2}  \frac{\widehat W_\eps(\eta',\iota k)  dk d\eta' d\eta''
}{\la/2-i\{\beta-\iota\om'(k) (\eta'+\eta'')/2\}} \nonumber\\
&&
\times \frac{|\nu(k)|^2
   }{\la/2+i\{\beta+\iota\om'(k) (\eta'-\eta'')/2\}} \times \frac{\hat
   G^\star(\eta,k) }{\la+i\om'(k)\eta}.
\end{eqnarray}
\end{lemma}
The last step will be to pass to the limit in $\widehat W_\eps(\eta',\iota k)$ and 
integrate in $\beta$, which is done in 
Section~\ref{sec:7.5}. 
%\begin{lemma}\label{lem-mar1204}
%We have
%\begin{eqnarray}\label{mar1206}
%&&\lim_{\eps\to 0}{\cal I}_\iota^{(4)}(\la,\eps)
%=-\frac{\gamma\la}{8\pi}\int_{\bbR\times \bbT_{2/\eps}}\frac{d\beta d\eta}{(\la/2)^2+\beta^2}
%\int_{ T_{\eps,\iota}^3}  \frac{\widehat W(\eta',\iota k)  dk d\eta' d\eta''
%}{\la/2-i\{\beta-\iota\om'(k) (\eta'+\eta'')/2\}} \nonumber\\
%&&
%\times \frac{|\nu(k)|^2
%   }{\la/2+i\{\beta+\iota\om'(k) (\eta'-\eta'')/2\}} \times \frac{\hat
%   G^\star(\eta,k) }{\la+i\om'(k)\eta}.
%\end{eqnarray}
%\end{lemma}
%
%
%The reader can easily check that the final limit after
%these passages and integration over $\beta$ is, indeed,
%\[
%{\cal I}_+(\la,\eps)\to \frac12{\cal I}_{tr}(\la),~~{\cal I}_-(\la,\eps)\to  \frac12{\cal I}_{ref}(\la),~~\hbox{as $\eps\to 0$},
%\]
%leading to the conclusion of Lemma~\ref{lem-feb1508}. 

\subsection{The proof of Lemmas~\ref{lem-mar202} and~\ref{lem-mar204}}
\label{fr}

We only present the proof of Lemma~\ref{lem-mar202} since the proof of 
Lemma~\ref{lem-mar204} is very similar except somewhat simpler. 
%As we have mentioned, in the first step we show that
%\begin{equation}
%\label{012811}
%\lim_{\eps\to0+} \{{\cal I}_\iota(\la,\eps)- {\cal
%    I}_\iota ^{(1)}(\la,\eps) \}=0, \quad \iota\in\{-,+\},
%\end{equation}
%where
%\begin{align}
%\label{022112}
%&  {\cal I}_{ \iota}^{(1)}(\la,\eps):=- \frac{\gamma\la }{8\pi }\int_{\bbR\times \bbT_{2/\eps}}\frac{d\beta
%d\eta}{(\la/2)^2+\beta^2}\int_{
%  T_{\eps,\iota}^3} \frac{\widehat W_\eps\left(\eta',\iota k+\eps \eta''/2\right) }{\la/2-i\{\beta-\iota \om'(k)( \eta'+ \eta'')/2\}}\\
%&
%\times 
%\frac{|\tilde g\left(\la\eps/2-i[\eps\beta+\om(k)]\right)|^2
%  }{\la/2+i\{\beta+\eps^{-1}[\om(k)-\om(k+\iota(- \eps \eta'/2+\eps
%  \eta''/2))]\}}\times \frac{\hat
%   G^\star(\eta,k+\eps\eta/2) }{\la+i\delta_\eps^+\om(k,\eta)}dkd\eta'
%  d\eta''.\nonumber
%\end{align}
We will also only consider $\iota=+$ (the transmission case) in the proof
of Lemma~\ref{lem-mar202}, as the reflection case can be treated in a similar
fashion. 

Let us  
drop the subscript $+$, setting 
\[
{\cal I}(\la,\eps):={\cal I}_+(\lambda,\eps),~~{\cal I}^{(1)}(\la,\eps):={\cal I}_+^{(1)}(\lambda,\eps),
\]
to reduce the number of subscripts. 
We 
%\subsubsection*{Proof of (\ref{012811}) for $\iota=+$}
split the domain of integration $T_{\eps,+}^3$ into four  regions
\begin{equation}\label{mar606}
T_{\eps,+,\iota_1,\iota_2}^3:=[(k,\eta',\eta'')\in
T_{\eps,+}^3:\,\iota_1
k>0,\,\iota_2(k-\eps\eta'/2+\eps\eta''/2)>0],\quad \iota_1,\iota_2\in\{-,+\},
\end{equation}
and write 
\[
{\cal I}(\la,\eps)=\sum_{\iota_1,\iota_2=\pm}{\cal
  I}_{\iota_1,\iota_2}(\la,\eps),
~~~~ {\cal I}^{(1)} (\la,\eps)=\sum_{\iota_1,\iota_2=\pm} {\cal I}_{\iota_1,\iota_2}^{(1)}(\la,\eps).
 \] 
We will only consider the case $\iota_1=\iota_2=+$, as the other cases
can be done similarly, and set 
\[
\tilde{\cal I}(\la,\eps)={\cal I}_{+,+}(\la,\eps),
~~\tilde{\cal I}^{(1)}(\la,\eps)={\cal I}^{(1)}_{+,+}(\la,\eps).
\] 
Our goal is to show that for any $\sigma>0$ we have %\red{maybe change this later}
 \begin{equation}
\label{012811i}
\limsup_{\eps\to0}\textcolor{blue}{|}\tilde{\cal
    I} (\la,\eps)-\tilde{\cal
    I}^{(1)} (\la,\eps)\textcolor{blue}{|}<\sigma. %\quad \iota_1,\iota_2\in\{-,+\}.
\end{equation}
% and
%   \begin{equation}
% \label{012811i1}
% \lim_{\eps\to0+}\left\{{\cal I}_{+,\iota,-\iota,\iota'}(\la,\eps)-\tilde{\cal I}_{\iota,-\iota,\iota'}^-(\la,\eps)\right\}=0,\quad \iota,\iota'=\pm.
% \end{equation}
%Choose an arbitrary $\si>0$. 
We perform the change of variables 
\begin{equation}
\label{cv}
w_0:=\om(k),\quad w_1:=\om(k-\eps \eta'/2+\eps\eta''/2),\quad w_2:=\om(k+\eps \eta'/2+\eps\eta''/2)
\end{equation}
in the integrals over $k,\eta',\eta''$, to get
\begin{align}
\label{012112}
&
\tilde {\cal I} (\la,\eps)-\tilde{\cal I}^{(1)}(\la,\eps)\\
&=\frac{\gamma\la i}{4\pi \eps^2}\int_{\bbR\times \bbT_{2/\eps}} \frac{ d\beta d\eta}{(\la/2)^2+\beta^2}
 \int_{D_\eps}\frac{\widehat W_\eps\left(\eps^{-1}[\om_+(w_2)-\om_+(w_1)],(1/2)[\om_+(w_2)+\om_+(w_1)]\right)   
}{\la/2-i\{\beta+\eps^{-1}(w_0-w_2)\}}\nonumber\\
&
\nonumber \\
&
\times \frac{\Delta_+^{(\eps)}(w_2,w_0,\beta)
|\tilde g\left(\la\eps/2-i(\eps\beta+w_0)\right)|^2 }
{\la/2+i\{\beta+\eps^{-1}(w_0-w_1)\}}\times \frac{\hat
G^*(\eta,\om_+(w_0)+\eps\eta/2) }{\la+i\delta_\eps^+\om(\om_+(w_0),\eta)}
\prod_{j=1}^2\frac{1}{\om'(\om_+(w_j))}dw_0 dw_1 dw_2,\nonumber
\end{align}
with $D_\eps\subset [\om_{\rm min},\om_{\rm max}]^3$ -- 
the image of $T_{\eps,+,+,+}^3$ under the change of
variables mapping,
\begin{equation}
\label{D2eps}
\Delta_\pm^{(\eps)}(w',w,\beta):=\frac{\eps^{-1}\delta\om_+ (w',w)}{\la/2\mp
  i\{\beta+\eps^{-1}(w-w')+
  \eps^{-1} \om'(\om_+(w)) \delta\om_+ (w',w)\}}, 
\end{equation} 
and
$$
\delta\om_+(w',w):=\om_+(w)-\om_+(w')-\om'_+(w)(w-w').
$$ 
%Clearly $D_\eps\subset [\om_{\rm min},\om_{\rm max}]^3$.

Let us explain some difficulties in passing to the limit in 
(\ref{012112}). Formally, we have a factor of~$\eps^{-2}$  in front of the
integral compensated by the terms of the order $\eps^{-1}$ in the first
two denominators. The factor of $\eps^{-1}$ in the first argument in $\widehat W_\eps$
seemingly would then bring a collapse of one variable of integration and show that
the overall expression is small in the limit. However, there are two obstacles:
first, the factors  $\omega'(\omega_+(w_j))$ have a square root singularity at 
$\omega_{\rm min}$ and $\omega_{\rm max}$, so that the effect of the $\eps^{-1}$ terms
in the first two denominators is reduced. Second, %the domain of integration 
%in the $\beta$-variable is of the size $O(\eps^{-1})$, hence 
the terms of the size
$\eps\beta$ are not necessarily small and may influence the limit since the domain of integration in $\beta$ is all of ${\mathbb R}$.

In order to deal with the first issue, using assumption \eqref{011812}, we 
see that there exists $\delta_0>0$ such that for all $(w_1,w_2)$, for
which we have
\begin{equation}
\label{011812a}
\widehat W_\eps\left(\frac{1}{\eps}\left[\om_+(w_2)-\om_+(w_1)\right], \frac{1}{2}\left[\om_+(w_2)+\om_+(w_1)\right]\right)     =0, 
 \end{equation} 
provided that either $ w_1,\,w_2\in[\om_{\rm min},\om_{\rm
  min}+\delta_0),\,\mbox{ or } w_1,\,w_2\in(\om_{\rm
  max}-\delta_0,\om_{\rm max}]$,  and $(w_0,w_1,w_2)\in D_\eps$ for some $w_0$.

%\red{\large{\bf DOTAD}}

We can further write
$$
{\cal I}(\la,\eps)-\tilde{\cal I}^{(1)}(\la,\eps)=\sum_{j=1}^2{\cal  J}_{j,\eps},
$$
where the integration is split into the regions $|w_1-w_2|\ge
\delta_0/4$ and otherwise. From \eqref{011812} and \eqref{012112} we
conclude that
$
\left|{\cal  J}_{1,\eps}\right|\preceq \eps$. If, on the other hand 
$|w_1-w_2|<
\delta_0/4$ we only need to be concerned with the integration over
$w_2\in I(\delta_0/2)$, where
$
I(\delta):=[\om_{\rm min}+\delta,\om_{\rm max}-\delta],
$
as otherwise the integrand
vanishes because of (\ref{011812a}). The above implies that $w_1\in I(\delta_0/4)$.
Since  $\om_+\in C^\infty(I(\delta_0/4))$ we can can find
$C>0$ such that, after integration in $\eta$, we have, with a constant depending on $\lambda$:
\begin{align*}
&
|{\cal  J}_{2,\eps}|\preceq {\cal R}_\eps:=\frac{1}{\eps^2}
\int_{\bbR\times \bbT_{2/\eps}} 
\frac{ %\sup_{k}|\hat G^\star(\eta,k)| 
d\beta}{1+\beta^2}
\int_{\om_{\rm min}}^{\om_{\rm max}}dw_0  \int_{I(\delta_0/4)}\int_{I(\delta_0/2)}\frac{\varphi\left(C\eps^{-1}(w_2-w_1) \right)  
}{1+|\beta+\eps^{-1}(w_0-w_2)|}%\nonumber
\\
%&
%\nonumber \\
&
\times 
\frac{|\Delta_+^{(\eps)}(w_2,w_0,\beta) |d  w_1 dw_2}{1+|\beta+\eps^{-1}(w_0-w_1)|}={\cal R}_\eps^1+ {\cal R}_\eps^2.
\end{align*}
The two terms above correspond to the integration in $w_0$ over the regions $I'(\rho):=[\om_{\rm min},\om_{\rm max}]\setminus I(\rho)$ 
and $I(\rho)$, with $\rho<\delta_0/8$.
% gives $ {\cal R}_\eps= {\cal R}_\eps^1+ {\cal R}_\eps^2$. 
We have
\begin{align}
\label{011701}
&
{\cal R}_\eps^1\preceq \frac{1}{\eps^3}
\int_{I'(\rho)}\left(1+\frac{1}{\om'(\om_+(w_0))}\right)dw_0
\int_{I(\delta_0/4)}\int_{I(\delta_0/2)}\varphi\left(C\eps^{-1}(w_2-w_1)\right)
  dw_1 dw_2\nonumber\\
&
\nonumber \\
&
\times \int_{\bbR} \frac{1 
}{1+|\beta+\eps^{-1}(w_0-w_2)|}\times \frac{ 1}{1+|\beta+\eps^{-1}(w_0-w_1)|} \times \frac{
  d\beta}{1+\beta^2}
 \\
&
\nonumber \\
&
\preceq \frac{1}{\eps^3}
\sum_{j=1}^2\int_{I'(\rho)}\left(1+\frac{1}{\om'(\om_+(w_0))}\right)dw_0
\int_{I(\delta_0/4)}\int_{I(\delta_0/2)}\varphi\left(C\eps^{-1}(w_2-w_1)\right)
  dw_1 dw_2\nonumber\\
&
\nonumber \\
&
\times \int_{\bbR} \frac{1 
}{1+|\beta+\eps^{-1}(w_0-w_j)|^2}\times \frac{
  d\beta}{1+\beta^2}.
\end{align}
An elementary estimate  
\begin{equation}
\label{Ca}
\int_{\bbR} \frac{1 
}{1+(\beta+a)^2}\times \frac{
  d\beta}{1+\beta^2}
\preceq \frac{1}{1+a^2},\quad a\in\bbR
\end{equation}
implies that 
\begin{align*}
&
{\cal R}_\eps^1\preceq \frac{1}{\eps^3}
\sum_{j=1}^2\int_{I'(\rho)}\left(1+\frac{1}{\om'(\om_+(w_0))}\right)dw_0
\int_{I(\delta_0/4)}\int_{I(\delta_0/2)}\varphi\left(C\eps^{-1}(w_2-w_1)\right)
  \frac{dw_1 dw_2}{1+\eps^{-2}(w_0-w_j)^2}.
\end{align*}
If $w_0\in
I'(\rho)$ we have $|w_0-w_j|\ge \delta_0/8$, thus
\begin{align*}
&
{\cal R}_\eps^1\preceq \frac{1}{\delta^2_0}
\int_{I'(\rho)}\left(1+\frac{1}{\om'(\om_+(w_0))}\right)dw_0
\int_{0}^{1}\int_{0}^{1}\frac{1}{\eps}\varphi\left(C\eps^{-1}(w_2-w_1)\right)
  dw_1 dw_2.
\end{align*}
We conclude that
\begin{equation}
\label{021701}
\limsup_{\eps\to0}{\cal R}_\eps^1\preceq \frac{1}{\delta^2_0}
\int_{I'(\rho)}\left(1+\frac{1}{\om'(\om_+(w_0))}\right)dw_0,
\quad \rho\in(0,1).
\end{equation}
Selecting $\rho>0$ sufficiently small, we 
deduce
\begin{equation}
\label{011601}
\limsup_{\eps\to0}{\cal R}_\eps^1<\si.
\end{equation}
Next, we fix $\rho>0$ sufficiently small, so that (\ref{011601}) holds and 
look at the term ${\cal R}_\eps^2$, that involves integration in $w_0$ over the region $I(\rho)$.
Note that $\om_+\in C^\infty(I(\rho))$ and  
\[
\inf_{w\in  I(\rho)}\om'(\om_+(w))>0,
\]
hence
$$
{\cal R}_\eps^2\preceq \frac{1}{\eps^2}\int_{\bbR} \frac{ d\beta }{1+\beta^2}
\int_{I(\rho)}  \int_{I(\delta_0/4)}\int_{I(\delta_0/2)}\frac{\varphi\left(C\eps^{-1}(w_2-w_1) \right)  
}{1+|\beta+\eps^{-1}(w_0-w_2)|}\times \frac{ |\Delta_+^{(\eps)}(w_2,w_0)| dw_0 d  w_1 dw_2}{1+|\beta+\eps^{-1}(w_0-w_1)|}.
$$
After the change of variables $w_1':=\eps^{-1}(w_1-w_0)$,
$w_2':=\eps^{-1}(w_2-w_0)$, $\beta':=\beta-w_2'$, the expression in
the right side can be estimated by
\begin{equation}\label{mar502}
{\cal I}_\eps:= \int_{I(\rho)} dw_0\int_{I_\eps(\delta_0)}%{[-C_1\eps^{-1}, C_1\eps^{-1}]^2} 
dw_1dw_2 \int_{\bbR} \frac{
  \varphi\left(C(w_2-w_1) \right)  }{1+|\beta+w_2-w_1|}\times \frac{ |\tilde\Delta_+^{(\eps)}(w_2,w_0,\beta)|  }{1+(\beta+w_2)^2 }\times
\frac{ d\beta 
}{1+|\beta|}={\cal I}_\eps^{(1)}+{\cal I}_\eps^{(2)},
\end{equation}
with %some $C_1>0$, 
$$
\tilde\Delta_\pm^{(\eps)}(w',w,\beta):=\frac{\eps^{-1}\tilde\delta_\eps\om_+(w',w)}{\la/2\mp
  i\{\beta+
  \eps^{-1}\om'(\om_+(w)) \tilde\delta_\eps\om_+ (w',w)\}}
$$
and
\begin{equation}\label{mar520}
\tilde\delta_\eps\om_+(w',w)= -\int_w^{w+\eps w'}(\om_+'(v)-\om'_+(w))dv=\omega_+(w)+\omega_+'(w)\eps w'-\omega_+(w+\eps w').
\end{equation}
The two terms in the right side of (\ref{mar502}) correspond to splitting the region
$I_\eps(\delta_0)\subset [-C_1\eps^{-1}, C_1\eps^{-1}]^2$ of integration in $w_1$, $w_2$ (the image 
of $I(\delta_0/4)\times I(\delta_0/2)$
under the above map) into two sub-regions,
corresponding to the integration over 
$$
{\cal B}_\eps(\rho'):=[w_2:\,|w_2|\le \rho'/\eps]
$$
and its complement, with $\rho'>0$ is to be determined later.
Note that in both regions we have the estimates
\begin{equation}
\label{031601}
\lim_{\eps\to0}\eps^{-1}\tilde\delta_\eps\om_+ (w',w)=0 \quad \mbox{ for
  each $w,w'$},
\end{equation}
and
\begin{equation}\label{mar504}
\int_{\bbR}\frac{
  \varphi\left(Cw \right) dw }{1+|\beta+w|}\preceq
\frac{1}{1+|\beta|},\quad\beta\in \bbR.
\end{equation}
As the domain of integration in (\ref{mar502}) depends on $\eps$, even with (\ref{031601}) in hand, we still can not apply the
Lebesgue dominated convergence theorem directly.
In addition, we have the estimate 
\begin{equation}
\label{011501aa}
 |\tilde\Delta_+^{(\eps)}(w,w_0,\beta)| \preceq|\eps^{-1}\tilde\delta_\eps\om_+(w,w_0)|\preceq \eps w^2,
%\quad w_0\in I(\rho),\,|w|\le C_1/\eps,
\end{equation}
for all $w_0$ and $w$ in the domain of integration in (\ref{mar502}).
Integrating out the $w_1$-variable using (\ref{mar504}) and (\ref{011501aa}), we obtain 
\begin{equation}\label{mar506}
{\cal I}_\eps^{(1)}\preceq  \int_{I(\rho)} dw_0 \int_{-\rho'/\eps}^{\rho'/\eps}dw\int_{\bbR} \frac{\eps w^2 }{1+(\beta+w)^2 }\times
\frac{  d\beta}{1+\beta^2} \preceq 
\int_{I(\rho)} dw_0 \int_{-\rho'/\eps}^{\rho'/\eps}\frac{\eps w^2dw}{1+ w^2 } \preceq \rho'.
\end{equation}
It follows that
\begin{equation}\label{mar612}
\lim_{\eps\to 0}{\cal I}_\eps^{(1)}\le\sigma,
\end{equation}
for a sufficiently small $\rho'\in(0,1)$.

%integrating out the $w_1$-variable in (\ref{mar502}), we obtain
%\begin{equation}\label{mar506}
%{\cal I}_\eps\preceq  \int_{I(\rho)} dw_0 \int_{-C_1/\eps}^{C_1/\eps}dw\int_{\bbR} \frac{ |\tilde\Delta_+^{(\eps)}(w,w_0,\beta)|  }{1+(\beta+w)^2 }\times
%\frac{  d\beta 
%}{1+\beta^2}={\cal I}_\eps^{(1)}+{\cal I}_\eps^{(2)},
%\end{equation}
%where the two expressions  in the right  side
%corresponding to integration over 
%$$
%{\cal B}_\eps(\rho'):=[w:\,|w|\le \rho'/\eps]
%$$
%and its complement. Here $\rho'>0$ is to be determined later on. For
%the first integral, we use the estimate
%\begin{equation}
%\label{011501aa}
%|\eps^{-1}\tilde\delta_\eps\om_+(w,w_0)|\preceq \eps w^2,\quad w_0\in
%I(\rho),\,|w|\le C_1/\eps,
%\end{equation}
%which leads to
%\begin{equation}
%\label{011501aab}
%|\tilde\Delta_+^{(\eps)}(w,w_0,\beta)|\preceq \eps w^2.
%\end{equation}
%From \eqref{011501aa} and \eqref{Ca} we conclude that
%\begin{equation}
%\label{011501}
%{\cal I}_\eps^{(1)}\preceq  \eps
%\int_{-\rho'/\eps}^{\rho'/\eps}\frac{w^2dw}{1+w^2}\preceq \rho'<\si,
%\end{equation}
%for a sufficiently small $\rho'\in(0,1)$.
%

For the second term in the right side of (\ref{mar502}), we use (\ref{mar504}) to integrate out the $w_1$-variable once again,
and write
\begin{equation}\label{mar512}
{\cal I}_\eps^{(2)}=\int_{I(\rho)} dw_0 \int_{I_\eps'}dw\int_{\bbR} \frac{ |\tilde\Delta_+^{(\eps)}(w,w_0)|  }{1+(\beta+w)^2 }\times
\frac{  d\beta 
}{1+\beta^2} ={\cal I}_\eps^{(2,1)}+{\cal I}_\eps^{(2,2)}.
\end{equation}
Here, $I_\eps'\subset [\rho'/\eps \le |w|\le C_1/\eps]$ is the projection of $I_\eps\cap {\cal B}_\eps^c(\rho')$ onto the $w_2$-axis. The
first integral in the right side of (\ref{mar512}) corresponds to integration over the set
\[
[(\beta,w)\in \bbR^2:\,|\beta+w|\le  |w|^{3/4}]
\]
and the second over its complement. %, with $\rho''>0$ to be chosen later.
We split again 
\[
{\cal I}_\eps^{(2,1)}={\cal I}_{\eps,+}^{(2,1)}+{\cal I}_{\eps,-}^{(2,1)},
\]
according to the integration in $w$ over $I_\eps^\pm=I_\eps'\cap[w>0]$ and its complement, so that
\begin{equation}
\label{011501b}
{\cal I}_{\eps,\pm}^{(2,1)} \preceq \int_{I(\rho)} dw_0 \int_{I_\eps^+}
%[\rho'/\eps\le\pm w\le C_1/\eps]}
  \frac{\eps^{-1}|\tilde\delta_\eps\om_+(w,w_0) |dw}{1+w^2}\int_{[|\beta-w|\le |w|^{3/4}]}\frac{d\beta}{1+|\beta-
  \eps^{-1}\om'(\om_+(w_0)) \tilde\delta_\eps\om_+(w,w_0)|}.
\end{equation}
Let us set 
\[
z_\eps(w,w_0):= \eps^{-1}\om'(\om_+(w_0)) \tilde\delta_\eps\om_+(w,w_0)
%=\farc{1}{\eps \om_+'(w_0)}[\omega_+(w_0)+\omega_+'(w_0)\eps w-\omega_+(w_0+\eps w)]
=w-\farc{\omega_+(w_0+\eps w)-\omega_+(w_0)}{\eps \omega_+'(w_0)},
\]
so that
%\begin{equation}\label{mar602}
%w-z\ge \delta w,\hbox{ for all $w_0\in I(\rho)$ and $w\in I_\eps^+$,}
%\end{equation}
%with some $\delta>0$ that depends on $\rho$ but not on $\eps$. In particular, we have
\begin{equation}\label{mar604}
w-z_\eps(w,w_0)>w^{4/5},\hbox{ for all $w_0\in I(\rho)$ and $w\in I_\eps^+$,} 
\end{equation}
for all $\eps>0$ sufficiently small. 
Then, we have
\[
\begin{aligned}
Z_\eps(w,w_0):=\int_{w-w^{3/4}}^{w+w^{3/4}}\frac{d\beta}{1+|\beta-z_\eps|}
=\int_{w-w^{3/4}-z_\eps}^{w+w^{3/4}-z_\eps}\frac{d\beta}{1+|\beta|}=
\log\Big(\farc{1+ w+w^{3/4}-z_\eps}{1+w-w^{3/4}-z_\eps}\Big). 
% ,\quad\hbox{ if $z< w-w^{3/4}$,}
\end{aligned}
\]
%and
%% then
%%\begin{equation}\label{mar530}
%%Z=\log\Big(\farc{1+ w+w^{3/4}-z}{1+w-w^{3/4}-z}\Big).
%%\end{equation}
%\begin{equation}\label{mar532}
%Z=\log\Big(\farc{1+z- w+w^{3/4}}{1+z-w-w^{3/4}}\Big),\quad \hbox{ if $z>w+w^{3/4}$.}
%\end{equation}
%And if $ w-w^{3/4}<z< w+w^{3/4}$, this is
%\begin{equation}\label{mar534}
%Z=\log(1+w+w^{3/4}-z)+\log(1+z-w+w^{3/4})).
%\end{equation}
%In both cases, 
It follows from (\ref{mar604}) that
%Note that if $|z-w|>w^{4/5}$, then we only need to consider (\ref{mar530}) and (\ref{mar532}). In that case, 
by taking~$\eps$ sufficiently
small we may ensure that 
\begin{equation}\label{mar536}
\limsup_{\eps\to 0}\sup_{w_0\in I(\rho),w\in I_\eps^+}|Z_\eps(w,w_0)|=0.
%\le\sigma\hbox{ for all $w\in I_\eps^+$ and $w_0\in I(\rho)$}.
\end{equation}
%On the other hand, when $|z-w|\le w^{4/5}$, we need to allow (\ref{mar534}), when we only have the bound
%\begin{equation}\label{mar538}
%|Z|\preceq|\log\eps|\hbox{ for all $z\in I_\eps^+$ s.t. $|z-w|\le w^{4/5}$}.
%\end{equation}
%However, to be in this situation we need to have
%\begin{equation}\label{mar540}
%|w-\eps^{-1}\om'(\om_+(w_0)) \tilde\delta_\eps\om_+(w,w_0)|\le w^{4/5},
%\end{equation}
%so that 
%\begin{equation}\label{mar542}
%\Big|\om_+'(w_0)-\frac{\om_+(w_0)-\om_+(w_0+\eps w)}{  \eps w}\Big|\preceq \eps^{1/5}.
%\end{equation}
%\red{If we assume that $\omega(k)$ is not linear on any piece, then the measure of $w\in I_\eps^+$ such (\ref{mar542}) holds is bounded
%by $\eps^{-4/5}$. } 
Then, using (\ref{011501aa}), we get
\begin{equation}
\label{mar544}
{\cal I}_{\eps,+}^{(2,1)} \preceq \int_{I(\rho)} dw_0 \int_{|w-z|\ge w^{4/5}}
%[\rho'/\eps\le\pm w\le C_1/\eps]}
|Z_\eps(w,w_0)|  \frac{\eps w^2dw}{1+w^2} \le\farc{\sigma}{2},
\end{equation}
for $\eps>0$ sufficiently small. 
%When is the equality
%\[
%\eps\beta\approx \omega'(\omega_+(w_0))\tilde\delta_\eps\om_+(w,w_0)
%\]
%possible? This is same as
%\[
%\eps\beta\omega_+'(w_0)\approx \omega_+(w_0)-\omega_+(w_0+\eps w).
%\]
%
%Integrating out the $\beta$-variable gives
%\begin{align*}
%{\cal I}_{\eps,+}^{(2,1)}=\int_{I(\rho)} dw_0 \int_{I_\eps^+}
%%[\rho'/\eps\le   w\le C_1/\eps]}
%\Phi_{\eps,\rho''} (w)\frac{\eps^{-1}|\tilde\delta_\eps\om_+(w,w_0)  |dw}{1+w^2},
%\end{align*}
%with
%$$
%\Phi_{\eps,\rho''} (w)=\log\left(\frac{[(1+\rho'')w]^{-1}+|(1-\rho'')(1+\rho'')^{-1}-(\eps w)^{-1}\tilde\delta_\eps\om_+(w,w_0)
% [ (1+\rho'')]^{-1}|}
%{[(1+\rho'')w]^{-1}+|1-(\eps w)^{-1}\tilde\delta_\eps\om_+(w,w_0) [(1+\rho'')]^{-1}|}\right).
%$$
%%Since  $\eps^{-1}|\tilde\delta_\eps\om_+(w,w_0)
%%  | \preceq \eps w^2
%%$ for all $w\in
%%[\rho'/\eps, C_1/\eps]$  
% 
%
%
%Using (\ref{011501aa}), we get
%$$
%{\cal I}_{\eps,+}^{(2,1)}\preceq \eps\int_{I_\eps^+}%[\rho'/\eps\le w\le C_1/\eps]}
%   \frac{w^2 dw}{1+w^2}\sup_{w\in I_\eps^+}%[\rho'/\eps, C_1/\eps]}
%\Phi_{\eps,\rho''} (w).
%$$
%Using the fact that
%$(\eps w)^{-1}|\tilde\delta_\eps\om_+(w,w_0)|\preceq 1$ and 
%$
%[(1+\rho'')w]^{-1}\preceq \eps $, we conclude further that
%\begin{equation}
%\label{021601}
%\limsup_{\eps\to0}{\cal I}_{\eps,+}^{(2,1)}<\si
%\end{equation}
%for a sufficiently small $\rho''\in(0,1)$. 
A similar calculation
yields the same estimate for ${\cal I}_{\eps,-}^{(2,1)}$, thus
\begin{equation}
\label{021601a}
\limsup_{\eps\to0}{\cal I}_{\eps}^{(2,1)}<\si.
\end{equation}
As for ${\cal I}_\eps^{(2,2)}$ we can write
\begin{align*}
&
{\cal I}_\eps^{(2,2)}\preceq \int_{I(\rho)} dw_0 \int_{I_\eps'}
%[\rho'/\eps \le  |w|\le C_1/\eps]}
\frac{\eps^{-1}|\tilde\delta_\eps\om_+(w,w_0) |dw}{1+w^{3/2}}\int_{\bbR} \frac{
  d\beta }{{(}1+|\beta+
  \eps^{-1}\om'(\om_+(w_0)) \tilde\delta_\eps\om_+(w,w_0)| {)}(1 +\beta^2)}.
\end{align*}
Using an elementary estimate
\begin{equation}
\label{Ca1}
\int_{\bbR} \frac{1 
}{1+|\beta+a|}\times \frac{
  d\beta}{1+\beta^2}
\preceq \frac{1}{1+|a|},\quad a\in\bbR,
\end{equation}
we obtain
\begin{align*}
&
{\cal I}_\eps^{(2,2)}\preceq \int_{I(\rho)} dw_0 \int_{I_\eps'}
%[\rho'/\eps \le |w|\le C_1/\eps]}
  \frac{\eps^{-1}|\tilde\delta_\eps\om_+(w,w_0) |dw}{[1+|
  \eps^{-1}\om'(\om_+(w_0)) \tilde\delta_\eps\om_+(w,w_0)|] (1+w^{3/2})}.
\end{align*}
Here, we can use the Lebesgue dominated convergence theorem and (\ref{031601}) to conclude that
\begin{align*}
&
\lim_{\eps\to0}{\cal I}_\eps^{(2,2)} =0.
\end{align*}
This finishes the proof of (\ref{012811i}).

\commentout{
%%%%%%%%%%%%%%%%%%%%%%%%%%%%%%%%%%%%%%%%
%%%%%%%%%%%%%%%%%This is the commentout of the proof of the second replacement

\subsection{The proof of Lemma~\ref{lem-mar204}}

Recall that we need to prove that
\begin{equation}
\label{012811xbis}
\lim_{\eps\to0+}\{{\cal I}_{\iota}^{(1)}(\la,\eps)- {\cal
    I}_{\iota}^{(2)}(\la,\eps)\vphantom{\int_0^1}\}=0,\quad \iota\in\{-,+\}.
\end{equation}
Here, $I_\iota^{(1)}$ and $I_\iota^{(2)}$ are given, respectively, by (\ref{022112}) and (\ref{022112bis}), and our goal is to replace
the difference  
\[
\om(k)-\om(k+\iota(- \eps \eta'/2+\eps \eta''/2))
\]
in  the second denominator in (\ref{022112}) by $\iota\om'(k)( \eta'-\eta'')/2$. 
%where
%\begin{align}
%\label{022112bis}
%& {\cal I}_{\iota}^{(2)}(\la,\eps):=- \frac{\gamma\la }{8\pi }\int_{\bbR\times \bbT_{2/\eps}}\frac{d\beta
%d\eta}{(\la/2)^2+\beta^2}\int_{
%  T_{\eps,\iota}^3} \frac{\widehat W_\eps\left(\eta',\iota k+\eps \eta''/2\right) }{\la/2-i\{\beta-\iota \om'(k)( \eta'+ \eta'')/2\}}\\
%&
%\times 
%\frac{|\tilde g\left(\la\eps/2-i[\eps\beta+\om(k)]\right)|^2
%  }{\la/2+i\{\beta+\iota\om'(k)( \eta'-\eta'')/2\}}\times \frac{\hat
%   G^\star(\eta,k+\eps\eta/2) }{\la+i\delta_\eps^+\om(k,\eta)}dkd\eta'
%  d\eta''.\nonumber
%\end{align}
We argue this point only  for $\iota=+$, and drop the subscript $+$ in the notation. 

Partitioning the domain $T_{\eps,+}^3$ into sub-domains $T_{\eps,+,\iota_1,\iota_2}^3$, as in (\ref{mar606}), 
we can introduce  $\tilde {\cal
  I}_{+,+,\iota_1,\iota_2}'(\la,\eps)$ corresponding to the
integration over the respective domain. Using change of variables
\eqref{cv} we obtain
\begin{align}
\label{012112xxx}
&
\tilde{\cal I}_{+,+,+,+}'(\la,\eps)-\tilde{\cal I}_{+,+,+,+}(\la,\eps)
=\frac{\gamma\la i}{4\pi \eps^2}\int_{\bbR\times \bbT_{2/\eps}} \frac{ d\beta d\eta}{(\la/2)^2+\beta^2}
 \int_{D_\eps}\frac{\widehat W_\eps\left(\eps^{-1}[\om_+(w_2)-\om_+(w_1)],(1/2)[\om_+(w_2)+\om_+(w_1)]\right)   
}{\la/2-i\{\beta-
  \eps^{-1}\om'(\om_+(w_0))(\om_+(w_2)-\om_+(w_0))\}}\nonumber\\
&
\nonumber \\
&
\times \frac{\Delta_-^{(\eps)}(w_1,w_0,\beta)|\tilde
  g\left(\la\eps/2-i(\eps\beta+w_0)\right)|^2
  }{\la/2+i\{\beta+\eps^{-1}(w_0-w_1)\}}
\times  \frac{\hat
   G^\star(\eta,\om_+(w_0)+\eps\eta/2) }{\la+i\delta_\eps^+\om(\om_+(w_0),\eta)}\prod_{j=1}^2\frac{1}{\om'(\om_+(w_j))}dw_0 d
   w_1 dw_2,\nonumber
\end{align}
with $\Delta_-^{(\eps)}(w_1,w_0,\beta)$ given by (\ref{D2eps}) and
$D_\eps$ - the image of $T_{\eps,+,+,+}^3$ under the transformation given
by (\ref{cv}).

We follow the steps made in the relevant part of Section
\ref{fr}. Suppose that $\si>0$ is arbitrary. We can focus our attention only on the expression
$\tilde{\cal J}_\eps$ that
correspons to the integration over the region $|w_1-w_2|<\delta_0/4$,
$w_1\in I(\delta_0/4)$ as the remaining part tends to $0$, as
$\eps\to0$. There exists $C>0$ such that for all $\eps\in(0,1)$ we have
\begin{align*}
&
|\tilde {\cal  J}_{\eps}|\preceq \tilde{\cal R}_\eps:=\frac{1}{\eps^2}\int_{\bbR\times \bbT_{2/\eps}} \frac{ \sup_{k}|\hat
   G^\star(\eta,k)| d\beta d\eta}{1+\beta^2}
\int_{\om_{\rm min}}^{\om_{\rm max}}dw_0  \int_{I(\delta_0/4)}\int_{I(\delta_0/2)}\frac{\varphi\left(C\eps^{-1}(w_2-w_1) \right)  
}{1+|\beta+\eps^{-1}(w_0-w_1)|}\nonumber\\
&
\nonumber \\
&
\times \frac{|\Delta_-^{(\eps)}(w_1,w_0,\beta) |d  w_1 dw_2}{1+|\beta-
  \eps^{-1}\om'(\om_+(w_0))(\om_+(w_2)-\om_+(w_0))|}
.
\end{align*}
Dividing the integration over $w_0$ into two parts that correspond to 
regions $I'(\rho)$ and $I(\rho)$, where $\rho<\delta_0/8$ we can write 
$ \tilde{\cal R}_\eps= \tilde{\cal R}_\eps^1+ \tilde{\cal R}_\eps^2$. We have
\begin{align*}
&
\tilde{\cal R}_\eps^1\preceq \frac{1}{\eps^3}
\int_{I'(\rho)}\left(1+\frac{1}{\om'(\om_+(w_0))}\right)dw_0
\int_{I(\delta_0/4)}\int_{I(\delta_0/2)}\varphi\left(C\eps^{-1}(w_2-w_1)\right)
  dw_1 dw_2\nonumber\\
&
\nonumber \\
&
\times \int_{\bbR} \frac{1 
}{1+|\beta+
  \eps^{-1}\om'(\om_+(w_0))(\om_+(w_0)-\om_+(w_2))|}\times \frac{ 1}{1+|\beta+\eps^{-1}(w_0-w_1)|} \times \frac{
  d\beta}{1+\beta^2}.
\end{align*}
From this point on the estimates can be carried out as in
(\ref{011701}) and we conclude that 
\begin{equation}
\label{011601}
\limsup_{\eps\to0+}\tilde{\cal R}_\eps^1<\si.
\end{equation}
Concerning $\tilde {\cal R}_\eps^2$
%$, since $\inf_{w\in
%   I(\rho)}\om'(\om_+(w))>0$ and $\om_+\in C^\infty(I(\rho))$,
we can write
\begin{equation}
\label{tr}
\tilde{\cal R}_\eps^2\preceq \frac{1}{\eps}\int_{\bbR} \frac{ d\beta }{1+\beta^2}
\int_{I(\rho)}  \int_{I(\delta_0/4)} \frac{ \tilde\Phi_\eps(w_1,w_0,\beta)|\Delta_-^{(\eps)}(w_1,w_0,\beta)| dw_0 d  w_1}{1+|\beta+\eps^{-1}(w_0-w_1)|},
\end{equation}
where
$$
\tilde\Phi_\eps(w_1,w_0,\beta):=\frac{1}{\eps}\int_{I(\delta_0/2)}\frac{\varphi\left(C\eps^{-1}(w_2-w_1) \right)  dw_2
}{1+|\tilde\beta+
  \eps^{-1}\om'(\om_+(w_0))(\om_+(w_1)-\om_+(w_2))|}.
$$
and
$$
\tilde\beta:=\beta+
  \eps^{-1}\om'(\om_+(w_0)) (\om_+(w_0)-\om_+(w_1)).
$$
\begin{lemma}
\label{lm011701}
We have
$$
\tilde\Phi_\eps(w_1,w_0,\beta)\preceq \frac{1}{1+|\tilde\beta|},\quad
(w_1,w_0,\beta)\in I(\rho)\times I(\delta_0/4)\times \bbR,\,\eps\in(0,1].
$$
\end{lemma}
\proof Note that there exists $C'>0$ such that
\begin{align*}
&
\varphi\left(C\eps^{-1}(w_2-w_1) \right)
  =\varphi\left(C\eps^{-1}\om'(\om_+(w_0))
  (\om_+(w_2)-\om_+(w_1))\frac{w_2-w_1}{\om'(\om_+(w_0))[\om_+(w_2)-\om_+(w_1)]}
  \right) \\
&\preceq \varphi\left(C'\eps^{-1}\om'(\om_+(w_0)) (\om_+(w_2)-\om_+(w_1))\right)
\end{align*}
 for $(w_1,w_0)\in I(\rho)\times I(\delta_0/4) $. Therefore 
$$
\tilde\Phi_\eps(w_1,w_0,\beta)\preceq\frac{1}{\eps}\int_{I(\delta_0/2)}\frac{ \varphi\left(C'\eps^{-1}\om'(\om_+(w_0))(\om_+(w_2)-\om_+(w_1))\right)dw_2
}{1+|\tilde\beta+
  \eps^{-1}\om'(\om_+(w_0))(\om_+(w_1)-\om_+(w_2))|},\quad
(w_1,w_0,\beta)\in I(\rho)\times I(\delta_0/4)\times \bbR,\,\eps\in(0,1].
$$
After the change of variables $v:=\eps^{-1}\om'(\om_+(w_0)) (\om_+(w_2)-\om_+(w_1))$, we obtain
$$
\tilde\Phi_\eps(w_1,w_0,\beta)\preceq \int_{\bbR}\frac{\varphi\left(C'v\right)  dv
}{1+|\tilde\beta-v|}\preceq\frac{1}{1+|\tilde\beta|}.
$$
\qed

\bigskip

Using the above lemma from \eqref{tr} we obtain 
\begin{equation}
\label{tr1}
\tilde{\cal R}_\eps^2\preceq \frac{1}{\eps}\int_{\bbR} \frac{ d\beta }{1+\beta^2}
\int_{I(\rho)}  \int_{I(\delta_0/4)} \frac{ |\Delta_-^{(\eps)}(w_1,w_0,\beta)| dw_0 d  w_1}{1+|\beta+
  \eps^{-1}\om'(\om_+(w_0)) (\om_+(w_0)-\om_+(w_1))|[1+|\beta+\eps^{-1}(w_0-w_1)|]},
\end{equation}
After the change of variables $w_1':=\eps^{-1}(w_1-w_0)$ the expression in
the right hand side can be estimated by
 $$
\tilde {\cal I}_\eps:= \int_{I(\rho)} dw_0\int_{\eps^{-1}(I(\delta_0/4)-w_0)} dw \int_{\bbR} \frac{
 |\eps^{-1}\om'(\om_+(w_0) )\tilde\delta_\eps\om_+(w,w_0)|
}{[1+|\beta-\eps^{-1}\om'(\om_+(w_0) )(\om_+(w_0+\eps w)-\om_+(w_0))|]^2}\times \frac{ 1}{1+|\beta-w| }\times
\frac{ d\beta 
}{1+\beta^2}.
 $$
We can write $\tilde {\cal I}_\eps = \tilde {\cal I}_\eps^1+ \tilde {\cal I}_\eps^2$, where the first and the second expression  in the right hand side
correspond to integration over 
$$
{\cal B}_\eps(\rho'):=[w:\,|w|\le \rho'/\eps]\cap \eps^{-1}(I(\delta_0/4)-w_0)
$$
and 
$$
{\cal B}_\eps'(\rho'):=
\eps^{-1}(I(\delta_0/4)-w_0)\setminus [w:\,|w|\le \rho'/\eps].
$$
 Here $\rho'>0$ is to be determined later on.
By the Cauchy-Schwarz inequality
\begin{align*}
&\tilde {\cal I}_\eps^1\preceq \int_{I(\rho)} dw_0\int_{{\cal B}_\eps(\rho')} dw \int_{\bbR} \frac{
 |\eps^{-1}\om'(\om_+(w_0) )\tilde\delta_\eps\om_+(w,w_0)|
}{[1+|\beta-\eps^{-1}\om'(\om_+(w_0) )(\om_+(w_0+\eps w)-\om_+(w_0))|]^4}\times
\frac{ d\beta 
}{1+\beta^2}\\
&
+\int_{I(\rho)} dw_0\int_{{\cal B}_\eps(\rho')} dw \int_{\bbR} \frac{
 |\eps^{-1}\om'(\om_+(w_0) )\tilde\delta_\eps\om_+(w,w_0)|
}{[1+|\beta-w|]^2}\times
\frac{ d\beta 
}{1+\beta^2}.
 \end{align*}
Denote the terms appearing in the right hand side by $\tilde {\cal
  I}_\eps^{(1,1)}$ and  $\tilde {\cal I}_\eps^{(1,2)}$ respectively.
Using \eqref{Ca} and \eqref{011501aa}
we show, as in \eqref{011501}, that
\begin{equation}
\label{031701}
\limsup_{\eps\to0}\tilde {\cal I}_\eps^{(1,2)}<\si,
\end{equation}
provided that $\rho'>0$ is sufficiently small. By the same token we
obtain
\begin{align*}
&\tilde {\cal I}_\eps^{(1,1)}\preceq  \eps\int_{I(\rho)} dw_0\int_{{\cal B}_\eps(\rho')} \frac{
 w^2 dw
}{1+[\eps^{-1}\om'(\om_+(w_0) )(\om_+(w_0+\eps w)-\om_+(w_0))]^2}\\
&
\preceq \eps\int_{I(\rho)} dw_0\int_{{\cal B}_\eps(\rho')} \frac{
 w^2 dw
}{1+w^2}
 \end{align*}
and again for a sufficiently small  $\rho'>0$ we get
\begin{equation}
\label{031701a}
\limsup_{\eps\to0}\tilde {\cal I}_\eps^{(1,1)}<\si.
\end{equation}
We have
$$
\tilde {\cal I}_\eps^2= \int_{I(\rho)} dw_0\int_{{\cal B}_\eps'(\rho')} dw \int_{\bbR} \frac{
 |\eps^{-1}\om'(\om_+(w_0) )\tilde\delta_\eps\om_+(w,w_0)|
}{1+[\beta-\eps^{-1}\om'(\om_+(w_0) )(\om_+(w_0+\eps w)-\om_+(w_0))]^2}\times \frac{ 1}{1+|\beta-w| }\times
\frac{ d\beta 
}{1+\beta^2}.
 $$
We can write  $\tilde {\cal I}_\eps^{2}=\tilde {\cal
  I}_{\eps}^{(2,1)}+\tilde {\cal
  I}_{\eps}^{(2,2)}$, where the first and second terms correspond to integration over
$$
{\cal D}_\eps(\rho''):=[(\beta,w)\in \bbR^2:\,|\beta-w|\le \rho''|w|]\cap {\cal B}_\eps'(\rho')
$$
and 
$$
{\cal D}_\eps'(\rho''):= {\cal B}_\eps'(\rho')\setminus [(\beta,w)\in \bbR^2:\,|\beta-w|\le \rho''|w|].
$$ 
Parameter $\rho''>0$ is to be chosen later
on. Again we write $\tilde {\cal
  I}_{\eps}^{(2,1)}=\tilde {\cal
  I}_{\eps,+}^{(2,1)}+\tilde {\cal
  I}_{\eps,-}^{(2,1)}$, where the terms correspond to the integration
over 
$$
{\cal D}_{\eps,+}(\rho''):= {\cal D}_\eps(\rho'')\cap
[w>0]\quad\mbox{and}\quad {\cal D}_{\eps,-}(\rho''):= {\cal D}_\eps(\rho'')\cap
[w<0].
$$ 
We only estimate $\tilde {\cal
  I}_{\eps,+}^{(2,1)}$, as the other case can be done anagously. Then,
\begin{align*}
&\tilde {\cal I}_{\eps,+}^{(2,1)}= \int_{I(\rho)}
  dw_0\int_{{\cal D}_{\eps,+}(\rho'')}
  \frac{|\eps^{-1}\om'(\om_+(w_0) )\tilde\delta_\eps\om_+(w,w_0)|
  dw}{1+w^2} \\
&
\times\int_{(1-\rho'')w}^{(1+\rho'')w} \frac{d\beta 
}{1+[\beta-\eps^{-1}\om'(\om_+(w_0) )(\om_+(w_0+\eps
  w)-\om_+(w_0))]^2}\\
&
\preceq
\eps\int_{I(\rho)}
  dw_0\int_{{\cal D}_{\eps,+}(\rho'')}
  \frac{w^2
  }{1+w^2} \left\{\arctan \left(A(w,w_0,1+\rho'')w\right)-\arctan \left(A(w,w_0,1-\rho'')w\right)\right\}dw
 \end{align*}
with
\begin{align*}
A(w,w_0,\ga):=\ga-(\eps w)^{-1}\om'(\om_+(w_0) )(\om_+(w_0+\eps
  w)-\om_+(w_0)).
\end{align*}
Using an elementary trigonometric identity we conclude that the
difference of the arctan functions equals
$$
\Psi_\rho(w,w_0):=\arctan\left(\frac{2\rho'' w}{1+A(w,w_0,1-\rho'')A(w,w_0,1+\rho'')w^2}\right)=\arctan\left(\frac{2\rho'' w}{1+A^2(w,w_0,1)+\rho''O(1)}\right),
$$
where $O(1)$ stays bounded for $\eps\in(0,1]$, $w\in {\cal
  D}_{\eps,+}(\rho'')$. We can choose therefore $\rho''>0$ sufficiently
small that
\begin{equation}
\label{051701}
\limsup_{\eps\to0+}\tilde {\cal I}_{\eps,+}^{(2,1)}<\si.
\end{equation}
Similarly,
\begin{equation}
\label{051701a}
\limsup_{\eps\to0+}\tilde {\cal I}_{\eps,-}^{(2,1)}<\si.
\end{equation}
Finally,
\begin{align*}
&\tilde {\cal I}_{\eps}^{(2,2)}\le  \int_{I(\rho)}
  dw_0\int_{{\cal D}'_\eps(\rho'')}
  \frac{|\eps^{-1}\om'(\om_+(w_0) )\tilde\delta_\eps\om_+(w,w_0)|
  dw}{1+|w|} \\
&
\times\int_{\bbR} \frac{1 
}{1+[\beta-\eps^{-1}\om'(\om_+(w_0) )(\om_+(w_0+\eps
  w)-\om_+(w_0))]^2}\times \frac{d\beta}{1+\beta^2}.
\end{align*}
Using \eqref{Ca} and \eqref{011501aa} we obtain
\begin{align*}
&\tilde {\cal I}_{\eps}^{(2,2)}
\preceq
\eps\int_{I(\rho)}
  dw_0\int_{{\cal D}'_\eps(\rho'')}\frac{1}{1+\theta_\eps(w,w_0)w^2}\times 
  \frac{w^2 dw
  }{1+|w|}, 
 \end{align*}
with 
$$
\theta_\eps(w,w_0):=[(\eps w)^{-1}\om'(\om_+(w_0) )(\om_+(w_0+\eps
  w)-\om_+(w_0))]^2
$$
that is bounded away from $0$ and infinity for $w_0\in I(\rho)$ and
$w\in{\cal D}'_\eps(\rho'')$.
From here we conclude that
\begin{equation}
\label{061701}
\limsup_{\eps\to0+}\tilde {\cal I}_{\eps}^{(2,2)}=0.
\end{equation}
Summarizing, from the above and \eqref{011601}, \eqref{031701} -  \eqref{051701} we conclude \eqref{012811x} for $\iota=+$.

}
%%%%%%%%%%%%%%%%%%%%%%%%%%%%%%
%%%%%%%%%%%%The end of the commentout of the proof of the second replacement
%%%%%%%%%%%%%%%%%%%%%%%%%%%%%%

\subsection{The proof of Lemma~\ref{lem-mar602}}

%We have shown so far that 
%$$
%{\cal I}_{+}(\la,\eps) ={\cal I}_{+,+}(\la,\eps)+ {\cal I}_{+,-}(\la,\eps)+o(1)
%$$
%as $\eps\ll1$, with
%\begin{eqnarray}
%\label{011811x}
%&&
%{\cal I}_{+,\iota}(\la,\eps)
%=-\frac{\gamma\la}{8\pi}\int_{\bbR\times \bbT_{2/\eps}}\frac{d\beta d\eta}{(\la/2)^2+\beta^2}
%\int_{ T_{\eps,\iota}^3}  \frac{\widehat W_\eps(\eta',\iota k+\eps\eta''/2)  dk d\eta' d\eta''
%}{\la/2-i\{\beta-\iota\om'(k) (\eta'+\eta'')/2\}} \nonumber\\
%&&
%\times \frac{|\tilde g\left(\la\eps/2-i[\eps\beta+\om(k)]\right)|^2
%   }{\la/2+i\{\beta+\iota\om'(k) (\eta'-\eta'')/2\}}\times \frac{\hat
%   G^\star(\eta,k+\eps\eta/2) }{\la+i\delta_\eps^+\om(k,\eta)},
%\end{eqnarray}
%where $T_{\eps,\iota}^3$ is defined in \eqref{T2ek1}. 
%In this step we prove that
%\begin{equation}
%\label{I}
%{\cal I}_{+}(\la,\eps) =\tilde{\cal I}_{+,+}(\la,\eps)+\tilde {\cal I}_{+,-}(\la,\eps)+o(1),
%\end{equation}
%with
%\begin{eqnarray}
%\label{011811xx}
%&&
%\tilde{\cal I}_{+,\iota}(\la,\eps)
%=-\frac{\gamma\la}{8\pi}\int_{\bbR\times \bbT_{2/\eps}}\frac{d\beta d\eta}{(\la/2)^2+\beta^2}
%\int_{ T_{\eps,\iota}^3}  \frac{\widehat W_\eps(\eta',\iota k+\eps\eta''/2)  dk d\eta' d\eta''
%}{\la/2-i\{\beta-\iota\om'(k) (\eta'+\eta'')/2\}} \nonumber\\
%&&
%\times \frac{|\nu(k)|^2
%   }{\la/2+i\{\beta+\iota\om'(k) (\eta'-\eta'')/2\}} \times \frac{\hat
%   G^\star(\eta,k+\eps\eta/2) }{\la+i\delta_\eps^+\om(k,\eta)}.
%\end{eqnarray}

Let us note that the integration in $\eta''$ both in expression (\ref{022112bis}) 
for ${\cal I}_\iota^{(2)}$ and (\ref{011811xx}) for ${\cal I}_\iota^{(3)}$ would bring
out the factor of $[\omega'(k)]^{-1}$ that is not integrable. This singularity should be compensated
by the~$\tilde g$-term in (\ref{022112bis}) and by its limit $|\nu(k)|^2$ in 
(\ref{011811xx}), as  can be seen from (\ref{nu}), (\ref{033110}) and (\ref{feb1420}).
The following auxiliary result will allow us to use this argument. 
\begin{lemma}
\label{lm5.4}
%Under the assumptions on the dispersion relation made at the beginning
%of Section \ref{sec5.2} 
For each  $k_*$ such that $\om'(k_*)=0$, we have
\begin{equation}
\label{012401}
\lim_{\delta'\to0}\limsup_{\eps\to0}\sup_{\beta\in(- \delta', \delta')}
|\tilde g\left(\eps -i[\beta+\om(k_*)]\right)|=0.
\end{equation}
\end{lemma}
\proof
As follows from \eqref{tg}, it suffices to show that 
\begin{equation}
\label{012401a}
\lim_{\delta'\to0}\liminf_{\eps\to0}\inf_{\beta\in(- \delta', \delta')}
|\tilde J\left( \eps -i[\beta+\om(k_*)]\right)|=+\infty,
\end{equation}
with $\tilde J(\cdot)$ as in \eqref{eq:2}. Consider the point $k_*=1/2$ where
$\om$ attains its maximum $\om_{\rm max}=\om(k_*)>0$, and write
\begin{align*}
\tilde J\left( \eps -i[\beta+\om(k_*)]\right)
=\frac{i}{2}\Big\{\int_{\bbT}\frac{ d\ell}{i\eps+\beta+\om(k_*)+\om(\ell)}
+\int_{\bbT}\frac{ d\ell}{i\eps+\beta+\om(k_*)-\om(\ell)}\Big\}.
\end{align*}
Hence, \eqref{012401a} would follow if we can show that for each $M>0$ there
exist $\eps_0,\delta_0\in(0,1)$ such that
\begin{equation}
\label{012401b}
\Big|\int_{\bbT}\frac{
    d\ell}{i\eps+\beta+\om(k_*)-\om(\ell)}\Big|>M,
    \quad \beta\in(-\delta_0,\delta_0),\,\eps\in(0,\eps_0).
\end{equation}
The real and imaginary parts of the expression under the absolute 
value in (\ref{012401b}) are
\begin{align*}
&
r_\eps(\beta):=\int_{\bbT}\frac{[\beta+\om(k_*) -\om(\ell)]
    d\ell}{\eps^2+[\beta+\om(k_*)-\om(\ell)]^2},~~~
j_\eps(\beta):=-\int_{\bbT}\frac{\eps
    d\ell}{\eps^2+[\beta+\om(k_*)-\om(\ell)]^2}.
\end{align*}
Changing variables $u:=\om(\ell)-\beta$, we obtain
\[
|j_\eps(\beta)|\ge\int_{\om_{\rm min}-\beta}^{\om_{\rm max}-\beta}\frac{\eps
    }{\eps^2+[\om_{\rm max}-u]^2}\times \frac{du}{|\om'(\om_+(u+\beta))|}.
\]
Choosing a sufficiently small $\delta_0>0$, we see that 
\[
\hbox{$|\om'(\om_+(u+\beta))|\le \pi/(2M)$ for $|\beta|<\delta_0$ and $u\in(
\om_{\rm max}-\delta_0,  \om_{\rm max}+\delta_0)$,}
\]
hence
\[
\inf_{\beta\in (-\delta_0,\delta_0)}|j_\eps(\beta)|\ge \frac{2 M}{\pi}\int_{\om_{\rm max}-\delta_0}^{\om_{\rm max}+\delta_0}\frac{\eps
    d u}{\eps^2+[\om_{\rm max}-u]^2}.
\]
%Since the limit of the integral in the right hand side, as $\eps\to0$ equals
%$\pi$ we conclude 
It follows that for a sufficiently small $\eps_0$ we have
\[
\inf_{\beta\in(-\delta_0,\delta_0)}|j_\eps(\beta)|\ge M,\quad \eps\in(0,\eps_0)
\]
and \eqref{012401b} follows.
\qed

We now turn to the proof of Lemma~\ref{lem-mar602}. Once again, we will only consider
$\iota=+$ and drop the subscript $+$ in the notation. 
Let $\si>0$ be arbitrary.  For  $\rho\in(0,\delta/4)$, with
$\delta>0$ as in \eqref{011812b}, we let 
\begin{equation}
\label{L-bis}
L(\rho):=[k:{\rm dist}(k,\Omega_*)<\rho],~~\Omega_*:=[k\in\bbT:\,\om'(k)=0]\subset \{0,1/2\}, 
\end{equation}
with $\rho$ to be specified further later. 
We can write
$$
{\cal I}^{(2)}(\la,\eps)-  {\cal I}^{(3)}(\la,\eps)=
\tilde{\cal I}^1(\la,\eps)+\tilde{\cal I}^2(\la,\eps),
$$
with the two terms corresponding to the integration
in (\ref{022112bis}) and (\ref{011811xx}) in the $k$-variable
over $L^c(\rho)$, the complement of $L(\rho)$, and  $L(\rho)$ itself, respectively. Since
$|\om'(k)|$ is bounded away from $0$ on  $L^c(\rho)$, an elementary
application of the Lebesgue dominated convergence theorem implies
that
\begin{equation}
\label{012001}
\lim_{\eps\to0}\tilde{\cal I}^1(\la,\eps)=0.
\end{equation}
Assumption (\ref{011812}), (\ref{011812b}) on the support of $\widehat W_\eps(\eta,k)$  
in $k$   allows us to write %(cf \eqref{T2ek1})
\begin{equation}
\label{012001aa}
|\tilde{\cal I}^{2}(\la,\eps)|\preceq  
\tilde{\cal I}^{2,1}(\la,\eps) +\tilde{\cal I}^{2,2}(\la,\eps),
\end{equation}
where
\[
\tilde{\cal I}^{2,j}(\la,\eps):=\!\!\int_{\bbR}\frac{d\beta
  }{(\la/2)^2+\beta^2}\int_{L(\rho)\times  B(\delta,\eps)\times A(\delta,\eps)}\!
  \frac{|\delta_{j,\eps}(\beta,k)|
   }{1+|\beta+\om'(k) (\eta'-\eta'')/2|}\times \frac{ \varphi(\eta')  dk d\eta' d\eta''
}{1+|\beta-\om'(k) (\eta'+\eta'')/2|},
\]
with 
\begin{equation}
\label{Ad}
A(\delta,\eps):=[\eta'':\,\delta/(2\eps)\le |\eta''|\le 6/\eps],\quad B(\delta,\eps):=[\eta':\, |\eta'|\le \delta/(2^{100}\eps)]
\end{equation}
and
\begin{equation}\label{mar802}
\delta_{1,\eps}(\beta,k):=|\tilde
g\left(\la\eps/2-i[\eps\beta+\om(k)]\right)|^2,\qquad \delta_{2,\eps}(\beta,k):=|\nu(k)|^2\preceq |\om'(k)|.
\end{equation}
The last inequality above follows from  (\ref{012410}) and (\ref{feb1402}). 
%we can estimate $\delta_{2,\eps}(\beta,k)\preceq |\om'(k)|$. 
It follows that
\begin{eqnarray} 
\label{mar804}
&&\!\!\!\!\!\!\!\!\!\!
\tilde {\cal I}^{2,2}(\la,\eps)\preceq \int_{\bbR}\frac{d\beta
  }{(\la/2)^2+\beta^2}\int_{L(\rho)\times  B(\delta,\eps)\times A(\delta,\eps)}
  \frac{|\om'(k)|
   }{1+|\beta+\om'(k) (\eta'-\eta'')/2|} \frac{ \varphi(\eta')  dk d\eta' d\eta''
}{1+|\beta-\om'(k) (\eta'+\eta'')/2|}\nonumber\\
&&~~~~~~~~\le C( J_++J_-),
\end{eqnarray}
with 
\begin{equation}
\label{012001a}
J_\pm:=\int_{L(\rho)\times  B(\delta,\eps)\times A(\delta,\eps)}  \frac{ \varphi(\eta') |\om'(k)| dk d\eta' d\eta''
   }{1+[\om'(k) (\eta'\pm \eta'')]^2}, %=J_\pm^1+J_\pm^2,
\end{equation}
and a constant $C>0$ independent of $\eps,\rho$.
We used the Cauchy-Schwarz inequality and   \eqref{Ca}  in the last inequality in (\ref{mar804}).
Note that
\begin{eqnarray}
\label{mar808}
J_\pm:=\int_{L(\rho)\times  B(\delta,\eps)\times A(\delta,\eps)}  \frac{ \varphi(\eta') |\om'(k)| dk d\eta' d\eta''
   }{1+[\om'(k) (\eta'\pm \eta'')]^2}.
\end{eqnarray}
Changing variables $\eta'':=\om'(k) (\eta'\pm \eta'')$ we conclude that
\begin{eqnarray}
\label{mar808-bis}
J_\pm\le \int_{L(\rho)\times \bbR^2}  \frac{ \varphi(\eta')  dk d\eta' d\eta''
   }{1+|\eta''|^2}\le\si,
\end{eqnarray}
provided that $\rho>0$ is sufficiently small.

%\textcolor{green}{TK comment: I do not follows the penultimate  estimate...}

% \blue{This was an old argument:
% The split in the right side of (\ref{012001a}) is according to integration over  the
% region~$\{|\eta'\pm\eta''|<|\eta''|/4\}$ and its complement.
% We only consider $J_+$.
% In the first term we have~$|\eta'|\ge |\eta''|/2$, hence
% \begin{equation}
% \label{012001aaa}
% J_+^1\preceq \int_{L(\rho)\times  \bbR^2}  |\widehat
% \varphi(\eta')|^{1/2} |\widehat \varphi(\eta'')|^{1/2}  dk d\eta'
% d\eta''\preceq \si,
% \end{equation}
% provided $\rho>0$ is sufficiently small (cf \eqref{011812}). In the
% other case, changing  variables $\tilde\eta'':=\om'(k)\eta''$, we can write
% \begin{equation}
% \label{012001c}
% J_+^2\preceq \int_{L(\rho)\times \bbR^2}  \frac{\widehat \varphi(\eta') dk d\eta' d\eta''
%    }{1+ (\eta'')^2}\preceq \si,
% \end{equation}
% provided that $\rho>0$ is sufficiently small. }

As for the term $\tilde{\cal I}^{2,1}(\la,\eps)$, 
using Cauchy-Schwarz inequality we obtain 
\[
\tilde{\cal I}^{2,1}(\la,\eps)\preceq K_{\eps,+}+K_{\eps,-},
\]
with
\begin{equation}\label{mar812}
\begin{aligned}
K_{\eps,\pm}(\rho)&:=\int_{\bbR}\frac{d\beta
  }{1+\beta^2}\int_{L(\rho)\times  B(\delta,\eps)\times A(\delta,\eps)}
  \frac{\delta_{1,\eps}(\beta,k) \varphi(\eta')  dk d\eta' d\eta''
   }{1+|\beta-\om'(k) (\eta''\pm\eta')/2|^2}\\
   &=K_{\eps,+}^1(\rho,\rho')+K_{\eps,+}^2(\rho,\rho').
\end{aligned}
\end{equation}
%We write 
%\[
%K_{\eps,+}(\rho)=K_{\eps,+}^1(\rho)+K_{\eps,+}^2(\rho,\rho')+K_{\eps,+}^3(\rho,\rho'),
%\]
%where 
The terms in
the right  side correspond to integration  over the regions
\begin{align*}
% &
% {\cal K}_{\eps,+}^1(\rho):=[(\beta,k,\eta',\eta'')\in \bbR\times  L(\rho)\times
%   \bbR\times A(\delta,\eps)\,:|\eta'|\ge \delta/(2^4\eps)],\\
&
{\cal K}_{\eps,+}^1(\rho,\rho'):=[(\beta,k,\eta',\eta'')\in \bbR\times  L(\rho)\times
  \bbR\times A(\delta,\eps)\,:|\eta'|< \delta/(2^{100}\eps),\,|\beta|\ge \rho'\eps^{-1}],\\
&
{\cal K}_{\eps,+}^2(\rho,\rho'):=[(\beta,k,\eta',\eta'')\in \bbR\times  L(\rho)\times
  \bbR\times A(\delta,\eps)\,:|\eta'|< \delta/(2^{100}\eps),\,|\beta|<\rho'\eps^{-1}],
\end{align*}
with $\rho'>0$ to be chosen later.
% As $\delta_{1,\eps}(\beta,k)\le 1$, see  (\ref{012410}), we can sacrifice the factor $\eps^{-1}$ for the integration
% over $\eta''$ in (\ref{mar812})  and~get
% \begin{equation}
% \label{K1}
% K_{\eps,+}^1(\rho)\preceq \frac{1}{\eps}\int_{[|\eta'|\ge \delta/(2^4\eps)]}
%  \varphi(\eta')  d\eta'\to0,\quad\mbox{as }\eps\to0,\,\mbox{
%    for each }\rho>0 .
% \end{equation}
Since $\om'(k_*)=0$, for each $\rho'>0$ we can find $\rho$ sufficiently small so that 
$$
|\beta-\om'(k) (\eta''+\eta')/2|\ge |\beta|/2,\quad \mbox{ on }{\cal
  K}_{\eps,+}^1(\rho,\rho').
$$
Therefore, for each $\rho'>0$ we can find $\rho>0$ sufficiently small so that
\begin{equation}
\label{K2}
K_{\eps,+}^1(\rho,\rho')\preceq \frac{1}{\eps}\int_{[|\beta|\ge \rho'\eps^{-1}]}\frac{d\beta
  }{(1+\beta^2)^2}\to0,\quad\mbox{as }\eps\to0,
\end{equation}
{with the pre-factor $\eps^{-1}$ coming again from the integration over $\eta''$ in (\ref{mar812}) .}
Finally, we can write
\[
K_{\eps,+}^2(\rho,\rho')\le m_\eps(\rho,\rho')\int_{\bbR}\frac{d\beta
  }{1+\beta^2}\int_{L(\rho)\times [|\eta'|< \delta/(2^{100}\eps)]\times A(\delta,\eps)}
  \frac{ \varphi(\eta')  dk d\eta' d\eta''
   }{1+|\beta-\om'(k) (\eta''+\eta')/2|^2},
\]
where 
\[
\delta':=\rho'+\sup_{k\in L(\rho)}|\om(k)-\om(k_*)|,~~m_\eps(\rho,\rho'):=\sup_{\beta'\in (-\delta',\delta')} |\tilde
g(\eps\la-i[\beta'+\om(k_*)] )|^2.
\]
Using \eqref{Ca} again gives
\begin{align*}
K_{\eps,+}^2(\rho,\rho')&\preceq m_\eps(\rho,\rho')\int_{L(\rho)\times[ |\eta'|< \delta/(2^{100}\eps)]\times A(\delta,\eps)}
  \frac{  \varphi(\eta')  dk d\eta' d\eta''
   }{1+|\om'(k) (\eta''+\eta')|^2}\\
&\preceq  m_\eps(\rho,\rho')\int_{L(\rho)\times A({\delta/10,\eps/2})}
  \frac{ dk d\eta''
   }{1+|\om'(k) \eta''|^2}\\
%\end{align*}
%%Changing variables $\eta''\to\om'(k) \eta''$ 
%We evaluate the
%integral over $\eta''$ and obtain
%\begin{align*}
%&K_{\eps,+}^3(\rho,\rho')
&\preceq m_\eps(\rho,\rho')\int_{L(\rho)}
  \frac{ dk}{|\om'(k)|} \left[\arctan\left(\frac{{12}|\om'(k)|}{\eps}\right)-\arctan\left(\frac{\delta|\om'(k)|}{\red{5}\eps}\right)\right].
\end{align*}
Using a well known trigonometric identity we write
$$
\arctan\Big(\frac{6|\om'(k)|}{\eps}\Big)-\arctan\Big(\frac{\delta|\om'(k)|}{2\eps}\Big)=
\arctan\Big(\frac{(12-\delta{/5})|\om'(k)|}{\eps}\Big\{1+\frac{{(12/5)}\delta|\om'(k)|^{{2}}}{\eps^2}\Big\}^{-1}\Big),
$$
therefore
\[
K_{\eps,+}^2(\rho,\rho')\preceq  {m_\eps(\rho,\rho')\int_{0}^{\rho}\frac{dk}{k}
\arctan\Big(\farc{(k/\eps)}{1+(k/\eps)^2}%\frac{k}{\eps}\Big\{1+\frac{k^2}{\eps^2}\Big\}^{-1}
\Big)
\preceq m_\eps(\rho,\rho')\int_{0}^{\infty}\frac{dk}{k}
\arctan\Big( 
\frac{k}{ 1+ k^2}\Big)\preceq m_\eps(\rho,\rho').
}
%\preceq  m(\rho,\rho')\int_{0}^{\rho}\frac{dk}{\eps+k+\eps^{-1}k^2}.
\]
% \blue{This is the old argument: Changing variables $k':=\eps k$ we obtain
% \begin{align*}
% &K_{\eps,+}^3(\rho,\rho')\preceq  m(\rho,\rho')\int_{0}^{\rho/\eps}\frac{dk}{1+k+k^2}.
% \end{align*}
% Thus,
% \begin{align*}
% &\limsup_{\eps\to0}K_{\eps,+}^3(\rho,\rho')\preceq \limsup_{\eps\to0}\sup_{\beta'\in (-\delta',\delta')}\left|\tilde
% g\left(\eps\la-i[\beta'+\om(k_*)]\vphantom{\int_0^1}\right)\right|^2.
% \end{align*}
% }
Lemma \ref{lm5.4} implies now that we can choose $\rho,\rho'$ so small
that 
\begin{equation}
\label{K3}
\limsup_{\eps\to0}K_{\eps,+}^2(\rho,\rho'){\preceq\sigma.}
\end{equation}
Combining \eqref{K2} and \eqref{K3} we conclude that for
each $\si>0$ there exists $\rho\in(0,1)$ such that
\begin{equation}
\label{K}
\limsup_{\eps\to0}K_{\eps,+}(\rho)\preceq \si.
\end{equation}
The analysis for $K_{\eps,-}(\rho)$ is very similar, finishing the proof of Lemma~\ref{lem-mar602}.  
%\begin{equation}
%\label{K-}
%\limsup_{\eps\to0}K_{\eps,-}(\rho)\preceq \si.
%\end{equation}
%This ends the proof of \eqref{I}.

\subsection{The proof of Lemma~\ref{lem-mar1202}}

As usual, we only consider $\iota=+$ and drop the corresponding subscript $+$.
A straightforward computation using (\ref{011812}), the regularity of the test function
$\hat G(\eta,k)$, and~(\ref{Ca}) shows that we can replace $\hat G^\star(\eta,k+\eps\eta/2)$
in (\ref{011811xx}) by $\hat G^\star(\eta,k)$, and $\delta_\eps^+\omega(k,\eta)$
by $\omega'(k)\eta$, so that
\begin{equation}\label{mar1302}
|{\cal I}^{(3)}(\la,\eps)-\tilde{\cal I}^{(3)}(\la,\eps)|\to 0,\hbox{ as
$\eps\to 0$},
\end{equation}
where
\begin{eqnarray}
\label{mar1208}
&&
\tilde{\cal I}^{(3)}(\la,\eps)
=-\frac{\gamma\la}{8\pi}\int_{\bbR\times \bbT_{2/\eps}}\frac{d\beta d\eta}{(\la/2)^2+\beta^2}
\int_{ T_{\eps,\iota}^3}  \frac{\widehat W_\eps(\eta',k+\eps\eta''/2)  dk d\eta' d\eta''
}{\la/2-i\{\beta-\iota\om'(k) (\eta'+\eta'')/2\}} \nonumber\\
&&
\times \frac{|\nu(k)|^2
   }{\la/2+i\{\beta+\iota\om'(k) (\eta'-\eta'')/2\}} \times \frac{\hat
   G^\star(\eta,k) }{\la+i\om'(k)\eta}.
\end{eqnarray}
We change variables $k':=k+\eps\eta''/2$
in  the right   side to obtain
\begin{align}\label{mar1306}
&
\tilde {\cal I}^{(3)}(\la,\eps)=-\frac{\gamma\la}{8\pi}
\int_{\bbR\times \bbT_{2/\eps}}\frac{d\beta d\eta}{(\la/2)^2+\beta^2}
\int_{U_{\eps}}  \frac{\widehat W_\eps(\eta', k)  dk d\eta' d\eta''
}{\la/2-i\{\beta-\om'(k- \eps\eta''/2) (\eta'+\eta'')/2\}} \nonumber\\
&
\times \frac{|\nu(k-\eps\eta''/2)|^2
   }{\la/2+i\{\beta+\om'(k-\eps\eta''/2) (\eta'-\eta'')/2\}} \times \frac{\hat
   G^\star(\eta, k) }{\la+i\om'(k)\eta}+o(1)\\
&
=-\frac{\gamma\la}{8\pi}\int_{\bbR\times \bbT_{2/\eps}}\frac{d\beta d\eta}{(\la/2)^2+\beta^2}
\int_{ U_{\eps}} 
\frac{\widehat W_\eps(\eta',k)  dk d\eta' d\eta''
}{\la/2-i\{\beta-\om'(k) (\eta'+\eta'')/2\}} \nonumber\\
&
\times \frac{|\nu(k)|^2
   }{\la/2+i\{\beta+\om'(k) (\eta'-\eta'')/2\}} \times \frac{\hat
   G^\star(\eta, k) }{\la+i\om'(k)\eta}+\Delta_{\eps}+o(1),\nonumber
\end{align}
with, cf \eqref{T2ek1},
\begin{equation}
\label{T2ek1a}
U_{\eps}:=\left[(k,\eta',\eta''):\,k\in\bbT,\,|\eta'|\le
  \frac{\delta}{2^{100}\eps},\,\left|k\right|\le \farc{1-\eps |\eta'|}{2},\, 
  {\rm sign}(k-\eps\eta''/2)={\rm sign} \left( k+\eps\eta'/2\right)\right].
\end{equation}
The term $o(1)$ in the right side of (\ref{mar1306}) appears because 
we have, once again, approximated the arguments in $\hat G^\star$ and
$\omega'$ by $k$ in the very last factor, despite the latest change of variables. 
The error~$\Delta_\eps$, that we now need to estimate,
appears in (\ref{mar1306})
because we have replaced the arguments of $\omega'$ by $k$ in the first
two factors.

Thanks to assumption \eqref{011812}, the
integration over $k$ in (\ref{mar1306}) is only over the complement of the set~$L(\delta)$, see \eqref{L}.
We can write then (cf \eqref{Ad})
\begin{equation}\label{mar1310}
|\Delta_{\eps}|\preceq \int_{L^c(\delta)\times
  \bbR\times B(\delta,\eps)\times \bbR} d_\eps(k,\eta',\eta'')\varphi(\eta')
  \frac{|\hat G^\star(\eta, k)| }{|\la+i\om'(k)\eta|} dk d\eta d\eta' d\eta'' 
  =\Delta_{\eps}'+\Delta_{\eps}''.
\end{equation}
The terms $\Delta_{\eps}'$ and $\Delta_{\eps}''$ correspond
to the integration in $\eta''$
over the domains
$
A'(\delta,\eps):=[|\eta''|\le \delta/(2\eps)],
$
and ${A''(\delta,\eps)=[|\eta''|\ge \delta/(2\eps)]}$, and 
\begin{align}
\label{de}
&
d_\eps(k,\eta',\eta''):=  \int_{\bbR}\frac{ d\beta }{(\la/2)^2+\beta^2} \left|\frac{1 
}{\la/2-i\{\beta-\om'(k- \eps\eta''/2) (\eta'+\eta'')/2\}} \right.\nonumber\\
&
\times \frac{|\nu(k- \eps\eta''/2)|^2
   }{\la/2+i\{\beta+\om'(k- \eps\eta''/2) (\eta'-\eta'')/2\}}  -
\frac{1
}{\la/2-i\{\beta-\om'(k) (\eta'+\eta'')/2\}} \nonumber\\
&
\left.\times \frac{|\nu(k)|^2
   }{\la/2+i\{\beta+\om'(k) (\eta'-\eta'')/2\}}  \right|.
\end{align}
Using \eqref{Ca} we can estimate, for $(k,\eta',\eta'')\in
L^c(\delta)\times B(\delta,\eps)\times  A'(\delta,\eps)$:
\begin{equation}
\label{del1}
d_\eps%1_{L(\delta)\times\bbR\times A'(\delta,\eps)}
(k,\eta',\eta'')\preceq
\frac{1}{1+(\eta'-\eta'')^2}+\frac{1}{1+(\eta'+\eta'')^2}.
%\varphi(\eta') \| G(\eta,  \cdot)\|_{L^\infty(\bbT)}.
\end{equation}
As $d_\eps(k,\eta',\eta'')\to 0$ pointwise, we can apply
the dominated convergence theorem in (\ref{mar1310}), to get
\begin{equation}\label{mar1312}
\lim_{\eps\to0}\Delta_{\eps}'=0.
%\preceq \int_{L(\delta)\times
%  \bbR^3} \lim_{\eps\to 0}d_\eps(k,\eta,\eta',\eta'') dk d\eta d\eta' d\eta'' =0,
\end{equation}
% as the integrands in the right hand
%side of \eqref{de} are uniformly  bounded and vanish, as $\eps\to0$.

To estimate $\Delta_{\eps}''$, observe that  \eqref{Ca} implies  
$$
d_\eps(k,\eta',\eta'')\preceq \sum_{\iota=\pm}\left(d_\eps^{1,\iota}(k,\eta',\eta'')+d_\eps^{2,\iota}(k,\eta',\eta'')\right),
$$
with %$d_\eps^j=\sum_{\iota=\pm}d_\eps^{j,\iota}$, $j=1,2$ and 
\begin{equation}
\label{de1}
d_\eps^{1,\iota}(k,\eta',\eta''):= \frac{|\nu(k-\eps\eta''/2)|^2
}{1+[\om'(k- \eps\eta''/2)(\eta'+\iota\eta'')]^2},~~
d_\eps^{2,\iota}(k,\eta',\eta''):= \frac{ |\nu(k)|^2 
}{1+[\om'(k)(\eta'+\iota\eta'')]^2},~~\iota\in\{-,+\}.
\end{equation}

 As $|\eta''|$ is  larger than
  $\delta/\eps$ on $A''(\delta,\eps)$ and $|\om'(k)|$ is
bounded away from $0$ on $L^c(\delta)$, the decay of~$\varphi(\eta')$ allows us to apply  
the dominated convergence theorem, to obtain 
 \begin{equation}
\label{dd2}
 \lim_{\eps\to 0}\int_{L^c(\delta)\times
  \bbR\times B(\delta,\eps) \times A''(\delta,\eps)}d_\eps^{2,\iota}(k,\eta',\eta'')  {\varphi(\eta')
  \frac{|\hat G^\star(\eta, k)| }{|\la+i\om'(k)\eta|}} dk d\eta d\eta' d\eta'' =0 ,
  \quad\iota\in\{-,+\}.
\end{equation}
{For the terms  $d_\eps^{1,\iota}$, we
  consider only the case $\iota=+$, as the other case can be done similarly.} 
Note that 
$$
B(\delta,\eps) \times A''(\delta,\eps) \subset A_1(\delta,\eps):=[(\eta',\eta'')\in\bbR\times
A''(\delta,\eps):\,|\eta'+\eta''|\ge |\eta''|/2].
$$
% $$
% A_2(\delta,\eps):=[(\eta',\eta'')\in\bbR\times
% A''(\delta,\eps):\,|\eta'+\eta''|\le |\eta''|/2],
% $$
% and split
Hence,
\begin{align}\label{mar1316}
% \blue{\lim_{\eps\to 0}}
& \int_{L^c(\delta)\times
  \bbR\times B(\delta,\eps) \times A''(\delta,\eps)}d_\eps^{1}(k,\eta',\eta'')  {\varphi(\eta')
  \frac{|\hat G^\star(\eta, k)| }{|\la+i\om'(k)\eta|}} dk
d\eta d\eta' d\eta'' \nonumber\\
&
\le {\cal D}_{\eps}: = \int_{L^c(\delta)\times
  \bbR\times A_1(\delta,\eps)}d_\eps^{1}(k,\eta',\eta'')  {\varphi(\eta')
  \frac{|\hat G^\star(\eta, k)| }{|\la+i\om'(k)\eta|}} dk
d\eta d\eta' d\eta'' .
\end{align}
% Here, the term ${\cal D}_{j,\eps}$ corresponds to the integration over
% $A_j(\delta,\eps)$ for $j=1,2$.
%\textcolor{green}{Using \eqref{011812c}}, 
For any $\kappa'\in(0,1)$ we can write
\begin{align}
\label{De1}
&
{\cal D}_{\eps}\preceq \int_{L^c(\delta)  
   \times A''(\delta,\eps)}\frac{|\nu(k-\eps\eta''/2)|^2
}{1+[\om'(k- \eps\eta''/2)\eta'']^2 }dk   d\eta''
\nonumber\\ &
\preceq \int_{L^c(\delta)\times
  A''(\delta,\eps)}\frac{|\nu(k-\eps\eta''/2)|^2}{|\om'(k-
  \eps\eta''/2)|}\times\frac{dk d\eta''}{|\om'(k-
  \eps\eta''/2)|^{\kappa'}|\eta''|^{1+\kappa'} }\\
&
\preceq \int_{L^c(\delta)\times
  A''(\delta,\eps)}\frac{dk d\eta''}{|\om'(k-
  \eps\eta''/2)|^{\kappa'}|\eta''|^{1+\kappa'} }\preceq \int_{
  A''(\delta,\eps)}\frac{d\eta''}{|\eta''|^{1+\kappa'} }\preceq
  \eps^{\kappa'}\to 0,\,\mbox{as $\eps\to0$.}\nonumber
\end{align}
% As for ${\cal D}_{2,\eps}$, note that $|\eta'|\ge |\eta''|/2\ge \delta/(4\eps)$ on
% $A_2(\delta,\eps)$. Therefore, we have 
% \begin{align}
% \label{De2}
% &{\cal D}_{2,\eps}\preceq \int_{L^c(\delta)\times
%     A_2(\delta,\eps)}\varphi(\eta')  
%    dk   d\eta' d\eta''
% \preceq \frac{1}{\eps}\int_{[|\eta'|\ge \delta/(4\eps)]}\varphi(\eta') d\eta'\preceq
%    {\eps^{2+2\kappa}}\to 0,\,\mbox{as $\eps\to0$,} 
% \end{align}
% with $\kappa>0$ as in \eqref{011812c}.
% Combining (\ref{De1}) with (\ref{De2}),
% We conclude therefore that
% \begin{equation}
% \label{dd1p}
%  \lim_{\eps\to 0}\int_{L^c(\delta)\times
%   \bbR^2\times A''(\delta,\eps)}d_\eps^{1,+}(k,\eta',\eta'')  {\varphi(\eta')
%   \frac{|\hat G^\star(\eta, k)| }{|\la+i\om'(k)\eta|}} dk{d\eta d\eta' d\eta''}=0.
% \end{equation}
% %Analogously,
% %\begin{equation}
% %\label{dd1n}
% % \lim_{\eps\to 0}\int_{L(\delta)\times
% %  \bbR^2\times A(\delta,
%\eps)}d_\eps^{1,-}(k,\eta,\eta',\eta'') dk=0.
%\end{equation}
We obtain from \eqref{dd2} and \eqref{De1} and its analog for $\iota=-$ that
\begin{equation}
\label{del2}
\lim_{\eps\to0}\Delta_{\eps}''=0,
\end{equation}
which, together with \eqref{mar1312} gives
\begin{equation}
\label{del+}
\lim_{\eps\to0}\Delta_{\eps}=0.
\end{equation}
We have shown that
\begin{align}\label{mar1320}
&
\tilde {\cal I}^{(3)}(\la,\eps)=-\frac{\gamma\la}{8\pi}\int_{\bbR\times \bbT_{2/\eps}}\frac{d\beta d\eta}{(\la/2)^2+\beta^2}
\int_{ U_{\eps}} 
\frac{\widehat W_\eps(\eta',k)  dk d\eta' d\eta''
}{\la/2-i\{\beta-\om'(k) (\eta'+\eta'')/2\}} \nonumber\\
&
\times \frac{|\nu(k)|^2
   }{\la/2+i\{\beta+\om'(k) (\eta'-\eta'')/2\}} \times \frac{\hat
   G^\star(\eta, k) }{\la+i\om'(k)\eta}+ o(1).\nonumber
\end{align}
Now, the dominated convergence theorem allows us to pass to the limit in the domains of integration in (\ref{mar1306}),
leading to (\ref{mar1204}).

%\red{\Large STOPPED HERE}
%Likewise, 
%$
%\lim_{\eps\to0}\Delta_{3,\eps}^{-}=0$, hence
%\begin{align*}
%&
%-\frac{\gamma\la}{8\pi}\int_{\bbR\times \bbT_{2/\eps}}\frac{d\beta d\eta}{(\la/2)^2+\beta^2}
%\int_{U_{\eps,\iota}}  \frac{\widehat W_\eps(\eta',\iota k+\eps\eta''/2)  dk d\eta' d\eta''
%}{\la/2-i\{\beta-\iota\om'(k-\iota \eps\eta''/2) (\eta'+\eta'')/2\}} \nonumber\\
%&
%\times \frac{|\nu(k-\iota \eps\eta''/2)|^2
%   }{\la/2+i\{\beta+\iota\om'(k-\iota \eps\eta''/2) (\eta'-\eta'')/2\}} \times \frac{\hat
%   G^\star(\eta, k-\iota \eps\eta'') }{\la+i\om'(k-\iota
%  \eps\eta'')\eta}\\
%&
%=-\frac{\gamma\la}{8\pi}\int_{\bbR\times \bbT_{2/\eps}}\frac{d\beta d\eta}{(\la/2)^2+\beta^2}
%\int_{ U_{\eps,\iota}} 
%\frac{\widehat W_\eps(\eta',\iota k)  dk d\eta' d\eta''
%}{\la/2-i\{\beta-\iota\om'(k) (\eta'+\eta'')/2\}} \nonumber\\
%&
%\times \frac{|\nu(k)|^2
%   }{\la/2+i\{\beta+\iota\om'(k) (\eta'-\eta'')/2\}} \times \frac{\hat
%   G^\star(\eta, k) }{\la+i\om'(k)\eta}+o(1).
%\end{align*}

\subsection{The end of the proof of Lemma~\ref{lem-feb1508}}\label{sec:7.5}

%Passing to the limit $\eps\to 0$ in $\hat W_\eps(\eta',k)$ in
%(\ref{mar1204}), we now obtain 
%Let us now restore the index $\pm$ by ${\cal H}$ defined in (\ref{Hpm}).
As a result of Lemmas~\ref{lem-mar202}-\ref{lem-mar1202}, together with (\ref{mar1406}),
we know that 
\begin{eqnarray}
\label{I1x}
&&{\cal H}_{+}(\la,\eps) =-\frac{\gamma\la}{8\pi}\sum_{\iota=\pm}\int_{\bbR^2}\frac{d\beta d\eta}{(\la/2)^2+\beta^2}
\int_{ \bbT\times\bbR^2} 
\frac{\widehat W_{\blue{\eps}}(\eta',\iota k)  dk d\eta' d\eta''
}{\la/2-i\{\beta-\iota\om'(k) (\eta'+\eta'')/2\}} \nonumber\\
&&
\times \frac{|\nu(k)|^2
   }{\la/2+i\{\beta+\iota\om'(k) (\eta'-\eta'')/2\}} \times \frac{\hat
   G^\star(\eta, k) }{\la+i\om'(k)\eta}+o(1),
\end{eqnarray}
as $\eps\ll1$.
Recall the elementary formula: for  $q_\pm\in\mathbb C$ such that  ${\rm
  Im}\,q_+>0> {\rm Im}\,q_-$ we have
\begin{equation}
\label{res}
\int_{\bbR}\frac{dq }{(q-q_+)(q-q_-)}= \frac{2\pi i }{q_+-q_-}.
\end{equation}
Performing the integral in the $\eta''$ variable  in
\eqref{I1x} we obtain
\begin{eqnarray}
\label{I1x1}
{\cal H}_{+}(\la,\eps) =-\frac{\gamma\la}{2}\sum_{\iota=\pm} 
\int_{\bbR^2}\frac{d\beta d\eta}{(\la/2)^2+\beta^2}
\int_{ \bbT\times\bbR} 
\frac{|\nu(k)|^2\widehat W_{\blue{\eps}}(\eta', \iota k)  dk d\eta' 
}{|\om'(k)|[\la+\iota i\om'(k)\eta']} \times\frac{\hat
   G^\star(\eta, k) }{\la+i\om'(k)\eta}+o(1).
\end{eqnarray}
%Using the identity
%\[
%\int_{\bbR}\frac{(\la/2)d\beta }{(\la/2)^2+\beta^2}=\pi
%\]
%and performing the integral over 
Integrating out the $\beta$-variable we get (recall that $\bar\om'(k)=\om'(k)/(2\pi)$)
\begin{eqnarray}
\label{I1x1-bis}
&&{\cal H}_{+}(\la,\eps) =-\frac{\gamma}{2}\sum_{\iota=\pm} 
\int_{ \bbT\times\bbR^2} 
\frac{|\nu(k)|^2\widehat W_\eps(\eta', \iota k) 
}{|\bar\om'(k)|[\la+\iota i\om'(k)\eta']} \times\frac{\hat
   G^\star(\eta, k) }{\la+i\om'(k)\eta} dk d\eta d\eta' +o(1).
\end{eqnarray}
An analogous formula holds for ${\cal H}_{-}(\la,\eps) $. % Using
% \eqref{041511a} and Lemma \ref{011311} we conclude therefore that 
% $$
% {\frak L}_{\eps,1}^2(\la)=-\frac{\gamma}{4}\left({\cal
%     I}_{+}(\la,\eps)+{\cal I}_{-}(\la,\eps)\right)+o(1),\quad \mbox{as }\eps\ll1.
% $$
Letting $\eps\to0$ we obtain  
\eqref{021511c}, finishing the proof of Lemma~\ref{lem-feb1508}.

\section{Proof of Lemma~\ref{lem-feb1502}: 
the limit of ${\frak L}_{scat,22}^{\eps}(\la)$}\label{sec:lem4.2}

We now turn to the computation that leads to (\ref{021511b}) the second and final ingredient in Lemma~\ref{lem-feb1502}:
\begin{equation}
\label{mar1412}
\lim_{\eps\to0+}{\frak L}_{scat,22}^\eps(\la)=0.
\end{equation}
Observe that, as follows from (\ref{041511}) and (\ref{feb1514}), we have
 \begin{eqnarray}
\label{mar1416}
{\frak L}_{scat,22}^\eps(\la)= -\frac{i\ga}{2}\int_{\bbR\times
  \bbT}[{\rm Im} {\frak
   d}_\eps^2\left(\la,k\right) ]\Big[ \frac{\hat
   G^*(\eta,k+\eps\eta/2)}{\la+i\delta_\eps^+\om(k,\eta)}-\frac{\hat
   G^*(\eta,k-\eps\eta/2)}{\la+i\delta_\eps^-\om(k,\eta)}\Big]d\eta dk
   %={\frak L}_{scat,21}^\eps(\la)+{\frak L}_{scat,22}^\eps(\la),\nonumber
\end{eqnarray}
with 
%the two terms corresponding to writing 
%\begin{equation}\label{feb1514}
%{\frak d}_\eps^2={\rm Re}{\frak d}_\eps^2+i{\rm Im}{\frak d}_\eps^2.
%\end{equation} 
%here
%\begin{equation}
%\label{041511a}
%{\frak L}_{scat,21}^\eps(\la)=-\frac{\ga}{2}\int_{\bbR\times
%  \bbT}{\rm Re}\,{\frak
%   d}_\eps^2\left(\la,k\right) \left[ \frac{\hat
%   G^\star(\eta,k+\eps\eta/2)}{\la+i\delta_\eps^+\om(k,\eta)}+\frac{\hat
%   G^\star(\eta,k-\eps\eta/2)}{\la+i\delta_\eps^-\om(k,\eta)}\right]d\eta dk
%\end{equation}
%and
%\begin{equation}
%\label{041511b}
%{\frak L}_{scat,22}^\eps(\la)=-\frac{i\ga}{2}\int_{\bbR\times
%  \bbT}{\rm Im}\,{\frak
%   d}_\eps^2\left(\la,k\right) \left[ \frac{\hat
%   G^\star(\eta,k+\eps\eta/2)}{\la+i\delta_\eps^+\om(k,\eta)}-\frac{\hat
%   G^\star(\eta,k-\eps\eta/2)}{\la+i\delta_\eps^-\om(k,\eta)}\right]d\eta dk.
%\end{equation}
%We recall that
\begin{equation}\label{mar1420}
{\frak d}_\eps^2(\la,k) %=\eps\int_0^{+\infty}e^{-\la\eps t}I\!I_\eps\left(t,k\right)dt
= -\gamma \eps\int_0^{+\infty}e^{-\la\eps t}dt \Big\{\int_0^{t}
  e^{i\om(k)(t-s)}\left\langle g\star {\frak p}_0^0(s)g\star {\frak p}_0^0(t)
\right\rangle_{\mu_\eps}\Big\} ds .
\end{equation}
 A lengthy calculation, similar to that at the beginning of Section~\ref{sec:6}, leads to an expression
\begin{eqnarray}
\label{010202b}
&&
i{\rm Im}\,{\frak d}_\eps^2(\la,k)
=-\frac{i\eps\la \gamma \om(k)  }{4\pi} \int_{\bbR}\frac{\,\beta\,|\tilde
   g(\eps\la/2-i\beta)|^2
   d\beta}{\{(\eps\la/2)^2+[\beta+\om(k)]^2\}\{(\eps\la/2)^2+[\beta-\om(k)]^2\}} \nonumber
\\
&&
\times  \int_{\bbT^2}\frac{\eps\langle
   \hat\psi(\ell)\hat\psi^*(\ell')\rangle_{\mu_\eps}d\ell d\ell'}{\{\eps\la/2-i[\beta-\om(\ell)]\}\{\eps\la/2+i[\beta-\om(\ell')]\}},
\end{eqnarray}
hence
\begin{eqnarray}
\label{020202}
&&
{\frak L}_{scat,22}^\eps(\la)=\frac{i\eps\la \ga^2}{8\pi}\int_{\bbR^2\times
  \bbT^3}\left[ \frac{\hat
   G^\star(\eta,k+\eps\eta/2)}{\la+i\delta_\eps^+\om(k,\eta)}-\frac{\hat
   G^\star(\eta,k-\eps\eta/2)}{\la+i\delta_\eps^-\om(k,\eta)}\right]\\
&&
\times 
\frac{\, \om(k) \beta\,|\tilde g(\eps\la/2-i\beta)|^2
}{\{(\eps\la/2)^2+[\beta+\om(k)]^2\}\{(\eps\la/2)^2+[\beta-\om(k)]^2\}}
%    \nonumber
%\\
%&&
%\times 
\frac{\eps\langle
   \hat\psi(\ell)\hat\psi^*(\ell')\rangle_{\mu_\eps}d\beta d\eta dk d\ell d\ell'}{\{\eps\la/2-i[\beta-\om(\ell)]\}\{\eps\la/2+i[\beta-\om(\ell')]\}}.\nonumber
\end{eqnarray}
% Change variables according to \eqref{ell-k}. We obtain then
% \begin{eqnarray}
% \label{020203}
% &&
% {\frak L}_{\eps,2}^2(\la)=\frac{i\la \ga^2}{4\pi}\int_{\bbR\times
%   \bbT}\left[ \frac{\hat
%    G^\star(\eta,k+\eps\eta/2)}{\la+i\delta_\eps^+\om(k,\eta)}-\frac{\hat
%    G^\star(\eta,k-\eps\eta/2)}{\la+i\delta_\eps^-\om(k,\eta)}\right]d\eta
%   dk\\
% &&
% \times 
% \int_{\bbR}\frac{\, \om(k) \beta\,|\tilde
%    g(\eps\la/2-i\beta)|^2
%    d\beta}{\{(\eps\la/2)^2+[\beta+\om(k)]^2\}\{(\eps\la/2)^2+[\beta-\om(k)]^2\}} \nonumber
% \\
% &&
% \times  \int_{T^2_\eps}\frac{\widehat{W}_\eps(\eta',k')d\eta' dk'}{\{\eps\la/2-i[\beta-\om(k'+\eps\eta'/2)]\}\{\eps\la/2+i[\beta-\om(k'-\eps\eta'/2)]\}}.\nonumber
% \end{eqnarray}
After the change of variables $\beta':=\eps^{-1}\beta$, we get
\begin{eqnarray}
\label{020203}
&&
{\frak L}_{scat,22}^\eps(\la)=-\frac{\la \ga^2}{8\pi\eps}\int_{
  \bbT}\om(k) {\cal G}_\eps(k) 
  dk
\int_{\bbR}\frac{\,\beta\,|\tilde
   g(\eps\la/2-i\eps\beta)|^2
   d\beta}{\{(\la/2)^2+[\beta+\eps^{-1}\om(k)]^2\}\{(\la/2)^2+[\beta-\eps^{-1}\om(k)]^2\}} 
\nonumber\\
&&
\times  \int_{T^2_\eps}\frac{\eps\langle
   \hat\psi(\ell)\hat\psi^*(\ell')\rangle_{\mu_\eps}d\ell
   d\ell'}{\{\la/2-i[\beta-\eps^{-1}\om(\ell)]\}\{\la/2+i[\beta-\eps^{-1}\om(\ell')]\}},
\end{eqnarray}
with
\begin{eqnarray}\label{mar1504}
&&
{\cal G}_\eps(k):=-i \int_{\bbR}\left[ \frac{\hat
   G^\star(\eta,k+\eps\eta/2)}{\la+i\delta_\eps^+\om(k,\eta)}-\frac{\hat
   G^\star(\eta,k-\eps\eta/2)}{\la+i\delta_\eps^-\om(k,\eta)}\right]d\eta \\
%&&
%=-i\left[ \int_{\bbR}\frac{\hat
%   G^\star(\eta,k+\eps\eta/2)}{\la+i\delta_\eps^+\om(k,\eta)}d\eta-\int_{\bbR}\frac{\hat
%   G^\star(\eta,k+\eps\eta/2)}{\la-i\delta_\eps^+\om(k,\eta)}d\eta\right]\\
&&
= \int_{\bbR}\frac{\hat
   G^\star(\eta,k+\eps\eta/2)[2\om(k)-\om(k+\eps\eta)-\om(k-\eps\eta)]}{\eps\{\la^2+[\delta_\eps^+\om(k,\eta)]^2\}}d\eta.\nonumber
\end{eqnarray}

%\subsubsection{The case $\om\in C^\infty(\bbT)$}

Let us first assume that $\om\in C^\infty(\bbT)$. Then
we can estimate 
\begin{equation}
\label{Ge}
|{\cal G}_\eps(k)|\preceq \eps\|\om''\|_{\infty}\int_{\bbR}\eta^2\|\hat
   G^\star(\eta,\cdot)\|_{\infty}d\eta\preceq \eps,
\end{equation}
while the last integral in the right side of~\eqref{020203} is bounded 
by
\begin{equation}
\label{040302}
\frac{4\eps
}{\la^2}\left\langle\left[\int_{\bbT^2}|\hat\psi(\ell)|d\ell\right]^2\right\rangle_{\mu_\eps}\le 
  \frac{4\eps
}{\la^2}\left\langle\|\hat\psi\|_{L^2(\bbT)}^2\right\rangle_{\mu_\eps} \preceq 1.
\end{equation}
Hence, we have
\begin{eqnarray}
\label{010204}
&&
|{\frak L}_{scat,22}^{\eps}(\la)|\preceq \eps\int_{
  \bbT}
  dk
\int_{0}^{+\infty}\frac{\, \eps^{-1}\om(k) \beta\,
   d\beta}{\{1+[\beta+\eps^{-1}\om(k)]^2\}\{1+[\beta-\eps^{-1}\om(k)]^2\}}
  \\
&&
=\eps\int_{
  \bbT}
  dk
\int_{0}^{\om(k)/\eps}\frac{\, \eps^{-1}\om(k) \beta\,
   d\beta}{\{1+[\beta+\eps^{-1}\om(k)]^2\}\{1+[\beta-\eps^{-1}\om(k)]^2\}}\nonumber\\
&&
+\eps\int_{
  \bbT}
  dk
\int_{\om(k)/\eps}^{+\infty}\frac{\, \eps^{-1}\om(k) \beta\,
   d\beta}{\{1+[\beta+\eps^{-1}\om(k)]^2\}\{1+[\beta-\eps^{-1}\om(k)]^2\}}\nonumber\\
&&
\!\!=\eps\int_{
  \bbT}
  dk
\int_{0}^{\om(k)/\eps}\frac{\, \eps^{-1}\om(k) \left(\eps^{-1}\om(k)-\beta\right)\,
   }{1+[2\eps^{-1}\om(k)-\beta]^2} \frac{d\beta}{1+\beta^2}+\eps\int_{
  \bbT}
  dk
\int_{0}^{+\infty}\frac{\, \eps^{-1}\om(k)  \left(\eps^{-1}\om(k)+\beta\right)\,
  }{1+[\beta+2\eps^{-1}\om(k)]^2} \frac{d\beta}{1+\beta^2}.\nonumber
\end{eqnarray}
Using the dominated convergence theorem, we conclude that
\begin{equation}\label{mar1502}
\lim_{\eps\to0}{\frak L}_{scat,22}^\eps(\la)=0.
\end{equation}

%\subsubsection{The case $\om\in C^\infty(\bbT\setminus\{0\})$}

Finally, consider (\ref{020203})-(\ref{mar1504}) when $\om\in C^\infty(\bbT\setminus\{0\})$.
Let $\si>0$ be arbitrary, and take
 $A>0$, to be chosen later.
We can write 
\[
{\frak L}_{scat,22}^\eps(\la)={\frak L}_{scat,22}^{\eps,1}(\la)
+{\frak L}_{scat,22}^{\eps,2}(\la),
\]
where the terms in the right hand
side correspond to the integration over $[k:|k|\le A\eps]$ and its
complement. 
As $\omega$ is Lipschitz, we have
$$
|{\cal G}_\eps(k)|\preceq \int_{\bbR}|\eta|\|\hat
   G^\star(\eta,\cdot)\|_{\infty}d\eta\preceq 1
$$
Using \eqref{040302} we  write
\begin{eqnarray}
\label{020203a}
&&
|{\frak L}_{scat,22}^{\eps,1}(\la)|\preceq \int_{[|k|\le A\eps]
  }
  dk
\int_{0}^{+\infty}\frac{\,\eps^{-1}\om(k)\beta
   d\beta}{\{1+[\beta+\eps^{-1}\om(k)]^2\}\{1+[\beta-\eps^{-1}\om(k)]^2\}} 
\nonumber\\
&&
\le
\int_{[|k|\le A\eps]
  }
  dk
\int_{0}^{+\infty}\frac{\,\eps^{-1}\om(k)\beta
   d\beta}{\{1+\eps^{-1}\om(k)\beta\}\{1+[\beta-\eps^{-1}\om(k)]^2\}}
   \preceq A\eps.
\end{eqnarray}
%Thus, $\lim_{\eps\to0}{\frak L}_{\eps,2}^{2,1}(\la)=0$.
Finally, we write 
\[
{\frak L}_{scat,22}^{\eps,2}(\la) ={\frak L}_{scat,22}^{\eps,21}(\la)+{\frak L}_{scat,22}^{\eps,22}(\la),
\]
corresponding to the partition of the integration domain in $\eta$
into  $[\eta:\,|\eta|<A/4]$ and its complement. In the first case, as $|k|>A\eps$ and $|\eta|<A/4$, we
can still use estimate \eqref{Ge}, hence
%use the fact that $\om\in C^2[0,1/2]$ and $\om\in C^2[-1/2,0]$,
%therefore we can use estimate \eqref{Ge}. 
%As a result
\[
\lim_{\eps\to0}{\frak L}_{scat,22}^{\eps,21}(\la)=0.
\]
In the other case, we can estimate
\begin{eqnarray}
\label{020203b}
&&|{\frak L}_{scat,22}^{\eps,22}(\la)|\preceq 
 \int_{[|\eta|>A/4]}|\eta|\|\hat
   G^\star(\eta,\cdot)\|_{\infty}d\eta\int_{[|k|> A\eps]
  }
  dk
\int_{0}^{+\infty}\frac{\,\eps^{-1}\om(k)\beta
   d\beta}{\{1+\eps^{-1}\om(k)\beta\}\{1+[\beta-\eps^{-1}\om(k)]^2\}}\nonumber\\
%   \preceq A\eps.
&&\preceq \int_{[|\eta|>A/4]}|\eta|\|\hat
   G^\star(\eta,\cdot)\|_{\infty}d\eta\le \si,
\end{eqnarray}
provided that $A$ is sufficiently large. This finishes the proof of \eqref{021511b}, and that of Lemma~\ref{lem-feb1502} as well.
%Thus the proof of
%\eqref{021511b} in the case of $\om\in C^\infty(\bbT\setminus\{0\})$
%has been finished.

\section{End of proof of Theorem \ref{main:thm}}\label{sec:end-proof}

\label{w-s}

In the present section we show Theorem \ref{main:thm} assuming that
the Fourier-Wigner transform
of the initial data  satisfies \eqref{011812aa} rather than the stronger assumption (\ref{011812}). 
Suppose that $\sigma>0$ and~$G\in {\cal S}(\bbR\times\bbT)$  are arbitrary.  Let us decompose the solution of 
\eqref{basic:sde:2aa} as
\[
\hat\psi(t,k)=\hat\psi^1(t,k)+\hat\psi^2(t,k),
\]
where
\begin{equation}
\begin{split}
 \label{basic:sde:2aa1}
 d\hat\psi^1(t,k) &= \Big\{-i\om(k)\hat\psi^1(t,k)
 - \frac{\ga}{2i} \int_{\bbT}[\hat\psi^1(t,k')-(\hat\psi^1(t,k'))^\star]dk'\Big\} dt
 +i\sqrt{2\ga T}dw(t),
 \\
\hat\psi^1(0,k) & =\hat\psi(k)\chi_\delta(k)
\end{split}
 \end{equation}
 and 
\begin{equation}
\begin{split}
 \label{basic:sde:2aa2}
 \frac{d\hat\psi^2(t,k)}{dt} &= -i\om(k)\hat\psi^2(t,k)
 -\frac{\ga}{2i} \int_{\bbT}\left[\hat\psi^2(t,k')-(\hat\psi^2(t,k'))^\star\right]dk' ,
 \\
\hat\psi^2(0,k) & = \hat\psi(k)[1- \chi_\delta(k)],
\end{split}
 \end{equation}
 with $\chi_\delta\in C(\bbT)$ such that $0\le\chi\le 1$,
 $\chi_\delta\equiv 0$ on $L(\delta)$ (see \eqref{L}), $\chi_\delta\equiv 1 $ on $L^c(2\delta)$
and $\delta$ chosen so small that
\begin{equation}
\label{021703-18}
\limsup_{\eps\to0+}\eps\bbE_\eps\|\hat \psi(1-\chi_\delta)\|^2_{L^2(\bbT)}<\sigma.
\end{equation}
Let $\widehat{
  w}_\varepsilon(\la,\eta,k)$ and $\widehat{
  w}_\varepsilon^1(\la,\eta,k)$ be the Laplace transforms of the
Fourier-Wigner functions corresponding to $\hat\psi(t,k)$ and
$\hat\psi^1(t,k)$ via \eqref{eq:20}.  Using estimates \eqref{011703-18}
and \eqref{021703-18}
we see that  
$$
\limsup_{\eps\to0+}\sup_{\eta\in\bbT_{2/\eps}}\int_{\bbT}\left|\widehat{
  w}_\varepsilon(\la,\eta,k)-\widehat{
  w}_\varepsilon^1(\la,\eta,k)\right|dk\preceq \si,\hbox{ for each $\la>0$}.
$$
It follows, in particular, that
\begin{equation}
\label{023110s1}
\limsup_{\eps\to0+}\left|\int_{\bbR\times\bbT}\hat G^*(\eta,k)\widehat{
  w}_\varepsilon(\la,\eta,k)d\eta dk-\int_{\bbR\times\bbT}\hat G^*(\eta,k)\widehat{
  w}_\varepsilon ^1(\la,\eta,k)d\eta dk\right|\preceq\sigma.
\end{equation}
In addition, the initial condition for  
$\hat\psi^1(t,k)$ satisfies assumption (I3') in \eqref{011812}. As we have  
already proved  Theorem \ref{main:thm} under this
hypothesis,
we conclude that
\begin{equation}
\label{023110-1}
\lim_{\eps\to0+}\int_{\bbR\times\bbT}\hat G^*(\eta,k)\widehat{
  w}_\varepsilon^1(\la,\eta,k)d\eta dk=\int_{\bbR\times\bbT}\hat G^*(\eta,k)\widehat{
  w}^1(\la,\eta,k)d\eta dk,
\end{equation}
with
$ \widehat{
  w}^1(\la,\eta,k) $ given by   \eqref{013110}, but with $\widehat
W_0(\eta,k)$ replaced by $\chi_\delta^2(k)\widehat W_0(\eta,k)$. Thus, for a sufficiently
small $\delta>0$ we have
\begin{equation}
\label{023110-2}
\left|\int_{\bbR\times\bbT}\hat G^*(\eta,k)\widehat{
  w}^1(\la,\eta,k)d\eta dk-\int_{\bbR\times\bbT}\hat G^*(\eta,k)\widehat{
  w}(\la,\eta,k)d\eta dk\right|<\sigma.
\end{equation}
We have thus shown that
\begin{equation}
\label{023110s}
\limsup_{\eps\to0+}\left|\int_{\bbR\times\bbT}\hat G^*(\eta,k)\widehat{
  w}_\varepsilon(\la,\eta,k)d\eta dk-\int_{\bbR\times\bbT}\hat G^*(\eta,k)\widehat{
  w}(\la,\eta,k)d\eta dk\right|\preceq\sigma,
\end{equation}
which ends
the proof of Theorem \ref{main:thm}.

\section{The properties of  $\nu(k)$}\label{sec8.3}

In this section, we prove relation (\ref{feb1402}).
The function
$$
\nu(k):=\lim_{\eps\to0}\tilde g(\eps -i\om(k))
$$
can be determined from the identity
$$
\nu(k)\left(1+\ga \lim_{\eps\to0 }\tilde J(\eps -i\om(k))\right)=1.
$$
Recalling (\ref{eq:2}), we write
%\begin{equation}
%\label{fsimple}
%\tilde J(\la)=\int_{\bbT}\frac{\la d\ell}{\la^2+\om^2(\ell)}=\frac12\left\{\int_{\bbT}\frac{ d\ell}{\la+i\om(\ell)}+\int_{\bbT}\frac{ d\ell}{\la-i\om(\ell)}\right\}
%\end{equation}
%implies
\begin{eqnarray*}
&&
 \lim_{\eps\to0}\tilde J(\eps -i\om(k))=\lim_{\eps\to 0}\int_{\bbT}\farc{(\eps-i\omega(k))d\ell}{(\eps-i\omega(k))^2+\omega^2(\ell)}
 =\frac12 \lim_{\eps\to0}
 \int_{\bbT}\frac{ d\ell}{\eps-i\om(k)+i\om(\ell)}\\
&&+ \farc 12\lim_{\eps\to0}\int_{\bbT}\frac{ d\ell}{\eps-i\om(k)-i\om(\ell)}
=\frac{i}{2} \int_{\bbT}\frac{ d\ell}{\om(k)+\om(\ell)}+\frac{i}{2}\lim_{\eps\to0}\int_{\bbT}\frac{ d\ell}{i\eps+\om(k)-\om(\ell)}  .
\end{eqnarray*}
Let us set
\[
G(u):=\farc 12\int_{\bbT}\frac{ d\ell}{u+\om(\ell)}=\int_0^{1/2}\frac{ d\ell}{u+\om(\ell)}=\int_{\omega_{min}}^{\omega_{max}}
\farc{dv}{|\omega'(\omega_+^{-1}(v))|(u+v)},
\]
and
\[
H(u):= \farc 12\lim_{\eps\to0}\int_{\bbT}\frac{ d\ell}{i\eps+u-\om(\ell)}= \lim_{\eps\to0}
\int_{\omega_{min}}^{\omega_{max}}\farc{dv}{|\omega'(\omega_+^{-1}(v))|(i\eps+u-v)},
\]
so that
\begin{equation}\label{mar1524}
\nu(k)=\farc{1}{1+i\gamma(G(\omega(k))+H(\omega(k))}.
\end{equation}
In our situation, with $u=\omega(k)\in(\omega_{min},\omega_{max})$, we  have
\begin{equation}\label{mar1520}
H(\omega(k))=\farc{1}{|\omega'(k)|}\left\{H_0(\omega(k))+
\lim_{\eps\to 0}\int_{\omega_{min}}^{\omega_{max}}\farc{dv}{i\eps+\omega(k)-v}\right\},
\end{equation}
with a continuous, bounded and real-valued function $H_0(u)$. For any $a,b\in\bbR$ and~$c\in(a,b)$, we have
\begin{eqnarray}\label{mar1522}
\int_{a}^{b}\farc{dv}{i\eps+c-v}=\int_{-(b-c)/\eps}^{(c-a)/\eps}\farc{dv}{i+v} = 
\int_{-(b-c)/\eps}^{(c-a)/\eps}\farc{(v-i)dv}{1+v^2}=-i\pi+\log\farc{c-a}{b-c}+o(1). 
\end{eqnarray}
As $G(u)$ and $H_0(u)$ are real-valued, using (\ref{mar1522}) in (\ref{mar1524}) immediately gives
 \begin{equation}\label{mar1526}
\hbox{Re}\,\nu(k)= \Big(1+\frac{\pi\ga}{|\om'(k)|}\Big)|\nu(k)|^2,
\end{equation}
which is  (\ref{feb1402}).
%\blue{But it should be this 
%\begin{equation}
%\label{020104}
%{\rm Re}\,\nu(k)=\left(1+\frac{\pi\ga}{|\om'(k)|}\right)|\nu(k)|^2.
%\end{equation}
%}

\commentout{
%%%%%%%%%%%%%%%%%%%%%%%%%%%%%%%%%%%%%%%%%%%%%%%%%%%

\section{Derivation of the equation on $W(t,x,k)$}

\label{sec6}

Consider first the case $\om'(k)>0$. In order to obtain a closed equation on $W(t,x,k)$ we use equation \eqref{032112d} that describes the Laplace-Fourier
transform of the Wigner function of the microscopic dynamics. We have shown so far, see \eqref{032112e1}, that
the limiting Wigner function $W(t,x,k)$ satisfies the following equation
\begin{align}
\label{032112e2}
&\partial_tW(t,x,k)+ \bar\om'(k) \partial_x W(t,x,k)
= \gamma T\delta_0(x)+ F(t,x,k),\\
&
W(0,x,k)=W_0(x,k),\nonumber
\end{align}
with
\begin{align}
\label{F}
& F(t,x,k):=
 |\bar\om'(k)|\delta_0(x)\left\{-\left(1-|\nu(k)|^2\right)\frac{\gamma T}{ |\bar\om'(k)|}+[p_+(k)-1]W_0\left(-\bar\om'(k)t,k\right)+\vphantom{\int_0^1}p_-(k) W_0\left(\bar\om'(k)t,-k\right)\right\}.
\end{align}
In order to derive the equation  we need to re-express $F(t,x,k)$ using $W(t,x,k)$. For that purpose we use equation \eqref{010304}.
We can write that then
\begin{align}
\label{010304x}
&W\left(t,0^+,k\right)
=\frac{\ga T|\nu(k)|^2}{|\bar\om'(k)|}+p_+(k)W_0\left(-\bar{\om}'(k)t,k\right)
+p_-(k) W_0\left(\bar\om'(k) t,-k\right),\nonumber\\
&W\left(t,0^+,-k\right)=W_0\left(\bar{\om}'(k)t,-k\right),
\\
&W\left(t,0^-,k\right)
=W_0\left(-\bar{\om}'(k)t,k\right) .\nonumber
\end{align}
In the case $\om'(k)<0$ we can write
\begin{align}
\label{010304xy}
&W\left(t,0^-,k\right)
=\frac{\ga T|\nu(k)|^2}{|\bar\om'(k)|}+p_+(k)W_0\left(-\bar{\om}'(k)t,k\right)
+p_-(k) W_0\left(\bar\om'(k) t,-k\right),\nonumber\\
&W\left(t,0^-,-k\right)=W_0\left(\bar{\om}'(k)t,-k\right),
\\
&W\left(t,0^+,k\right)
=W_0\left(-\bar{\om}'(k)t,k\right) \nonumber
\end{align}
and
\begin{align}
\label{010304xxx}
\frac{\ga T}{|\bar\om'(k)|}
=\frac{1}{|\nu(k)|^2} \sum_{\iota=\pm}1_{(0,+\infty)}(\iota\bar\om'(k))\left\{W\left(t,0^\iota,k\right)-p_+(k)W\left(t,0^{-\iota},k\right)
-p_-(k) W\left(t,0^\iota,-k\right)\right\}.
\end{align}
This allows us to rewrite equation \eqref{032112e2} in an alternative form 
\begin{align}
\label{010512ax}
&\partial_tW(t,x,k)+ \bar\om'(k) \partial_x W(t,x,k)
=\ga T\delta_0(x)+ \frac{|\bar\om'(k)|}{|\nu(k)|^2}\delta_0(x) \sum_{\iota=\pm}1_{(0,+\infty)}(\iota\bar\om'(k))\left\{\vphantom{\int_0^1}p_+(k)W(t,0^{-\iota},k)\right.
\nonumber\\
&
\left. \vphantom{\int_0^1} +p_-(k)W(t,0^{\iota},-k)-(1-|\nu(k)|^2)W(t,0^{\iota},k)-|\nu(k)|^2W(t,0^{-\iota},k)\right\} ,\\
&
W(0,x,k)=W_0(x,k).\nonumber
\end{align}
%From the first equations in \eqref{010304x} and \eqref{010304xy} we obtain
%$$
%\sum_{\iota=\pm}1_{(0,+\infty)}(\iota\bar\om'(k))W\left(t,0^\iota,k\right)
%=\frac{\ga T|\nu(k)|^2}{|\bar\om'(k)|}+\sum_{\iota=\pm}1_{(0,+\infty)}(\iota\bar\om'(k))\left[p_+(k)W\left(t,0^\iota,-k\right)
%+p_-(k) W\left(t,0^{-\iota},k\right)\right].
%$$
%Substituting into \eqref{010512} we obtain
%\begin{align}
%\label{010512aa}
%&\partial_tW(t,x,k)+ \bar\om'(k) \partial_x W(t,x,k)
%=T\delta_0(x)+ \frac{|\bar\om'(k)|}{|\nu(k)|^2}\delta_0(x) \sum_{\iota=\pm}1_{(0,+\infty)}(\iota\bar\om'(k))\left\{\vphantom{\int_0^1}p_+(k)W(t,0^{-\iota},k)\right.
%\nonumber\\
%&
%\left. \vphantom{\int_0^1} +p_-(k)W(t,0^{\iota},-k)-(1-|\nu(k)|^2)W(t,0^{\iota},k)-|\nu(k)|^2W(t,0^{-\iota},k)\right\} ,\\
%&
%W(0,x,k)=W_0(x,k).\nonumber
%\end{align}
%

\subsection*{Uniqueness of solution to equations (\ref{010512}) and (\ref{010512ax})}

We show uniqueness of solutions to \eqref{010512}. The argument for the uniqueness of solutions for  (\ref{010512ax}) proceeds along the same lines.
Suppose that $\om'(k)>0$ and consider
$$
V(t,x,k):=W(t,x+\bar\om'(k)t,k).
$$
From \eqref{010512} we obtain
\begin{align*}
&\frac{dV}{dt}(t,x,k)
=T|\bar\om'(k)|\fgeeszett(k)\delta_0(x+\bar\om'(k)t)
\nonumber\\
&
+|\bar\om'(k)|\delta_0(x+\bar\om'(k)t) \sum_{\iota=\pm}1_{(0,+\infty)}(\iota\bar\om'(k))\left\{\vphantom{\int_0^1}[p_+(k)-1]W\left(t,0^{-\iota},k\right) \vphantom{\int_0^1} +p_-(k)W(t,0^{\iota},-k)\right\} ,\\
&
W(0,x,k)=W_0(x,k).\nonumber
\end{align*}
Integrating over $t$ we obtain
\begin{align*}
&
W(t,x+\bar\om'(k)t,k)=W_0(x,k)+T\fgeeszett(k)1_{[-\bar\om'(k)t,0]}(x)\nonumber\\
&
+ 1_{[-\bar\om'(k)t,0]}(x)\left\{[p_+(k)-1]W\left(-\frac{x}{\bar\om'(k)},0^{-},k\right) \vphantom{\int_0^1} +p_-(k)W\left(-\frac{x}{\bar\om'(k)},0^{+},-k\right)\right\},
\end{align*}
or equivalently
\begin{align*}
&
W(t,x,k)=W_0(x-\bar\om'(k)t,k)+T\fgeeszett(k)1_{[0,\bar\om'(k)t]}(x)\nonumber\\
&
+ 1_{[0,\bar\om'(k)t]}(x)\left\{[p_+(k)-1]W\left(t-\frac{x}{\bar\om'(k)},0^{-},k\right) \vphantom{\int_0^1} +p_-(k)W\left(t-\frac{x}{\bar\om'(k)},0^{+},-k\right)\right\},
\end{align*}
Hence,
\begin{align*}
W(t,0^-,k)=W_0(-\bar\om'(k)t,k),\quad W(t,0^+,-k)=W_0(\bar\om'(k)t,-k)
\end{align*}
and, combining the above equalities, we get
 \begin{align*}
&
W(t,x,k)=W_0(x-\bar\om'(k)t,k)+T\fgeeszett(k)1_{[0,\bar\om'(k)t]}(x)\nonumber\\
&
+ 1_{[0,\bar\om'(k)t]}(x)\left\{[p_+(k)-1]W_0\left(x-\bar\om'(k)t,k\right) \vphantom{\int_0^1} +p_-(k)W_0\left(-x+\bar\om'(k)t,-k\right) \right\},
\end{align*}
that is equivalent with \eqref{010304}.

}%%%%%%%%%%%end of commentout

% \noindent
% {Tomasz Komorowski}\\
% { Institute of Mathematics, Polish Academy Of Sciences\\Warsaw, Poland.} 
% {{\footnotesize \tt komorow@hektor.umcs.lublin.pl}}
% \bigskip

% \noindent
% {Stefano Olla\\
% CEREMADE, UMR-CNRS, Universit\'e de Paris Dauphine, PSL Research University}\\
% {\footnotesize Place du Mar\'echal De Lattre De Tassigny, 75016 Paris, France}\\
% {\footnotesize \tt olla@ceremade.dauphine.fr}\\
% \\

% \noindent
% {Lenya Ryzhik}\\
% { Mathematics Department, Stanford University\\Stanford, CA 94305, USA} 
% {{\footnotesize \tt ryzhik@stanford.edu}}
% \bigskip

% \noindent
% {Herbert Spohn}\\
% { Zentrum Mathematik and Physik Department, Technische Universit ̈at Munchen\\
% {\footnotesize Boltzmannstraße 3, 85747 Munich, Germany}} 
% {{\footnotesize \tt spohn@ma.tum.de}}
% \bigskip

\end{document}